\documentclass[amssymb,aps,prli,preprint]{revtex4}
\usepackage{latexsym,amsmath,graphics,graphicx,cmbright,exscale,colordvi}

\begin{document}
\title{Non-collinear magnetism induced by frustration in transition-metal nanostructures deposited on surfaces}

\author{S. Lounis}
\affiliation{Peter Gr\"unberg Institut and Institute for Advanced Simulation, Forschungszentrum J\"ulich \& JARA, 52425 J\"ulich, Germany}
\date{\today}

\begin{abstract}

  How does magnetism behave when the physical dimension is reduced to
  the size of nanostructures? The multiplicity of magnetic states in
  these systems can be very rich, in that their properties depend on
  the atomic species, the cluster size, shape and symmetry or choice
  of the substrate. Small variations of the cluster parameters may
  change the properties dramatically. Research in this field has
  gained much by the many novel experimental methods and techniques
  exhibiting atomic resolution. Here I review the ab-initio approach,
  focusing on recent calculations on magnetic frustration and
  occurrence of non-collinear magnetism in antiferromagnetic
  nanostructures deposited on surfaces.

\end{abstract}

\maketitle

\section{Introduction}

Atomic and nanometer scale magnetism - in short called nanomagnetism -
stands as one of the frontier fields in magnetism. Nanomagnetism opens
on one hand new vistas to magnetic storage media, on the other hand it
is a largely unexplored area of physics where novel effects ought to
be expected. Controlling the flow of magnetic and charge information
between increasingly smaller structures hinges on the meticulous
control of the coupling between spins. Obviously, this is of central
importance for the design of novel devices engineered on the level of
individual atomic spins~\cite{sarma} whose functionality is geared
towards computing speed, storage capacity and energy saving.
Unprecedented opportunities for atomic engineering of future
spintronics and quantum information devices arise thanks to
fundamental explorations of magnetic nanochains and nanostructures
using advanced experimental
methods (see e.g. Refs.~\cite{crommie,mirkovic,rusponi,stroscio,wurth,hirjibehedin,wiebe,wiebe2,
wiebe3,wulfhekel,heinze,loth}).

By means of the scanning tunneling microscope (STM), nanostructures
are built atom by atom on different kind of substrates in controlled
processes, resulting in well-defined magnetic units on the nano-scale.
For example, logic gates based on magnetic nanochains were recently
built~\cite{alex_science} while even the magnetic exchange
interactions between adatoms could be evaluated quantitatively using
STM~\cite{wiebe,wiebe3}. Moreover a recently developed technique, inelastic 
scanning tunneling spectroscopy (ISTS), allowed access of spin excitations with STM. 
Thus values of the magnetic anisotropy energy and magnetic properties of nanostructures down 
to single adatoms are measured experimentally and simulated theoretically (see for example 
Refs.~\cite{hirjibehedin,wulfhekel2,khajetoorians,persson,fernandez-rossier,lorente,lounis_ists}). 
The properties of the magnetic nano-objects
crucially depend on the atom species, the cluster shape and size and
on the substrate material, magnetization and surface orientation.
Therefore, there are numerous properties and effects, the
understanding of which requires an interdisciplinary theoretical
approach by {\it ab-initio} electronic structure methods, simplifying
models, and statistical-mechanical methods, which together with
experiment serve the goal of a description and a qualitative
understanding of magnetic nano-structures (see e.g. Refs.~\cite{smogunov,carbone,
  lazarovitz,mokrousov,lagoute,ebert,mavropoulos,klautau}

Among the manifestations of magnetic complexity, perhaps most striking
is the phenomenon of non-collinear magnetism, i.e., the case when the magnetic
moments of atoms in a system are oriented in different
directions. Far from paramagnetism, which occurs in the limit of
vanishing inter-atomic interactions, here we are faced with
particularly strong nearest-neighbor magnetic-moment coupling,
reaching the order of magnitude of 0.5~eV, with non-collinearity being
the result of competition among interactions. Basically, there are two
types of competition. The first comes from direct antiferromagnetic
exchange, with the competing interactions being of the same order of
magnitude and the non-collinearity arising in a nearest-neighbor length
scale. This effect is usually termed as \emph{frustration} and will be
the main topic of discussion in the present paper. The second is a
competition of direct exchange with anisotropic exchange which arises
from spin-orbit coupling and is typically at least an order of
magnitude weaker. It leads to longer-range manifestation of
non-collinear magnetism, typically on a length-scale of a few
interatomic distances or more. Here we will simply reference works that
have studied the latter, while our focus is on the former.

\begin{figure*}[!h]
\begin{center}
\hspace{-1cm}
\includegraphics*[width=0.3\linewidth]{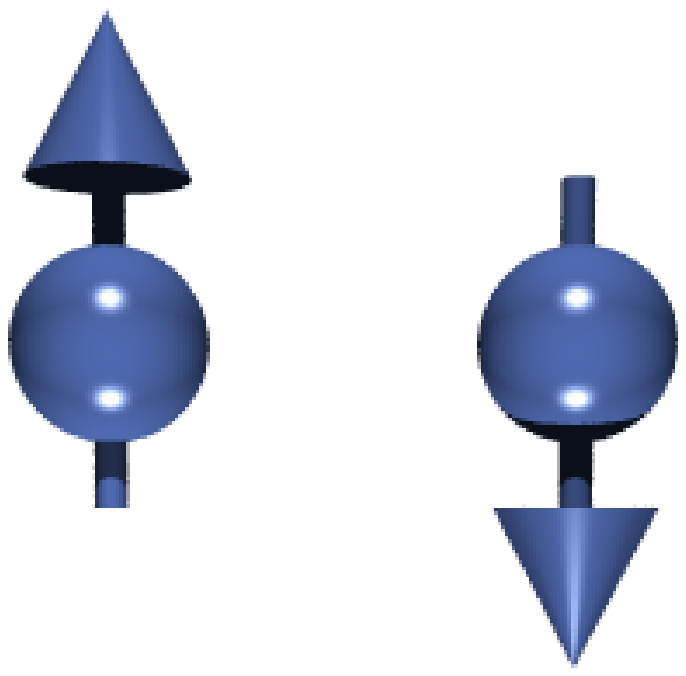}
\hspace{0cm}
\includegraphics*[width=0.2\linewidth]{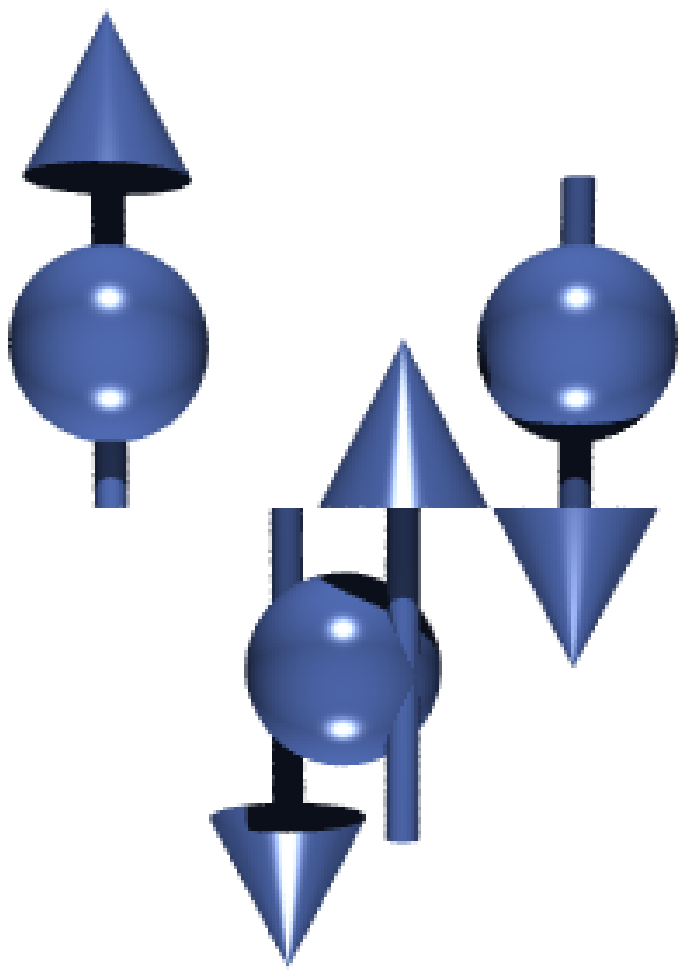}
\hspace{0cm}
\includegraphics*[width=0.3\linewidth]{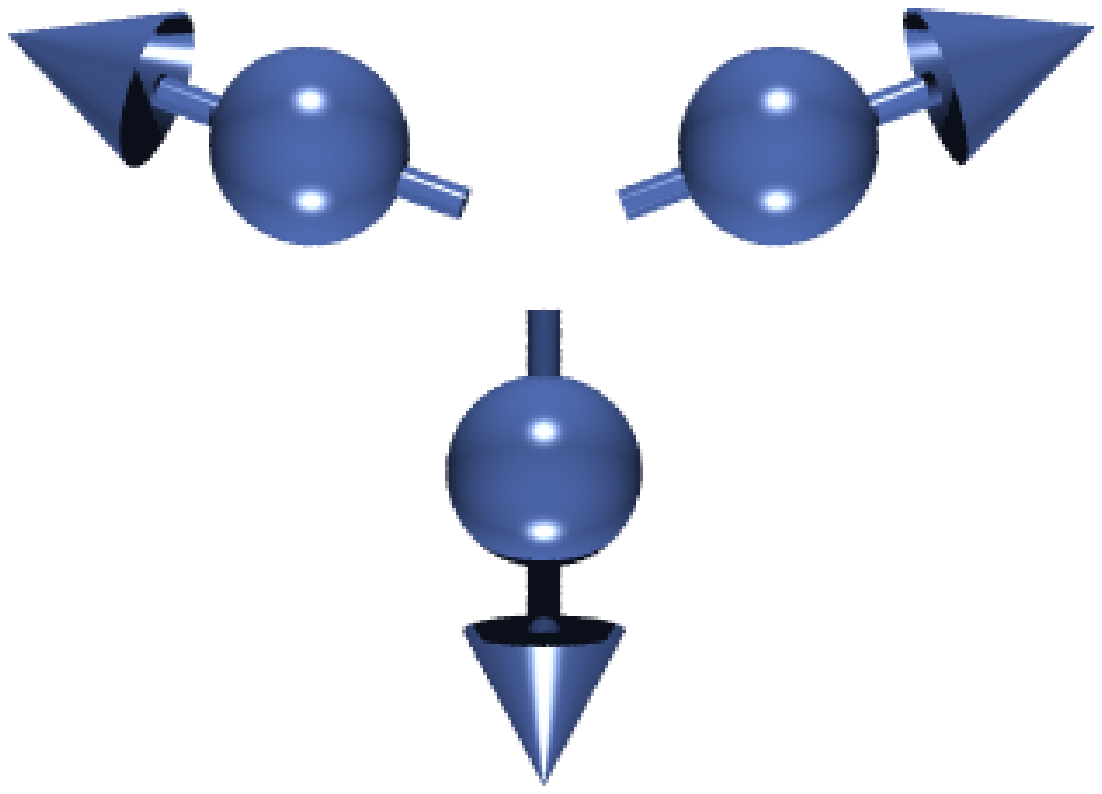}\\
\hspace{-1cm}{\huge A} \ \ \ \ \ \  \ \ \ \ \ \ \ \ \ \ \ \  \ \ \ \ \ \ \ \ \ \ {\huge B}  \ \ \ \ \ \  \ \ \ \ \ \ \ \ \ \ \ \  \ \ \ \ \ \ \ \ \ \ {\huge C}
\caption{Example of frustration seen with an antiferromagnetic Cr
  dimer ({\bf A}) to which we add a third equidistant atom. A magnetically
  frustrated state is obtained ({\bf B}) because all three atoms tend to couple
  antiferromagnetically to each other leading then to a non-collinear
  magnetic structure ({\bf C}) as a ground state.}
\label{exp-frustration}
\end{center}
\end{figure*}

Magnetic frustration denotes the inability to satisfy competing
exchange interactions between neighboring atoms. A simple model for
frustration is the following (see Fig.~\ref{exp-frustration}), based
on the antiferromagnetic (AF) interaction among neighboring Cr atoms.
Starting with an antiferromagnetic (AF) Cr dimer, the addition of a
third Cr atom to form an equilateral triangle leads to a frustrated
geometry. Each atomic moment tends to be aligned AF to both its
neighbors. Since this is impossible, the moments of the three atoms
relax in a state of compromise. The ground state is then
non-collinear, characterized by an angle of 120$^{\circ}$ between each
two atoms. The number of non-collinear solutions that share this
property is infinite, since if all moments are rotated by the same
angle their relative orientation to each other does not change. On the
other hand, interaction with a magnetic substrate of a fixed-moment
orientation stabilizes only one or perhaps a few of these infinitely
many states.

This Neel state is an example of intra-cluster frustration that can
occur in clusters deposited on non-magnetic surfaces with a triangular
symmetry such as fcc(111)
surfaces~\cite{bergman,demangeat,costa,stocks,gotsis,antal,bergman_gold,klautau_pt,nikos}
(e.g. Cu(111), Au(111) or Pt(111)). On magnetic surfaces, however, the 
non-collinear state can be also realized without intra-cluster
frustration if there is a competition between the intra-cluster
interactions, on the one hand, and the cluster-substrate interactions,
on the other~\cite{lounis05,lounis07,lounis08,lounis_prl,wulfhekel}. We call such a situation a
cluster-substrate frustration, which as we shall see can also lead to
complex magnetic behavior.

In fact, it is helpful in general to distinguish between three factors
contributing to the equilibrium magnetic state: \\
(i) the pair
interaction among the atoms in the cluster, \\
(ii) the interaction of
the cluster atoms with the substrate, and \\
(iii) the geometry of the
cluster (which is fixed by the substrate). \\
This separation is
meaningful because the nearest-neighbors exchange interaction is
energetically dominant compared to second, third, etc.~neighbors, and
because in different cluster sizes or shapes the type of pair
interaction (ferro- or antiferromagnetic) does not change
qualitatively. Quantitatively, however, this is only an approximation,
and effects beyond this occur which are computed during the
self-consistent calculations presented in this Highlight.

A first approximation to a description of magnetic frustration
phenomena can be achieved by employing a classical Heisenberg
Hamiltonian of the form
\begin{equation}
H = - \frac{1}{2}\sum_{i \neq j}{J}_{ij}{\vec{e}_i\cdot \vec{e}_j}.
\label{eq}
\end{equation}
Here, $\vec{e}_i$ is a unit vector defining the direction of the
magnetic moment and $i$ and $j$ indicate the magnetically involved
atoms, including the substrate atoms. The sign and strength of the
terms $J_{ij}$ (where the magnitude of the moments has been absorbed)
define the ground state.  Spin-orbit interaction could lead to
non-collinear magnetism and recently it has been shown that
anti-symmetric type of interactions, called the Dzyaloshinskii-Moriya
interactions~\cite{dzialoshinskii} could occur on
surfaces~\cite{heide_nature,heide_prb,szunyogh_prl} or even
nanostructures on surfaces~\cite{ebert_prl,antal,ebert,Bauer}. These kind of
interactions, reviewed in Ref.~\cite{heide_highlight}, can of course easily be included in the
previous Heisenberg Hamiltonian by terms of the form
$\vec{D}_{ij}\cdot \vec{e}_i\times\vec{e}_j$.  Henceforth, however, we
limit our discussion to the physics of finite nanostructures where the
spin-orbit coupling is negligible.

One way to proceed is to derive the values of $J_{ij}$ from
density-functional calculations at a particular (e.g. collinear
ferromagnetic) state \cite{lichtenstein} (see also Refs.\cite{bruno,lounis_jij,szilva}), and then find the energy
minimum using Eq.~(\ref{eq}). This is probably a good approximation
under the condition that the magnitude of the moments does not depend
strongly on the relative direction to the neighboring moments, and
that no higher-order corrections to the energy are necessary. These
conditions are usually met in systems where the direction of the
moments varies in a length-scale of several inter-atomic distances,
for example in the case of spin-orbit-induced non-collinear states, or
in the case of low-energy magnetic excitations. Here, however, we are
faced with strong directional fluctuations between neighboring moments,
and it turns out that the conditions are not met. In addition, as we
shall see, there occur more than one energy minima that are not
reproduced by the Heisenberg model. Therefore one has to proceed by
doing a full self-consistent calculation of the non-collinear state,
using the Heisenberg model only as guideline.

Part of the reason that the Heisenberg Hamiltonian (\ref{eq}) fails is
that we have in mind $3d$ transition elements. These are characterized
by $d$ orbitals localized enough to produce a magnetic moment, but
also delocalized enough to provide strong exchange
interactions. Precisely this delocalization of the orbitals, creating
itinerant states, has as a consequence that the electronic structure
of an atom is affected by the magnetic orientation of its
neighbors. Most susceptible to changes are actually the early and
middle transition elements, e.g. V or Cr, due to their more
delocalized $d$ orbitals compared to the Mn or later elements where
the $d$ orbitals are deeper in the potential well; and also Ni,
because of the low-energy scale of its magnetic moment.

\begin{figure}[h!]
\begin{center}
\includegraphics*[width=.6\linewidth]{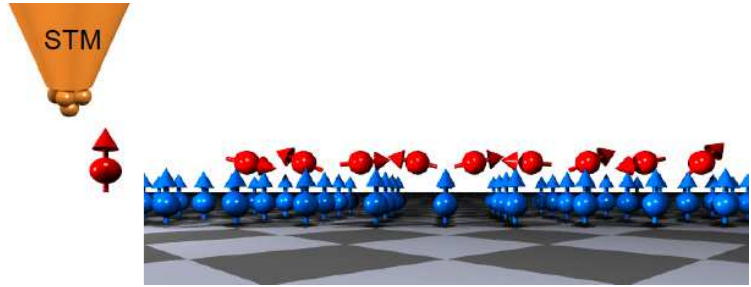}
\caption{A STM tip moving an adatom towards a nanochain.}
\label{stm}
\end{center}
\end{figure}

The correlation function $\chi_{ij}$, i.e., the response of the moment
at site $i$ to a rotation of the moment at $j$, is long-ranged. As a
consequence, adding one magnetic atom at the boundary of a
non-collinear nanostructure can change the whole state. Experimentally
this can be achieved by moving a surface-adsorbed atom by an STM tip,
as shown schematically in Fig.~\ref{stm}. We will see that such
manipulations can lead to interesting even-odd effects, depending on
the size and shape of the nanostructure.

Most of the presented results are on magnetic surfaces, although some
of the work carried out on non-magnetic surfaces is briefly discussed.
The choice of cluster-atoms, substrate materials and surface
orientations is motivated mainly according to points (i)-(iii)
discussed earlier. Basically Cr and Mn are excellent candidates for
the cluster because of their antiferromagnetic nature. Fe and Ni
provide substrates with different strength of exchange interaction
with the cluster. Finally, the fcc(111) and (001) surfaces provide
different geometry types, the former inducing an intra-cluster
frustration due to its triangular geometry, the latter not.

Two of the studied magnetic surfaces have non-triangular symmetry,
thus frustration is induced by the interaction with the magnetic
substrate. These are of fcc(001) type: Ni(001) and
Fe$_{3\mathrm{ML}}$/Cu(001) surface. The former surface provides a
smaller magnetic coupling to the ad-clusters compared to the latter
one. Fe$_{3\mathrm{ML}}$/Cu(001) substrate, known to be ferromagnetic
up to four Fe monolayers~\cite{thomassen,asada,stepanyuk1,moroni}, was
chosen since it was used for x-ray magnetic circular dichroism
measurements on Cr ad-clusters. The third surface is Ni(111) where the
surface geometry is triangular, meaning, in terms of magnetic
coupling, that a compact trimer with antiferromagnetic interactions
resting on the surface necessarily suffers magnetic frustration. This
frustration leads to the well-known non-collinear Neel states being
characterized by 120$^\circ$ angles between the moments. Hence, in
such a system we face an interplay between the non-collinear coupling
tendencies arising from the interaction among the adatoms in the
cluster and the collinear tendencies arising from the additional
coupling to the substrate atoms. This is very different to the Ni(001)
or Fe$_{3\mathrm{ML}}$/Cu(001) surfaces where the frustration and
non-collinear state arises from the competition between the coupling
in the cluster and with the substrate.

\begin{figure}[h!] \begin{center} (a)
    \includegraphics*[width=.4\linewidth]{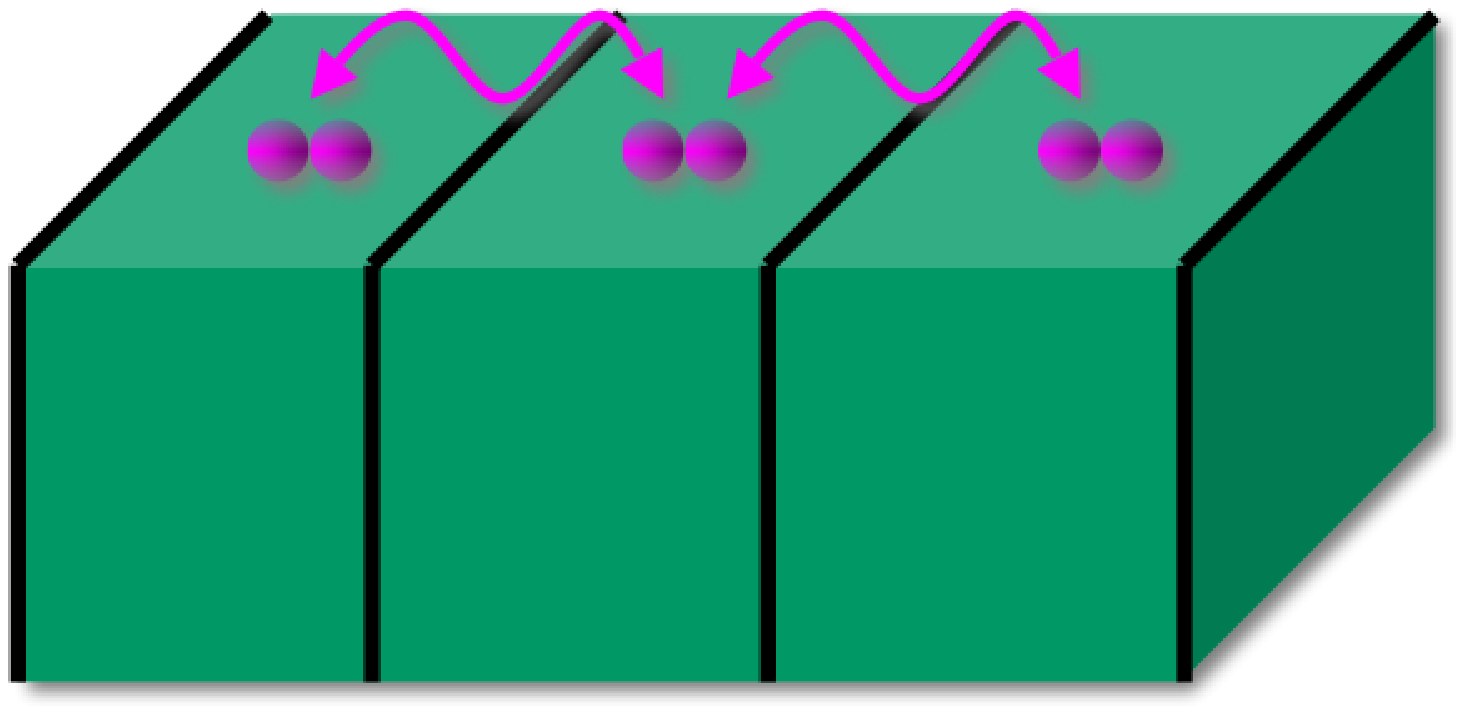} \hspace{-0.1cm}
    (b)\includegraphics*[width=.4\linewidth]{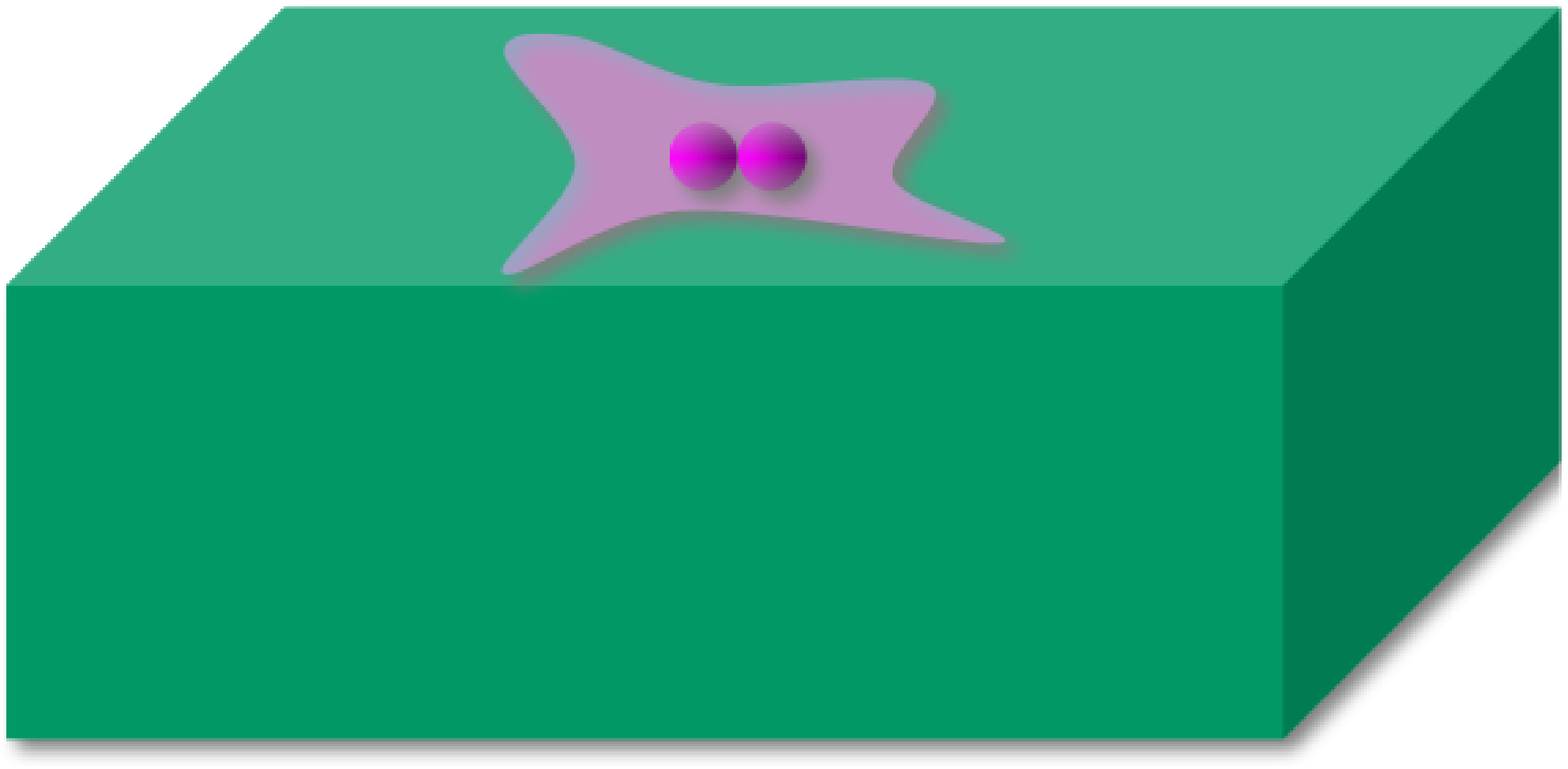}
    \caption{Adatoms on surface. In (a) the supercell approach is
      depicted with arrows showing the spurious interaction between
      nanostructures of adjacent supercells. This unwanted interaction
      is avoided in the real-space KKR formalism (b).} \label{slab}
  \end{center} \end{figure}

The majority of the {\it ab-initio} methods available for the
treatment of non-collinear magnetism make explicit use of Bloch's
theorem and are thus restricted to periodic systems (bulk or films).
Then one needs large supercells to simulate impurities in a given host
(bulk or film) in order to avoid spurious interactions of the
impurities from adjacent supercells. In contrast, the
Korringa-Kohn-Rostoker Green function (KKR) method does not require a
supercell which makes it an ideal tool to nanostructures on surfaces.
Indeed since the method is based on Green functions, a real-space
approach can elegantly be used~\cite{papanikolaou,ebert_highlight}
(see Fig.~\ref{slab}).

First non-collinear calculations by the KKR Green function method,
though not self--consistent, were already performed in 1985. Oswald
{\it et al.}~\cite{oswald0} could show by using the method of
constraints that the exchange interaction between the moments of Mn
and Fe impurity pairs in Cu is in good approximation described by
the $\cos\theta$--dependence of the Heisenberg model.

Sandratskii {\it et al.}~\cite{sandratskii1} and K\"ubler {\it et
al.}~\cite{kuebler1,kuebler2} pioneered the investigation of
non-collinear magnetic structures using self--consistent density
functional theory and investigated the spin spiral of bcc Fe with the
KKR method. Later on, $\Delta$-Fe was a hot topic, and the appearance
of the experimental work of Tsunoda {\it et
al.}~\cite{tsunoda1,tsunoda2} led to the development of other
first--principles methods able to deal with non-collinear magnetism
such as LMTO~\cite{mryasov}, ASW~\cite{knoepfle} and
FLAPW.~\cite{kurz,nordstroem,sjoestedt}

Several papers~\cite{sandratskii2,sandratskii3} describe how symmetry
simplifies the computational effort for the spiral magnetic structures
in the case of perfect periodic systems---this involves the
generalized Bloch theorem. In {\it ab-initio} methods, this principle
is used together with the constrained density functional
theory~\cite{dederichs,grotheer} giving the opportunity of studying
arbitrary magnetic configurations where the orientations of the local
moments are constrained to nonequilibrium directions.

Concerning unsupported clusters, few methods are developed. For example, Oda
{\it et al.}~\cite{oda} developed a plane-wave pseudopotential scheme
for non-collinear magnetic structures. They applied it to small Fe
clusters for which they found non-collinear magnetic structures for
Fe$_{5}$ and linear-shape Fe$_{3}$. This last result was in
contradiction with the work of Hobbs {\it et al.}~\cite{hobbs} who
found only a collinear ferromagnetic configuration using a projector
augmented-wave method. Small Cr clusters were found magnetically
non-collinear,~\cite{oda} as shown also by Kohl and Bertsch~\cite{kohl}
using a relativistic nonlocal pseudopotential method.

One main result of Oda {\it et al.}~\cite{oda} and Hobbs {\it et
al.}~\cite{hobbs} concerns the variation of the magnetization density
with the position. The spin direction changes in the interstitial
region between the atoms where the charge and magnetization densities
are small, while the magnetization is practically collinear within the
atomic spheres. This supports the use of a single spin direction for
each atomic sphere as an approximation in order to accelerate the
computation; this approximation is followed also here.

\section{Theory: Non-collinear KKR formalism}
The KKR method uses multiple-scattering theory in order to determine
the one-electron Green function in a mixed site and angular--momentum
representation. In the simple case of collinear magnetism, 
the retarded Green function is spin-diagonal,
$G=\text{diag}(G_{\uparrow},G_{\downarrow})$, and is expanded as:
\begin{equation}
{G}_s(\vec{R}_n+\vec{r},\vec{R}_{n'}+\vec{r}';E) = 
-i \sqrt{E} \sum_{L}{R}_{Ls}^n({\vec r_<};E){H}_{Ls}^n({\vec r_>};E)\delta_{nn'}
+\sum_{LL'}{R}_{Ls}^n({\vec r};E) {G}_{s;LL'}^{nn'}(E)
{R}_{L'}^{n'}({\vec r'};E)
\label{eq:1}
\end{equation}
Here, $E$ is the energy and $\vec{R}_{n}$, $\vec{R}_{n'}$ refer
to the atomic nuclei positions. By $\vec{r}_<$ and $\vec{r}_>$ we
denote respectively the shorter and longer of the vectors $\vec{r}$
and $\vec{r}'$ which define the position in each Wigner--Seitz cell
relative to the position $\vec{R}_{n}$ or $\vec{R}_{n'}$.  
The wavefunctions ${R}_{sL}^n(\vec{r};E)$ and
${H}_{sL}^n(\vec{r};E)$ are, respectively, the regular and irregular
solutions of the Schr\"odinger equation for the potential $V_{sn}$ at
site $n$, being embedded in free space; $L=(l,m)$ is a combined index
for angular momentum quantum numbers; $l$ is truncated at a maximum
value of $l_{\mathrm{max}}$. The first term on the RHS of
Equation~(\ref{eq:1}) is the so-called \emph{single site scattering
term}, which describes the behavior of an atom $n$ in free space. All
multiple-scattering information is contained in the second
\emph{back-scattering} term via the structural Green functions
${G}_{s;LL'}^{nn'}(E)$ which are obtained by solving the algebraic Dyson
equation:
\begin{equation}
{G}_{s;LL'}^{nn'}(E) = \mathring{G}^{nn'}_{s;LL'}(E) 
+ \sum_{n'',L''L'''}\!\!\!\!\! \mathring{G}_{s;LL''}^{nn''}(E) 
{\Delta t}_{s;L''L'''}^{n''}(E) {G}_{s;L'''L'}^{n''n'}(E)
\label{eq:2}
\end{equation}
Equation~(\ref{eq:2}) follows directly from the usual Dyson eq.~of the
form $G_s=\mathring{G}_s+\mathring{G}_s\,\Delta V_s\,G_s$, with $\Delta V_s$ the 
perturbation in the
potential for spin $s$ and $\mathring{G}_s$ the Green function for some
already solved reference system. The 
summation in (\ref{eq:2}) is over all lattice sites
$n''$ and angular momenta $L''$ for which the perturbation ${\Delta
t}_{L''L'''}^{n''}(E)={t}_{L''L'''}^{n''}(E) -\mathring{t}_{L''L'''}^{n''}(E)
$ between the ${t}$ matrices of the real and the reference system is
significant (the $t$-matrix gives the scattering amplitude of the
atomic potential). The quantities $\mathring{G}_{s;LL'}^{nn'}(E)$ are the
structural Green functions of the reference system. For the
calculation of a crystal bulk or surface, the reference system can be
free space, or, within the screened KKR formulation \cite{SKKR}, a
system of periodically arrayed repulsive potentials. After the host
(bulk or surface) Green function is found, it can be used in a second
step as a reference for the calculation in real space of the Green function of an
impurity or a cluster of impurities embedded in the host.

The algebraic Dyson equation (\ref{eq:2}) is solved by 
matrix inversion, as we will
see later on in Equation~(\ref{eq:15}). In case of spin-dependent
electronic structure, spin indices enter in the $t$-matrix, the Green
functions and in Eq.~(\ref{eq:2}). Especially in the case of
non-collinear magnetism, these quantities become $2\times 2$ matrices
in spin space, denoted by $\boldsymbol{t}$ and $\boldsymbol{G}$ (see for example Refs.~\cite{lounis05} or~\cite{mertig}).

Once the spin-dependent Green function is known, all physical
properties can be derived from it. In particular, the charge density
$n(\vec{r})$ and spin density $\vec{m}(\vec{r})$ are given
by an integration of the imaginary part of $\boldsymbol{G}$ up to the Fermi level
$E_F$ and a trace over spin indices $s$ (putting the Green
function in a matrix form in spin space):
\begin{eqnarray}
n(\vec{r})&=&
-\frac{1}{\pi}\mathrm{Im}\mathrm{Tr}_{s}
\int^{E_F}\boldsymbol{G}(\vec{r},\vec{r};E)\,dE
\label{eq:2.5} \\
\vec{m}(\vec{r})&=&
-\frac{1}{\pi}\mathrm{Im}\mathrm{Tr}_{s}
\int^{E_F}\vec{\boldsymbol{\sigma}}\,\boldsymbol{G}(\vec{r},\vec{r};E)\,dE.
\label{eq:2.6} 
\end{eqnarray}
Here, $\vec{\boldsymbol{\sigma}}=(\boldsymbol{\sigma}_x,\boldsymbol{\sigma}_y,\boldsymbol{\sigma}_z)$
are the Pauli matrices.

The basic difference between non-collinear and collinear magnetism is
the absence of a natural spin quantization axis common to the whole
crystal. The density matrix is not anymore diagonal in spin space as
in the case of collinear magnetism. Instead, in any fixed frame of
reference it has the form
\begin{equation}
\boldsymbol{\rho}(\vec{r})=
\begin{bmatrix}
{\rho}_{\uparrow\uparrow}(\vec{r}) & {\rho}_{\uparrow\downarrow}(\vec{r}) \\
{\rho}_{\downarrow\uparrow}(\vec{r}) & {\rho}_{\downarrow\downarrow}(\vec{r})
\end{bmatrix}
= 
\frac{1}{2}\left[n(\vec{r}) + \vec{\boldsymbol{\sigma}} \cdot  \vec{m}(\vec{r})\right]
\label{eq:3}
\end{equation}
At any \emph{particular point} in space, of course, a \emph{local}
frame of reference can be found in which $\boldsymbol{\rho}$ is diagonal, but
this local frame can change from point to point.

The KKR Green function {\it ansatz} for non-collinear magnetism is
analogous to (\ref{eq:1}), but including non-spin-diagonal elements
$ss'$~\cite{lounis05}. A simplification is achieved by an approximation to the
exchange-correlation potential which is assumed to be collinear within
each atomic cell [by averaging the direction of the non-collinear
exchange-correlation potential $\vec{B}_{\rm xc}(\vec{r})$],
accelerating computational time of the single-site solutions and
reducing the number of iterations. Then for each cell we define a
local reference frame with respect to which the local solutions of the
Schr\"odinger equation and the $t$-matrix, $\boldsymbol{t}^{\rm
  loc}_n$, are spin-diagonal. After the local Schr\"odinger equation
is solved, the $t$-matrix of each atom is rotated in spin-space to a
pre-defined global frame by a site-dependent transformation in spin
space, $\boldsymbol{t}^{\rm
  glob}_n=\boldsymbol{U}_n\boldsymbol{t}^{\rm loc}_n
\boldsymbol{U}_n^\dagger$. The resulting matrix $t_{n;ss'}$ is not any
more spin-diagonal (but always site-diagonal), with the non-diagonal
terms containing the information on spin-flip scattering by the atomic
potential. From $\boldsymbol{t}^{\rm glob}_n$, and from the reference-system
structural Green function, we calculate, just as in the collinear
case, the structural Green function of the perturbed system by solving
the algebraic Dyson equation, where now all objects are matrices in
terms of site, angular momentum, and spin index: \begin{equation}
  \boldsymbol{G}_{\mathrm{str}}(E) =
  \mathring{\boldsymbol{G}}_{\mathrm{str}}(E)[1 - \Delta
  \boldsymbol{t}^{\mathrm{\text{glob}}}(E)\,
  \mathring{\boldsymbol{G}}_{\mathrm{str}}(E)]^{-1}. \label{eq:15}
\end{equation}

In order to obtain the output charge- and spin-density, the local
wavefunctions ${R}_{sL}^n(\vec{r};E)$ and ${H}_{sL}^n(\vec{r};E)$ are
also projected to the global frame using the projection matrices
$\boldsymbol{\sigma}_{ns}$ for the local spin--up ($\uparrow$) and
spin--down ($\downarrow$) directions: \begin{equation}
  \boldsymbol{\sigma}_{ns} = \frac{1}{2} \boldsymbol{U}_n
  (\boldsymbol{1} \pm \boldsymbol{\sigma}_z) \boldsymbol{U}^{\dag}_n =
  (\boldsymbol{\sigma}_{ns})^2 \ \ \ (\mbox{$+$ for $s=\uparrow$, $-$
    for $s=\downarrow$}) \label{eq:12} \end{equation}
Then we have:
\begin{eqnarray}
\boldsymbol{G}^{\text{glob}}(\vec{R}_n+{\vec r},\vec{R}_{n'}+{\vec r'};E) =
-i \sqrt{E}\sum_{Ls}
{R}^{\text{loc}}_{nLs}({\vec r_<};E)
{H}^{\text{loc}}_{nLs}({\vec r_>};E)
\boldsymbol{\sigma}_{ns}
\nonumber\\
+\sum_{LL'ss'}
{R}^{\text{loc}}_{nLs}({\vec r};E) 
\boldsymbol{\sigma}_{ns} 
\boldsymbol{G}^{\text{glob}}_{LL'nn'}(E)
\boldsymbol{\sigma}_{n's'} 
{R}^{\text{loc}}_{n'L's'}({\vec r'};E).
\label{eq:20}
\end{eqnarray}

At the end, given the spin-density, an average is made
in order to define the new site-dependent local axis
$(\theta_n,\phi_n)$ with respect to the global reference frame:
\begin{equation}
{\tan} {\theta}_{n}= \frac
{\int_{\mathrm{WS}}{m}^z_{n}({\vec r}) d{\vec r}}
{\int_{\mathrm{WS}}{m}_{n}({\vec r}) d{\vec r} } 
, \ \ \ 
{\tan} {\phi}_{n} = \frac
{\int_{\mathrm{WS}}{m}^y_{n}({\vec r}) d{\vec r}}
{\int_{\mathrm{WS}}{m}^x_{n}({\vec r}) d{\vec r}} 
.
\label{eq:25}
\end{equation}
In order to find the output exchange-correlation potential within the local
spin-density approximation, the spin density of each atom is projected
on its local-frame direction $(\theta_n,\phi_n)$ and the
self-consistency cycle is repeated in the usual density-functional
theory sense.

\section{3{\it d} single adatoms and inatoms}

In order to understand the behavior of complex nanostructures it is
necessary to investigate their building block that are adatoms and
inatoms (i.e., impurity atoms in the first surface layer). Here
we would like to review the behavior of 3$d$ adatoms on the three
chosen ferromagnetic surfaces, Ni(001) (see for example
Refs.~\cite{nonas,nonas1,nonas2,nonas3,lounis05}), 
Fe$_{3\mathrm{ML}}$/Cu(001)~\cite{lounis08}
and Ni(111)~\cite{lounis07}.

\begin{figure}[ht!]
\begin{center}
\includegraphics*[width=.6\linewidth]{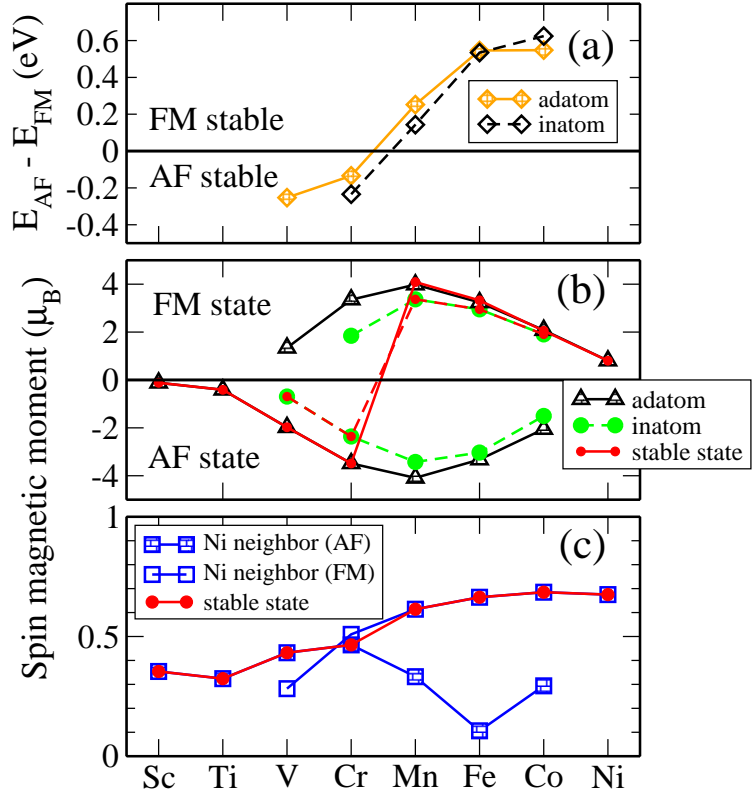}
\caption{3d adatoms and inatoms on Ni(001): (a) Energy difference between the
AF and FM solutions. The values related to adatom and inatoms are described by 
respectively full and empty black diamonds. (b) 
Magnetic moments of the adatoms (black triangles) and inatoms (green circles) 
within the two possible magnetic configurations FM and AF. (c)  
Variation of the 
magnetic moments of Ni nearest neighbors of the adatoms. Adapted with permission from Ref.\cite{lounis05}.}
\label{adatom-energy}
\end{center}
\end{figure}

By comparing the energies of the FM solution, where the adatom moment
is parallel to the surface-atom moments, with the AF solution, where the 
relative orientation is of antiferromagnetic type, we find the first elements of 
the 3d series (Sc, Ti, V, Cr) are AF whereas Mn, Fe, Co and Ni are FM. 
This is is shown in Fig.~\ref{adatom-energy}(a) where the energy difference between the 
AF and FM solutions is plotted.

\begin{figure*}[!ht]
\begin{center}
\includegraphics*[angle=270,width=\linewidth]{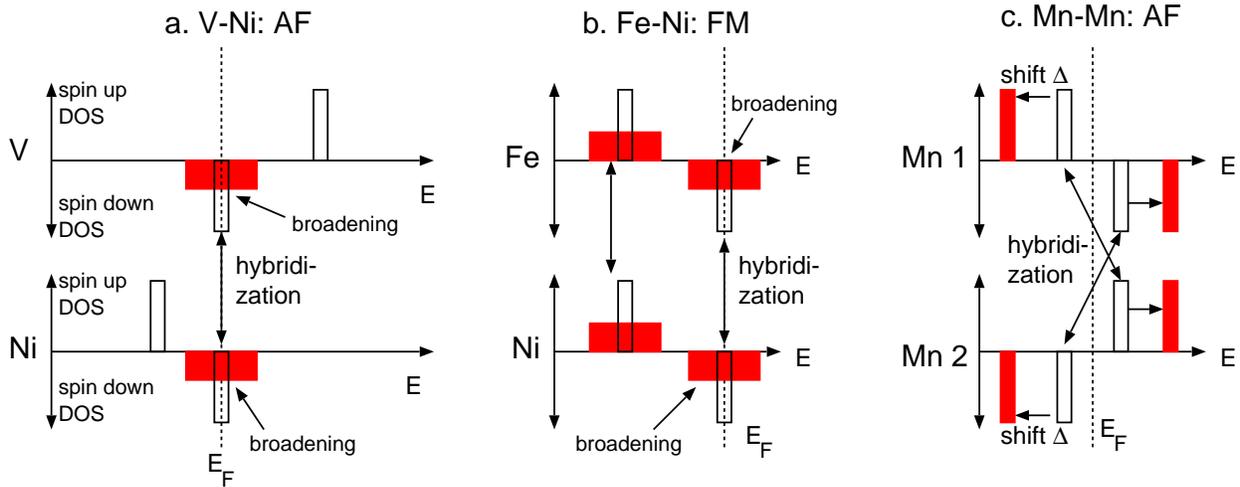}
\caption{Alexander--Anderson model for neighboring magnetic atoms: (a) Early 
3{\it d} transition elements in interaction with Ni surface atoms
(double exchange); (b) Late 
3{\it d} transition elements in interaction with Ni surface atoms
(double exchange); (c) Cr or Mn dimer (kinetic exchange).  Adapted with permission from Ref.\cite{lounis05}.}
\label{alexander-anderson}
\end{center}
\end{figure*}

Clearly, the AF-FM transition occurs when the adatom atomic number
changes from Cr ($Z=24$) to Mn ($Z=25$).~\cite{lounis05} This transition
can be interpreted as in the case of the interatomic interaction of
magnetic dimers~\cite{anderson,oswald}, in terms of the energy gain
due to the formation of hybrid states with the Ni substrate as the
$3d$ virtual bound state (VBS) comes lower in energy with increasing
$Z$  (see Fig.~\ref{alexander-anderson}). Energy is gained when a
half-occupied {\it d} VBS at $E_F$ is broadened
by hybridization with the Ni minority 3{\it d} states, which lie at
$E_F$ (the Ni majority {\it d} states are fully occupied and
positioned below $E_F$). This mechanism is called \emph{double
  exchange} (the term is borrowed from the magnetism of
transition-element impurities in oxides, since the mechanism is
similar). For the early 3{\it d} adatoms
(Fig.~\ref{alexander-anderson}a), it is the majority {\it d} VBS which
is at $E_F$, thus the majority--spin direction of the adatom is
favorably aligned with the minority--spin direction of Ni, and an AF
coupling arises. For the late 3{\it d} adatoms
(Fig.~\ref{alexander-anderson}b), on the contrary, the minority {\it
  d} VBS is at $E_F$, and this aligns with the Ni minority {\it d}
states; then a FM coupling arises. For our purposes we keep in mind
that, since Cr and Mn are in the intermediate region, {\it i.e.}, near
the AF-FM transition point, their magnetic coupling to the Ni
substrate is weak; this has consequences to be seen in the behavior
of dimers, trimers, etc., in the next sections.

We should also stress the importance of \emph{kinetic exchange}, which
produces antiferromagnetic coupling, and occurs when occupied states
of one atom hybridize with unoccupied states of its neighbor. This
situation, demonstrated in Fig.~\ref{alexander-anderson}, leads to a
down-shifting of the occupied levels, gaining energy.  Contrary to
this, a parallel alignment does not lower the energy, since there is
no level shifting, but only level broadening of majority VBS. Since
these are fully occupied, the broadening brings no energy gain.  This
is the reason that Cr and Mn neighboring atoms couple
antiferromagnetically~\cite{anderson,oswald}.

The magnetic moments of the adatoms and Ni nearest neighbors in the
surface layer are shown in Fig.~\ref{adatom-energy}(b) and (c).
Evidently the moment of the Ni nearest neighbors is strongly affected
by the adatoms. Especially for Mn, Fe, and Co adatoms, where the FM
configuration is stable, the Ni moment is strongly in the AF state. As
regards the adatom moments, the half filled $d$ VBS, together with
Hund's rule, cause the Mn adatom to carry the highest magnetic moment
(4.09 $\mu_B$) followed by Cr (3.48 $\mu_B$) and Fe (3.24$\mu_B$).

To understand the effect of coordination and stronger hybridization on
the magnetic behavior of the adatoms, we take the case of impurities
sitting in the first surface layer (inatoms). We carried out the
calculations for V, Cr, Mn, Fe and Co impurities. The corresponding
spin moments are shown in Fig.~\ref{adatom-energy}b (green circles),
and the FM-AF energy differences are shown in
Fig.~\ref{adatom-energy}a (open diamonds and dashed line).  Compared
to the adatom case, the spin moments are reduced, especially for V and
Cr. This effect is expected due to the increase of the coordination
number from 4 to 8 and the subsequent stronger hybridization of the
impurity 3{\it d} levels with the host wavefunctions, especially for V
and Cr where the $d$-levels are more extended. Moreover, the energy
difference $\Delta E$ between the AF and FM solutions is affected, but
in a non-uniform way. The energy trend has two origins. First, one has
stronger total coupling simply due to the increased number of
neighbors.  Second, the interaction to each neighbor changes because
the reduction of the local magnetic moment $M$ is accompanied by a
reduction of the exchange splitting $\Delta E_X$ as $\Delta E_X
\approx I \cdot M$, where $I \approx$~1eV is the intra-atomic exchange
integral.  This means that, for the inatom, the occupied 3{\it d}
states are closer to $E_F$ than for the adatom. In turn, this
intensifies the hybridization of these states with the Ni 3{\it d}
states (which are close to $E_F$). The hybridization-induced level
shift or broadening (see Fig.\ref{alexander-anderson}) increases, and
the coupling energy is affected. Depending on the position of the VBS,
this effect can have the same or opposite sign compared to the effect
of more neighbors. Thus, in Cr we have a strengthening of the AF
coupling, while in Mn we have a competition leading to the weakening
of the FM coupling of Mn inatom compared to the adatom. Similarly, the
stronger hybridization of the Co--inatom {\it d}--states stabilizes
even more its FM configuration due to the energy gain from the
broadening of the {\it d} virtual bound state.

We turn now to the Fe$_{3\mathrm{ML}}$/Cu(001) substrate, known to be
ferromagnetic~\cite{thomassen,asada,stepanyuk1,moroni}. The motivation
comes partly from experiments carried out using for x-ray magnetic
circular dichroism measurements on Cr ad-clusters \cite{lounis07}. Also
it has the advantage that it keeps the fcc structure, so that the
adatoms can be placed in nearest-neighbor positions, while for
example adatoms on Fe bcc(001) would be placed in second-nearest
neighbor positions. Finally, Fe$_{3\mathrm{ML}}$/Cu(001) is expected
to exert a much stronger exchange coupling on adatoms compared to the
Ni(001) surface~\cite{lounis08}. However, for Mn which is at the edge
between FM and AF coupling, the net result is a (weak) AF coupling to
the substrate, contrary to the weak FM coupling obtained on Ni
surface. The difference in energy ($\Delta E_{AF} - E_{FM} = -49$~meV)
is in the same order of magnitude with previously published
results~\cite{stepanyuk1} ($\Delta E_{AF} - E_{FM} = -34$~meV) . As
regards the adatom moments, due to its half filled $d$ VBS the Mn
adatom carries the highest magnetic moment (3.81 $\mu_B$) followed by
Cr (3.30 $\mu_B$) and Fe (2.95$\mu_B$).

Cr adatoms are antiferromagnetically (AF) coupled to the
Fe$_{3\mathrm{ML}}$/Cu(001) substrate, as on Ni(001), but on a four
times stronger energy scale of $\Delta E_{FM-AF} = 565$~meV. This
shows the strength of the interaction on Fe. The preference of the
antiferromagnetic configuration shows up also in a considerably larger
AF Cr-moment of $3.30 \mu_B$ compared to the metastable ferromagnetic
configuration ($2.80 \mu_B$).

Changing the substrate geometry to triangular, we examine adatoms on
the Ni(111) surface. The adatoms have three first neighboring atoms
instead of four on fcc(001) surfaces. Our calculations show that the
single Cr adatom is AF coupled to the surface with an increase of the
magnetic moments ($M_{\mathrm{AF}}$ = 3.77 $\mu_B$ and
$M_{\mathrm{FM}}$ = 3.70 $\mu_B$) compared to the results obtained for
Ni(001)~\cite{lounis07}.  This increase arises from the weaker
hybridization of the $3d$ wavefunctions with the substrate---the
adatom has three neighbors on the (111) surface and four on the
(001).  The calculated energy difference between the FM and AF
configurations is high enough that the AF configuration is stable at
room temperature ($\Delta E_{\mathrm{AF}-\mathrm{FM}}=-94$~meV,
corresponding to 1085~K). Also our results for the Mn adatom on
Ni(111) are similar to what we found on Ni(001). The single Mn adatom
prefers to couple ferromagnetically to the substrate at an energy
scale of $\Delta E_{\mathrm{AF}-\mathrm{FM}}=208$~meV~\cite{lounis07}.
For the (001) surfaces the energy differences are both for Cr and Mn
larger, roughly scaling with the coordination number ($\Delta
E_{\mathrm{AF}-\mathrm{FM}}^{\mathrm{Cr}}=-134$~meV, $\Delta
E_{\mathrm{AF}-\mathrm{FM}}^{\mathrm{Mn}}=252$~meV). The magnetic
moment of Mn is high and reach a value of 4.17 $\mu_B$ for the FM
configuration and 4.25 $\mu_B$ for the AF configuration. The moments
are higher than for the Mn adatoms on Ni(001) ($M_{\text{AF}}$ = 4.09
$\mu_B$ and $M_{\text{FM}}$ = 3.92 $\mu_B$), again due to the lower
coordination and hybridization of the $3d$ levels.

\section{Dimers}

Having established the single adatom behavior, we turn to adatom
dimers, where frustration effects can be already witnessed. Here we
discuss only the most interesting case, that is when the two adatoms
are nearest neighbors and antiferromagnetically coupled to each
other.  In this situation, the interaction is strong enough to allow a
frustration and thus non-collinear magnetism~\cite{lounis05} either in
the presence of ferromagnetic or antiferromagnetic coupling to the
substrate.

\begin{figure}[h!]  \begin{center} (a) \includegraphics*[width=0.2\linewidth]{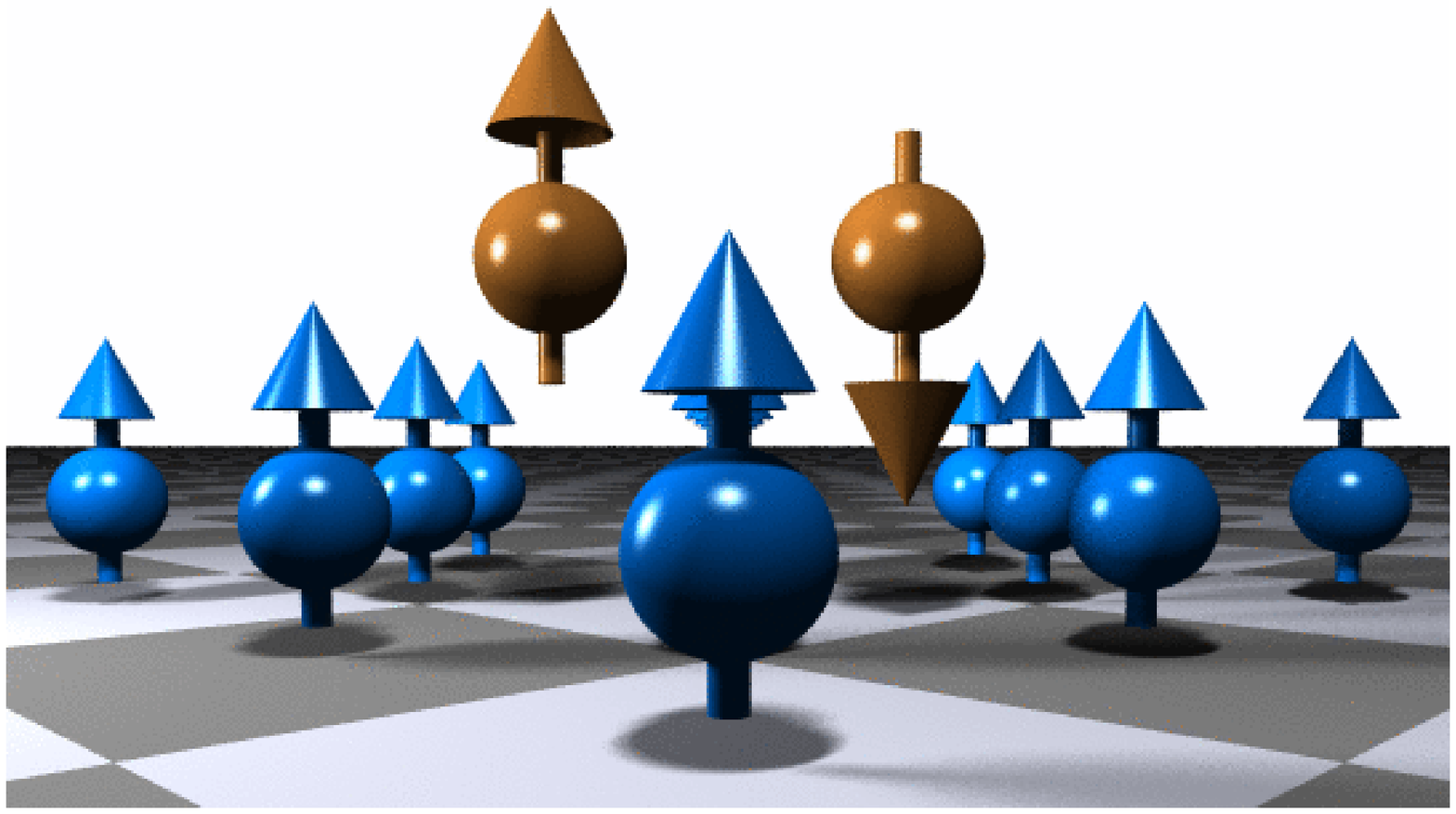} \hspace{-0.1cm} (b)
    \includegraphics*[width=0.2\linewidth]{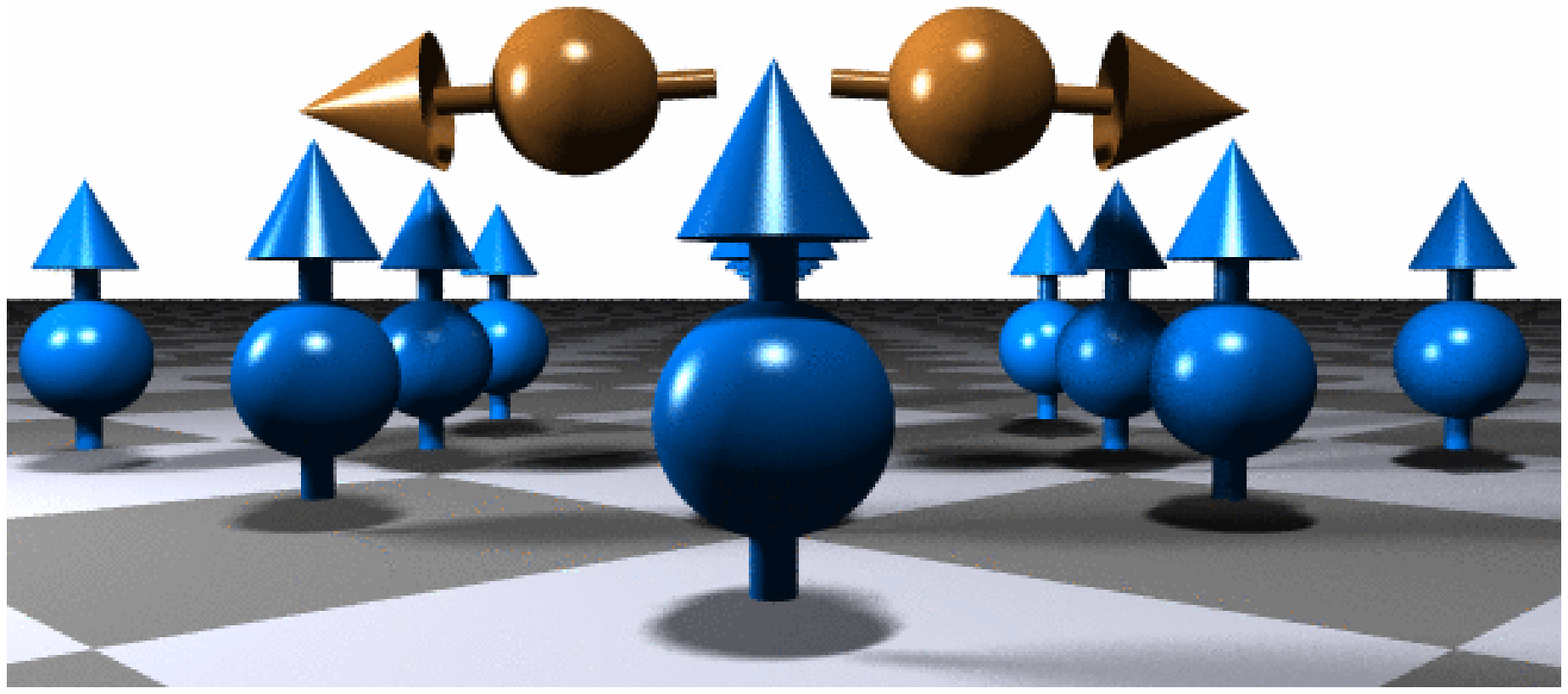}\hspace{-0.1cm}    (c)
    \includegraphics*[width=0.2\linewidth]{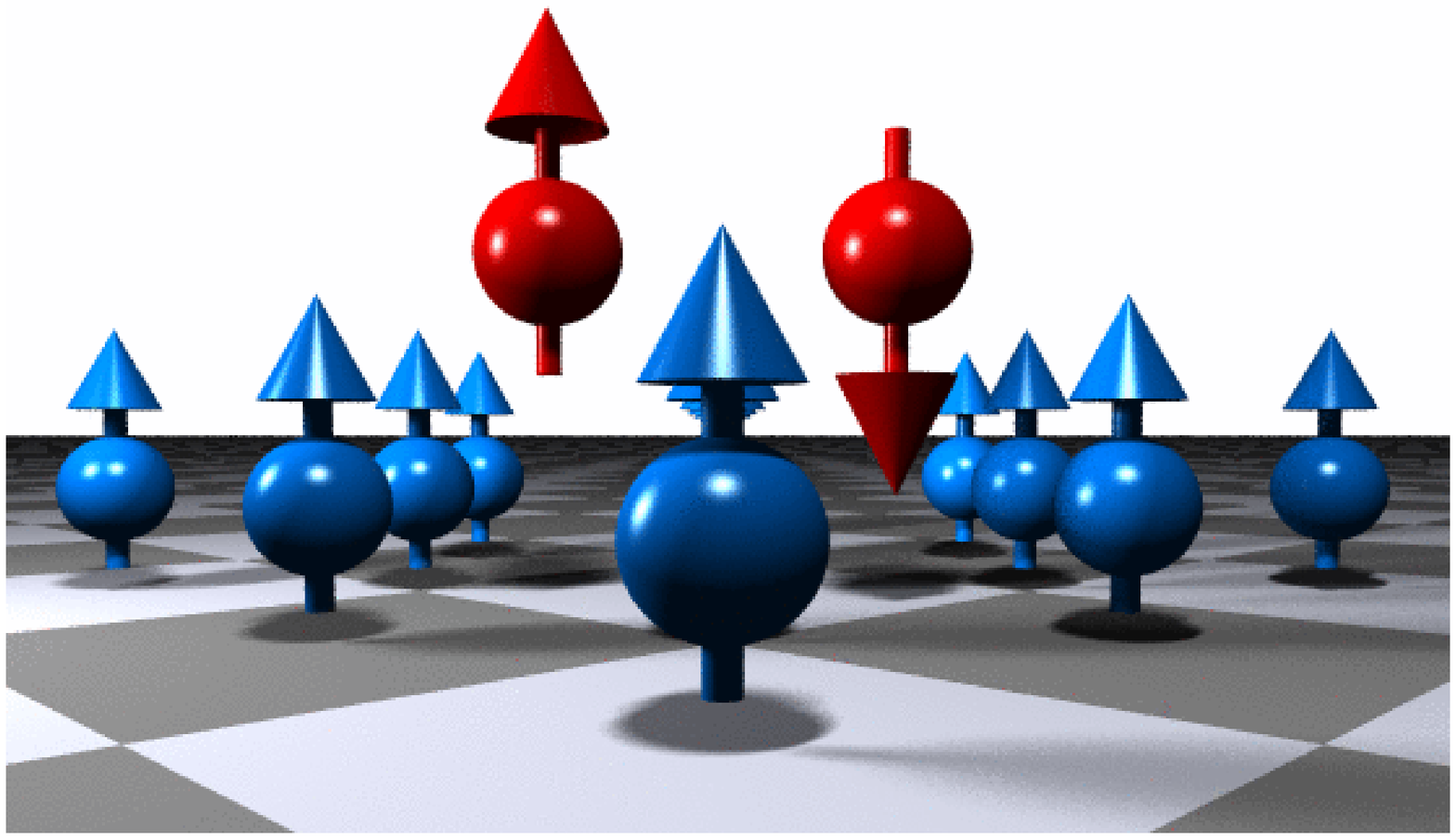}
    \hspace{-0.1cm} (d)
    \includegraphics*[width=0.2\linewidth]{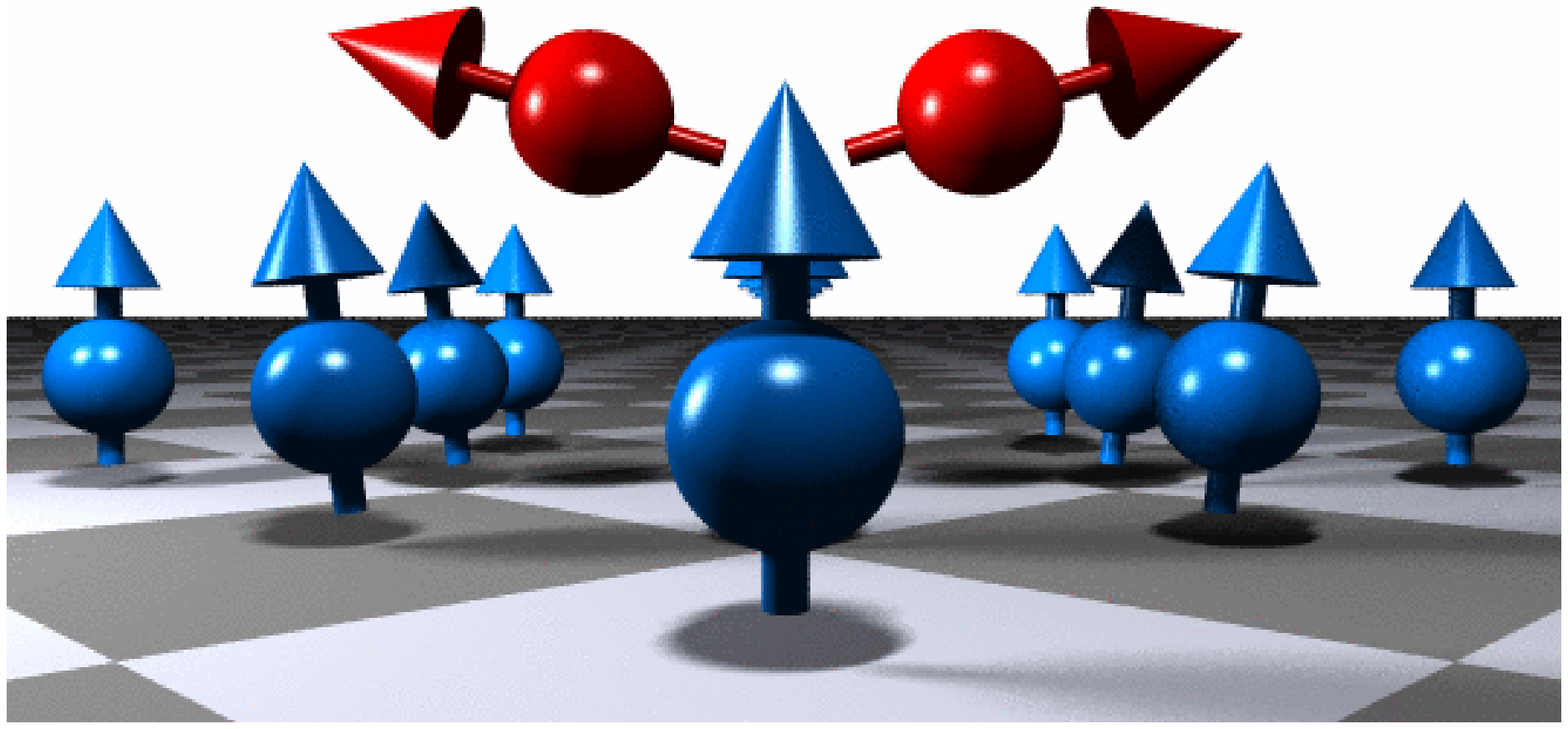}
    \caption{Most stable configurations of Cr/Mn dimer obtained with
      (a/c) the collinear KKR method and (b/c) the non-collinear KKR
      method. The rotation angle with respect to the $z$ axis is equal
      to 94.2$^{\circ}$/72.6$^{\circ}$. While the non-collinear state
      is the ground state for Mn-dimer it is only a local minimum for
      Cr-dimer. Atoms in blue correspond to the Ni substrate. Adapted with permission  
from Ref.\cite{lounis05}.}
\label{cr-dimer-ncol}
\end{center}
\end{figure}

The starting, frustrated collinear configuration is the ferrimagnetic
(FI) state, to be compared to the non-collinear configurations for Cr
and Mn dimers on Ni(001).  When we allow for a rotation of the
magnetic moments, non-collinear solutions are obtained for the Cr- and
Mn-dimer/Ni(001) systems. Fig.~\ref{cr-dimer-ncol}(a) represents the
collinear magnetic ground state of the Cr system. As one expects from
the adatom picture, both adatoms forming the dimer tend to couple AF
to the substrate but due to their half filled {\it d} band they also
tend to couple AF to each other.

Thus there is a competition between the interatomic coupling within
the dimer, which drives it to a FI state, and the exchange interaction
with the substrate, which drives the moments of both atoms in the same
direction: AF for Cr and FM for Mn. As discussed in the previous
section, the magnetic exchange interaction (MEI) to the substrate is
relatively weak for Cr and Mn. Thus, the \emph{intra--dimer} MEI is
stronger than the MEI with the substrate, and in the collinear
approximation the ground state is found FI. Removing the collinear
constraint, a compromise can be found such that both adatom moments
are oriented almost to each other and at the same time (for Cr)
slightly AF to the substrate. This is shown in
Fig.~\ref{cr-dimer-ncol}(b): the Cr adatom moments are aligned
antiparallel to each other and basically perpendicular to the
substrate moments. However, the weak AF interaction with the substrate
causes a slight tilting, leading to an angle of 94.2$^{\circ}$ with
respect to the surface normal, instead of 90$^{\circ}$. We also
observe a very small tilting ($\approx 0.3^{\circ}$) of the magnetic
moments of the four outer Ni atoms neighboring the Cr dimer (the two
inner Ni atoms do not tilt for symmetry reasons).

Despite the above considerations, the collinear FI state
(Fig.~\ref{cr-dimer-ncol}(a)) is also a self-consistent solution of the
Kohn-Sham equations, even if the collinear constraint is
removed. Total energy calculations are needed in order to determine if
the non-collinear state (NC) is the true ground state, or if it represents a
local minimum of energy with the collinear result representing the
true ground state. After performing such calculations we find that the
ground state is actually collinear with an energy difference of $\Delta
E_{\mathrm{NC}-\mathrm{FI}} = 40$~meV to the non-collinear
state. This delicate balance cannot be captured by the Heisenberg
model with fitted exchange interactions, but requires self-consistent
density-functional calculations.

The result is different for a Mn dimer. Fig.~\ref{cr-dimer-ncol}(c)
and (d) show the collinear and the non-collinear solutions. The dimer
atoms couple strongly antiferromagnetically to each other but, just as
for the single Mn adatoms, they also couple (weakly) ferromagnetically
to the substrate. Both adatom moments, while aligned AF with respect
to each other, are tilted in the direction of the substrate
magnetization. With a rotation angle of $\approx 72.6^{\circ}$, the
deviation from the $90^{\circ}$ configuration is rather large. Also
the Ni moments are tilted by $7.4^{\circ}$. Finally total-energy
calculations show that for the Mn-dimer the non-collinear solution is
the ground state (total energy calculations yield $\Delta
E_{\mathrm{NC}-\mathrm{FI}} = -13$~meV).

In both cases (Cr and Mn dimers) the frustrated collinear solution is
asymmetric, while the non-collinear ground state restores the twofold
symmetry of the system. The differences in energy between the FI 
and the non-collinear solutions are small and can be altered either by
using a different type of exchange and correlation functional such as
GGA or LSDA$+U$, or after relaxing the atoms. We note, however, that
in a test calculation we found the Cr single-adatom relaxation
to be small (3.23 \% inward with respect to the interlayer distance), and thus
we believe that the relaxation cannot affect the exchange interaction
considerably.

As a cross-check, it is interesting to compare these non-collinear {\it
ab-initio} results to model calculations based on the previously defined 
Heisenberg model~\ref{eq}
with the exchange parameters fitted to the total energy
results. Taking into account only nearest-neighbor interactions
and neglecting the rotation of Ni moments, we rewrite the Hamiltonian
for the dimer in terms of the tilting angles $\theta_1$ and $\theta_2$
of the two Cr (or Mn) atoms (the azimuthal angles $\phi$ do not enter
the expression because of symmetry reasons):
\begin{equation}
{H}  = - J_{\mathrm{Cr-Cr}} \cos(\theta_1 - \theta_2) 
 - 4 J_{\mathrm{Cr-Ni}} (\cos\theta_1+\cos\theta_2) +
 \mathrm{const}. 
\label{eq:28}
\end{equation}

The interatomic exchange constants
$J_{\mathrm{Cr-Ni}}$, $J_{\mathrm{Mn-Ni}}$, $J_{\mathrm{Mn-Mn}}$ and
$J_{\mathrm{Cr-Cr}}$ are evaluated via a fit to the total energy obtained from
collinear LSDA calculations of the FM, AF, and FI 
configurations.

We note the two extreme cases arising from this Heisenberg
Hamiltonian: (i) $|J_{\mathrm{Cr-Ni}}|\gg|J_{\mathrm{Cr-Cr}}|$ leads
to the stabilization of the collinear FM or AF configuration
(adatom-like behavior) and (ii)
$|J_{\mathrm{Cr-Ni}}|\ll|J_{\mathrm{Cr-Cr}}|$ leads to
antiferromagnetic coupling within the dimer if $J_{\mathrm{Cr-Cr}} < 0$.
Within the Heisenberg model the FI solution and the non-collinear solution 
with $\theta = 90^{\circ}$ have the same energy.

Table~{\ref{table-heisenberg1}} summarizes the estimated exchange
parameters. 
\begin{table}
\begin{center}
\begin{tabular}{c|rrrr}
\hline
$J_{ij}$ (meV)&$J_{\mathrm{Cr-Ni}}$&$J_{\mathrm{Cr-Cr}}$&$J_{\mathrm{Mn-Ni}}$&$J_{\mathrm{Mn-Mn}}$\\
\hline
(a)&$-1.3$&$-189.1$&$13.0$&$-138.2$\\
(b)&$-11.6$&$-221.3$&$27.0$&$-140.2$\\
\hline
\end{tabular}
\end{center}
\caption{Values of magnetic exchange parameters $J_{ij}$ for Cr and Mn
  dimers on Ni(001), and obtained by the Lichtenstein
  formula~\cite{lichtenstein} (a) and extracted from collinear first-principles total
  energy calculations (b)  ($J_{\mathrm{Cr-Ni}}$ and
  $J_{\mathrm{Mn-Ni}}$ are averaged over the different Ni nearest
  neighbors of the dimer atoms). Positive $J_{ij}$ correspond to
  ferromagnetic interactions, negative $J_{ij}$ to antiferromagnetic
  ones. Adapted with permission from Ref.~\cite{lounis05}}.
\label{table-heisenberg1}
\end{table}
It is striking that the strong antiferromagnetic Cr-Cr and
Mn-Mn interaction for the dimer (nearest neighbors) are more than an order
of magnitude larger than the exchange interactions with the substrate,
and responsible for the stabilization of the non-collinear states
 shown in Fig.~\ref{cr-dimer-ncol}.
 The exchange constants $J_{ij}$ fitted to total energy results can be
 compared to the ones obtained by starting from the FI state and using
 the Lichtenstein formula~\cite{lichtenstein}, having in mind also its
 restriction to low-angle
 rotations~\cite{bruno,lichtenstein2,lounis_jij} . This rests on the
 force theorem, and yields the exchange constants corresponding to an
 infinitesimal rotation of the moments. The results of the two methods
 agree best for the Mn-Mn interaction, and reasonably well for the
 Cr-Cr interaction, but not for Mn-Ni and Cr-Ni. This is expected,
 since a rotation causes a significant change in the magnitude of the
 Ni moments, so that the force theorem is not applicable any more.

With the parameters from Table~{\ref{table-heisenberg1}} one can also
recalculate the non-collinear structure of the ground state. The
agreement with the {\it ab-initio} results is quite reasonable. For
the Cr dimer, one finds a slightly smaller tilting, i.e. 96$^{\circ}$
instead of 94.2$^{\circ}$, while for the Mn dimer the angle is
67.3$^{\circ}$ instead of 70.6$^{\circ}$.

The differences in energy calculated within this simple model suggest
that the Cr-dimer has a non-collinear ground state ($\Delta
E_{\mathrm{NC}-\mathrm{FI}} = -9.7$~meV) as well as the Mn-dimer
($\Delta E_{\mathrm{NC}-\mathrm{FI}} = -42$~meV). The discrepancy
obtained for the case of Cr-dimer (the LSDA calculation gives the
collinear FI ground state) can be attributed to the restrictions of
the Heisenberg model. For instance, for the FI and non-collinear
configurations, the Cr moments are slightly different, and also the
reduction of the Ni moments as a function of the rotation angle cannot
be described by the Heisenberg model, where the absolute values of the
moments are assumed to be constant.  Within the Heisenberg model, the
FI solution (with $\theta_1= 0^{\circ}$ and $\theta_2 = 180^{\circ}$)
is degenerate with the non-collinear solution ($\theta_{1,2} =
90^{\circ}$ with AF coupling within the dimer).

To evaluate the effect of change in coordination and hybridization, we
have undertaken a study of inatom dimers (i.e., embedded in the
surface layer), where we found that the ground state is of FI type for
both Cr and Mn systems. It is interesting to note that recent simulations on Mn dimers deposited on Ni(001) surface were presented in Ref.~\cite{stepanyuk} 
considering the impact of an external electric field. It was shown that 
depending on the magnitude of the field, switching of the nature of
 magnetic ground state can be achieved.

On the fcc Fe$_{3\mathrm{ML}}$/Cu(001) surface, the magnetic coupling
between the surface atoms and the adatoms is expected to be stronger
than on Ni(001) surface. Because of the AF coupling preference of the
single Mn adatom to the substrate, the Mn--dimer is characterized by a
non-collinear magnetic solution where the two moments are slightly
tilted AF to the substrate ($\theta = 115^\circ$). The Cr-dimer is,
however, FI and no NC solution was found. The reason is the very
strong AF coupling of the single Cr adatom to the Fe surface moments,
that cannot be overcome by the MEI between the adatom and the
substrate moments.  In order to explain this we use the Heisenberg
model, Eq.~(\ref{eq:28}), to calculate the angle
$\theta_1=\theta_2=\theta/2$ defining the non-collinear solution:
\begin{equation}
\cos(\theta)=-2\frac{J_{\mathrm{Cr-Fe}}}{ J_{\mathrm{Cr-Cr}}} \ \mathrm{for}\ 
2J_{\mathrm{Cr-Fe}}<J_{\mathrm{Cr-Cr}}
\label{eq:3-3}
\end{equation}
If $2|{J_{\mathrm{Cr-Fe}}|>|J_{\mathrm{Cr-Cr}}}|$, the angle is not
defined and the solution is collinear. This is clearly
realized in the present case: $2|{J_{\mathrm{Cr-Fe}}}|=2\times 80.8 >
|{J_{\mathrm{Cr-Cr}}}| = 78$~meV.  Note that the Cr-Fe coupling
constants are considerably smaller than for the single adatom.

On Ni(111) surface, the magnetic behavior for the Cr and Mn dimers is
similar to the behavior on Ni(001) surface with the difference of a
reduction of the adatom-substrate MEI due to the lower coordination on
Ni(111).  This leads once more to a FI ground state for Cr-dimer
whereas non-collinearity is a metastable state for
Mn-dimer. Energetically, this state characterized with a rotation
angle of $\theta=79^{\circ}$ is slightly higher, by 4.4 meV/adatom,
than the energy of the FI solution.

\section{Chains}
Now that we found the presence of both collinear and non-collinear
states in dimers, it is reasonable to ask what happens in larger
systems, such as antiferromagnetic clusters or chains. Within the
Heisenberg model, the infinite antiferromagnetic chain deposited on a
ferromagnetic surface is predicted to be non-collinear with atomic
moments tilted in a similar fashion as the dimer moments at the
condition that dimer is non-collinear~\cite{lounis_prl}. But what is the ground state of
finite-length chains? For a preliminary answer one can employ again
the classical Heisenberg Hamiltonian that we rewrite as
follows: \begin{eqnarray} H&=&-J_1\sum_{i = 1}^{N-1} \cos(\theta_i
  -\theta_{i+1}) - J_2\sum_{i = 1}^{N} \cos(\theta_i).\label{Eq:2}
\end{eqnarray} 
$N$ is the number of atoms in the chain and $\theta$ is the rotation
angle of the chain atom moment with respect to the magnetization of the
surface.  $J_1$($<0$) stands for an (antiferromagnetic) exchange
interaction between two neighboring chain atoms at sites $i$ and $i\pm
1$ in the chain, while $J_2$ is the interaction between a given chain
atom and the substrate.
\begin{figure}[!ht]
\hspace{-1cm}
{(a)}
\includegraphics*[width=0.25\linewidth]{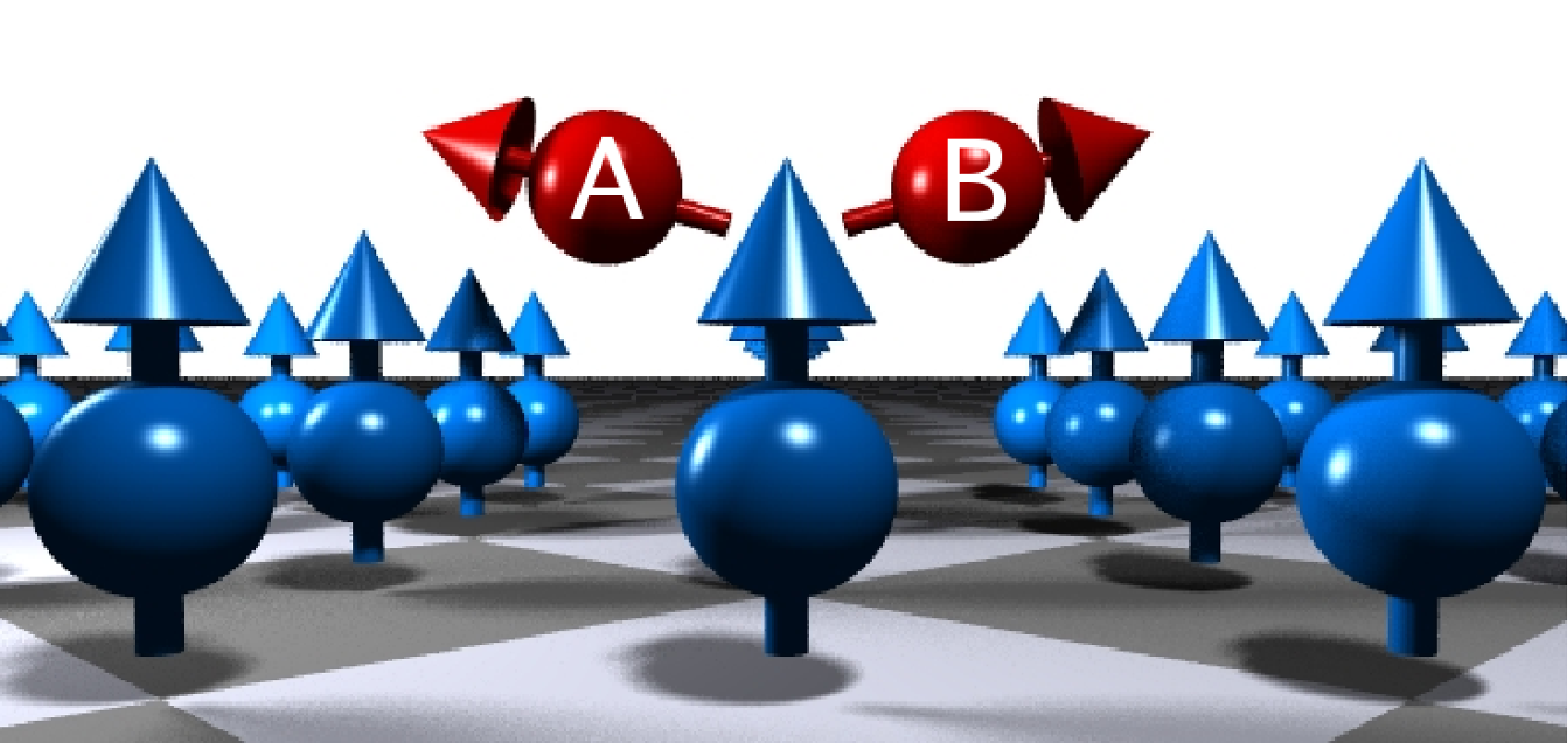}
\hspace{-0.2cm}
{(b)}\hspace{-0.1cm}
\includegraphics*[width=0.25\linewidth]{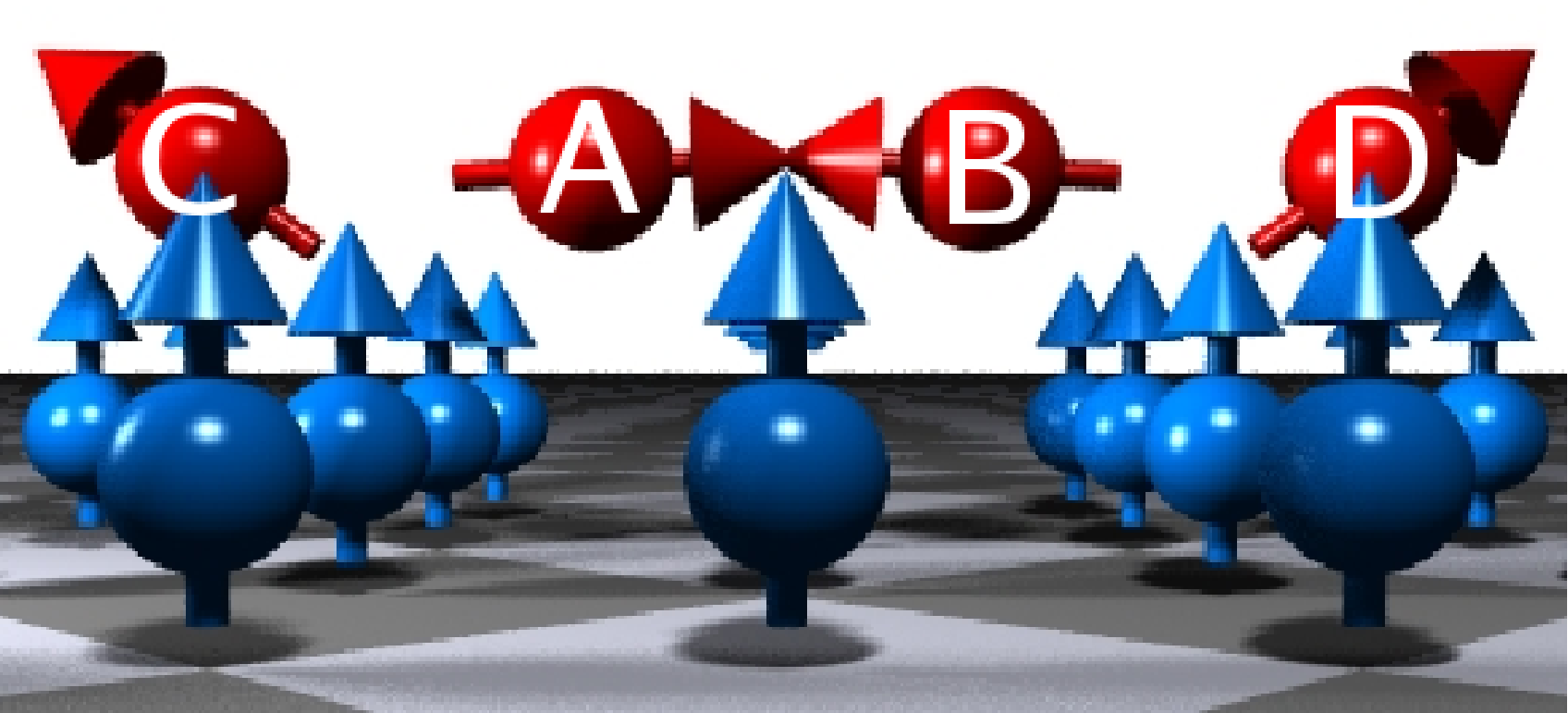}
\hspace{-0.2cm}{(c)}\hspace{-0.1cm}
\includegraphics*[width=0.4\linewidth]{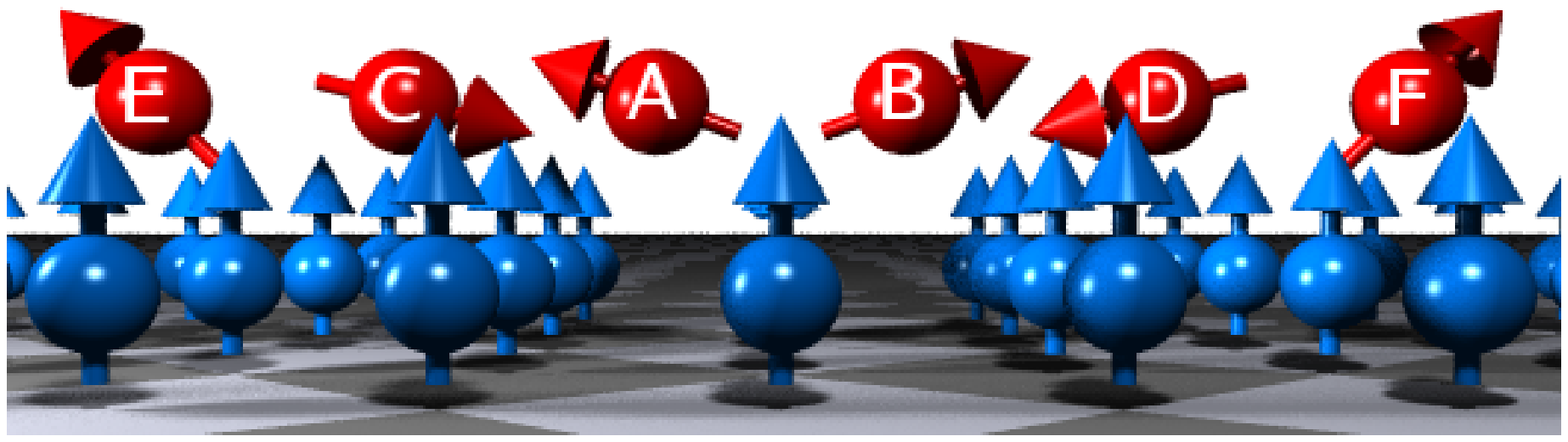}
\\
\begin{center}{(d)}
\includegraphics*[width=0.4\linewidth]{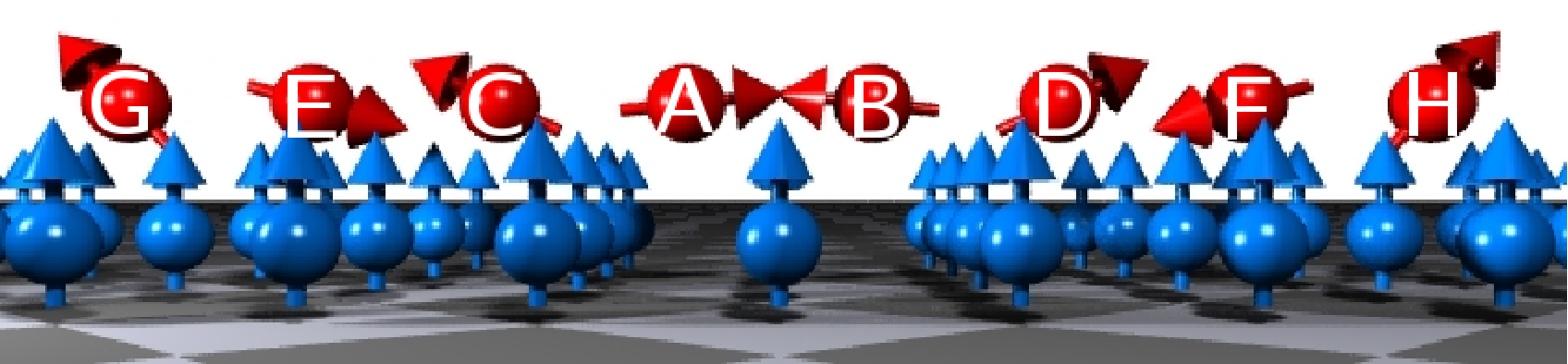}
\hspace{-0.1cm}{(e)}\hspace{-0.1cm}
\includegraphics*[width=0.5\linewidth]{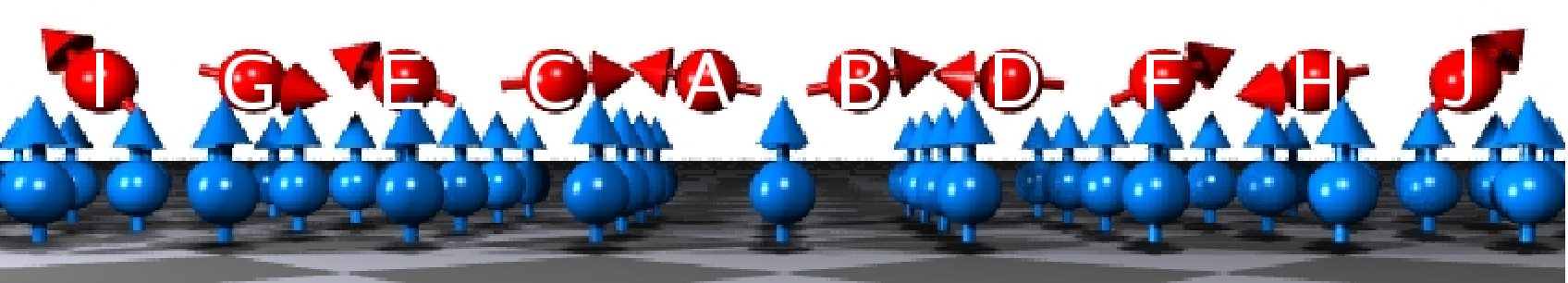}
\end{center}
\begin{center}{(f)}
\includegraphics*[width=0.3\linewidth]{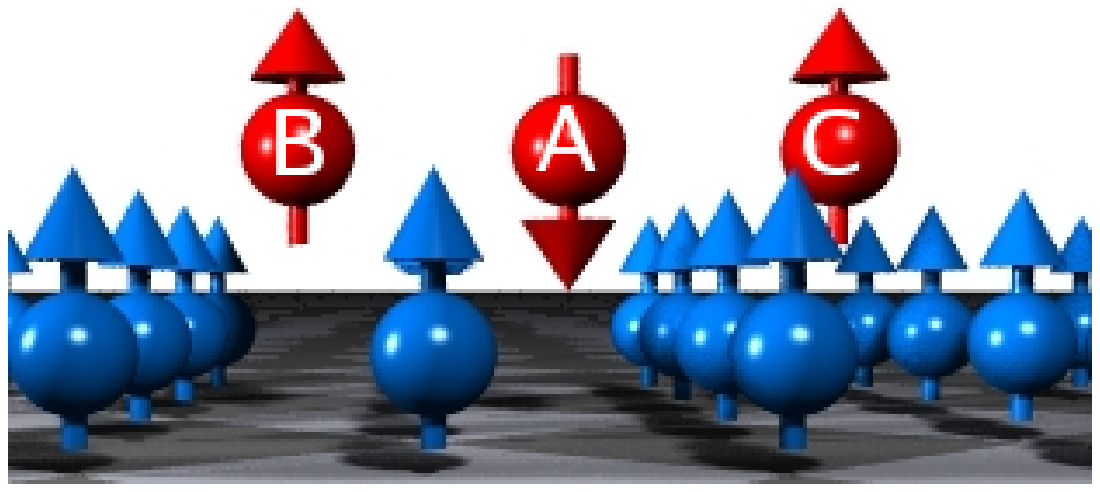}
\hspace{-0.cm}{(g)}\hspace{-0.cm}
\includegraphics*[width=0.35\linewidth]{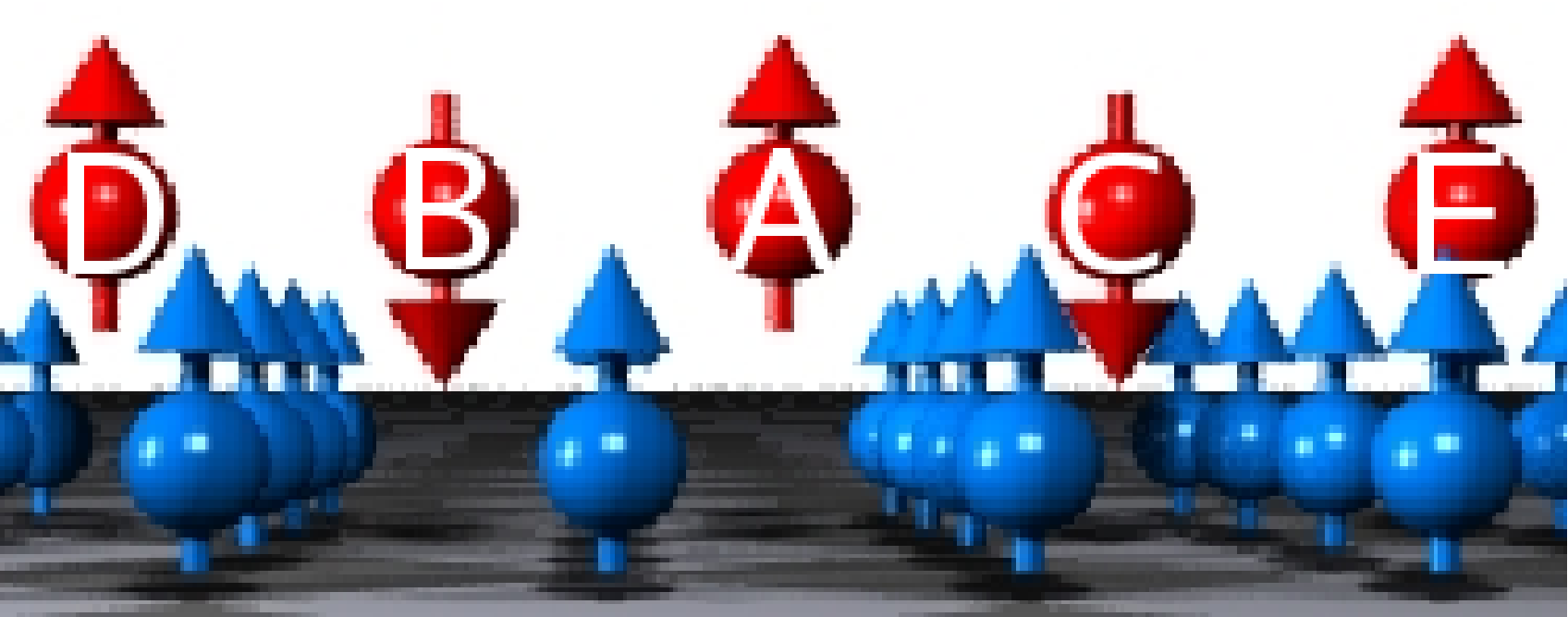}\end{center}
\hspace{-0.2cm}
{(h)}\hspace{-0.1cm}
\includegraphics*[width=0.46\linewidth]{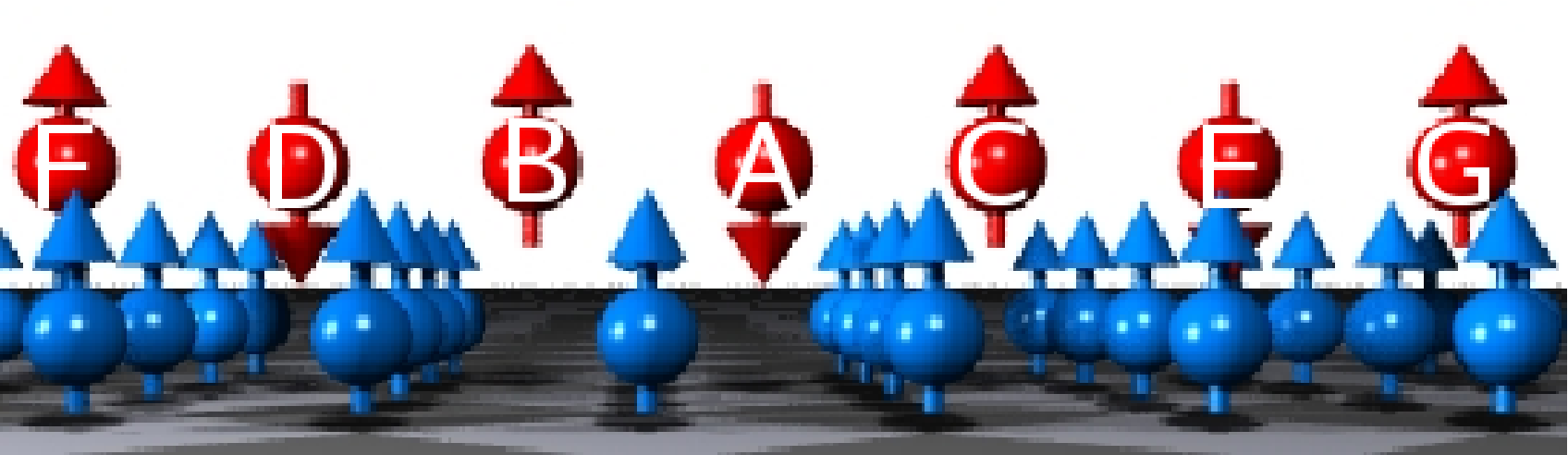}
\hspace{-0.1cm}{(i)}\hspace{-0.1cm}
\includegraphics*[width=0.46\linewidth]{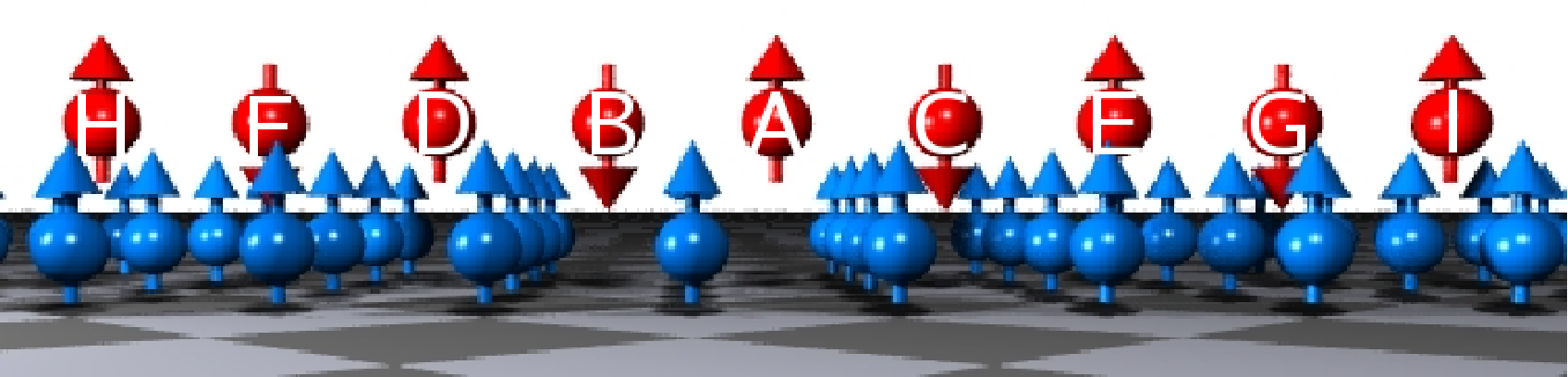}
\caption{Magnetic ground configurations 
of the Mn nanochains on Ni(001) calculated with the KKR method. 
The nanochains with even number of Mn 
atoms (2, 4, 6, 8, 10) prefer a non-collinear 
ground state, the odd ones a collinear one.  Atoms in blue correspond to the Ni substrate.  Adapted with permission from Ref.\cite{lounis_prl}.}
\label{fig:chains-Ni001}
\end{figure}

\begin{table}[ht!]
\begin{center}
\caption{\label{table:noncol-even} {\em Ab initio} results for 
even-numbered nanochains: size and angle of the magnetic moments
as well as total energy differences between the NC and FI solutions.
In every pair of chain atoms connected by a ``-'' sign, the  azimuthal 
angles $\phi$ are equal to 0$^\circ$-180$^\circ$, while the 
magnetic moments and rotation angles $\theta$ are the same. Adapted with permission from Ref.~\cite{lounis_prl}}
\begin{tabular}{lcccc}
\hline
 Length        & $E_{\mathrm{NC}} - E_{\mathrm{FI}}$   & Adatom             & $\theta $($^\circ$) &  $M$ ($\mu_B$) \\
 (adatoms)     & (meV/adatom)    &                    &                 &             \\
\hline
 2             & $-$11.16   &A-B                 &  73             &   3.71       \\
 4             & $-$8.48   &A-B,C-D             &  87, 54         &   3.55, 3.72 \\
 6             & $-$7.82   &A-B, C-D,           &  70, 104,       &   3.46, 3.52,\\
               &   &E-F                 &         45      &         3.67 \\
 8             & $-$5.46   &A-B, C-D,           &  84, 65,        &   3.47, 3.46,\\
               &    &          E-F, G-H  &  106, 44,       &   3.52, 3.66 \\
10             & $-$3.53   &A-B, C-D,           &  78, 85,        &  3.47, 3.47  \\
               &   &        E-F, G-H,   &  67, 102,       &  3.46, 3.52  \\
               &   &                I-J &  48             &  3.67        \\
\hline
\end{tabular}
\end{center}
\end{table}

As an example, Mn-chains on Ni(001) surfaces are discussed; the
obtained results are rather general ~\cite{lounis_prl}.  For the
dimer-case, the energy of the FI solution depends only on $J_1$
($E_{\mathrm{FI}}=J_1$), because the contributions $J_2$ of both
adatoms cancel out due to their antiparallel alignment. On the other
hand, the energy of the NC solution (Fig.~\ref{fig:chains-Ni001}(a))
depends also on the magnetic interaction with the substrate in terms
of $J_2$ ($E_{\mathrm{NC}}=-J_1 \cos(2\theta) - 2 J_2 \cos(\theta)$).
For three Mn adatoms (Fig.~\ref{fig:chains-Ni001}(f)), we find the FI
solution to be the ground state.  Contrary to the dimer, the energy of
the collinear solution of the trimer depends on $J_2$
($E_{\mathrm{FI}}=2J_1 - J_2$) due to the additional third adatom,
which in fact allows the FI solution to be the ground state.

One sees here the premise of an odd-even effect on the nature of the
magnetic ground state. On this basis one can conjecture that chains
with even number of atoms would behave similarly to the dimer, because
an additional energy with the substrate proportional to $J_2$ can be
gained in the NC state by the small tilting off the $90^\circ$ angle shown
in Fig.~\ref{fig:chains-Ni001}(a), while odd-numbered chains would
behave similarly to the trimer. They can always gain energy in the
collinear state due to one $J_2$ interaction term which does not
cancel out.

Investigating the longer nanochains with even number of atoms shows
that their ground state is always NC. Examples, calculated with the
KKR method, are presented in
Fig.~\ref{fig:chains-Ni001}(b)-(c)-(d)-(e) and in
Table~\ref{table:noncol-even}. In a first approximation, the magnetic
moments are always in the plane perpendicular to the substrate
magnetization keeping the magnetic picture seen for the dimer almost
unchanged. Moreover, the neighboring magnetic moments are coupled
almost AF. The atoms at both ends of the chains are closest to a FM
orientation to the substrate (see Table~\ref{table:noncol-even}). The
two central chain atoms A-B (see Fig.~\ref{fig:chains-Ni001} for the
notation) are the ones which keep their rotation angles almost
unaltered with respect to the dimer. The angle $\theta$ oscillates
between 70$^\circ$ obtained for the chain with 6 atoms up to
87$^\circ$ obtained for the chain with 4 atoms. Note that the angle
between two successive moments is about 150$^\circ$, similar to the
dimer result.

The considered odd-numbered nanochains are characterized by a FI
ground state in which the majority of atoms are coupled FM to the
surface.  The total energy differences to the lowest lying metastable,
i.e.\ NC state, first increases with respect to the length of the
chain (see Table~\ref{table:col-even-odd}) up to a maximum for a chain
with 7 atoms (10.5 meV/adatom) followed by a decrease for longer
chains.  This behavior is the property of the metastable NC state and
arises from a competition between the edge and inner atoms of the
chain.  Edge atoms in odd chains favor collinear moment alignment to
the substrate. For short chains, trimer and 5 atoms chains, they
dominate the total magnetic behavior permitting only a slight tilting
of the moments away from the FI state. For longer chains, however, the
inner atoms experience basically the same local environment as the
atoms in even chains resulting in similar moment orientations.

When increasing the length of the chains, both kinds of chains should
converge to the same magnetic ground state since the even-odd parity
is expected to be obsolete for infinite systems. Within the DFT
framework, the investigation of longer chains is computationally very
demanding. Thus, the Heisenberg model is used to investigate this
magnetic transition.
\begin{table}[ht!]
\begin{center}
  \caption{\label{table:col-even-odd} {\em Ab initio} results for
    odd-numbered nanochains: size of the magnetic moments and total
    energy differences between the NC and FI state.  Atoms connected
    by a hyphen have the same magnetic moment.}
\begin{tabular}{lccc}
\hline
 Length           & Adatom             & $M$ ($\mu_B$)       & $E_{\mathrm{NC}} - E_{\mathrm{FI}}$\\
 (adatoms)        &                    &  for FI             &  (meV/adatom)  \\
\hline
 3                &A, B-C              &  $-$3.78, 3.65      & 8.85\\
 5                &A, B-C, D-E         &  3.43, $-$3.56, 3.64& 9.52\\
 7                &A, B-C, D-E,        &  $-$3.54, 3.43, $-$3.56& 10.48\\
                  &F-G                 &  3.64               &\\
 9                &A, B-C, D-E,        &  3.43, $-$3.54, 3.43& 9.29 \\
                  &F-G, H-I            &  $-$3.56, 3.64      &\\
11   &A, ..., K-L &$-$3.59, ..., 3.67&6.49\\
\hline
\end{tabular}
\end{center}
\end{table}

\begin{figure}
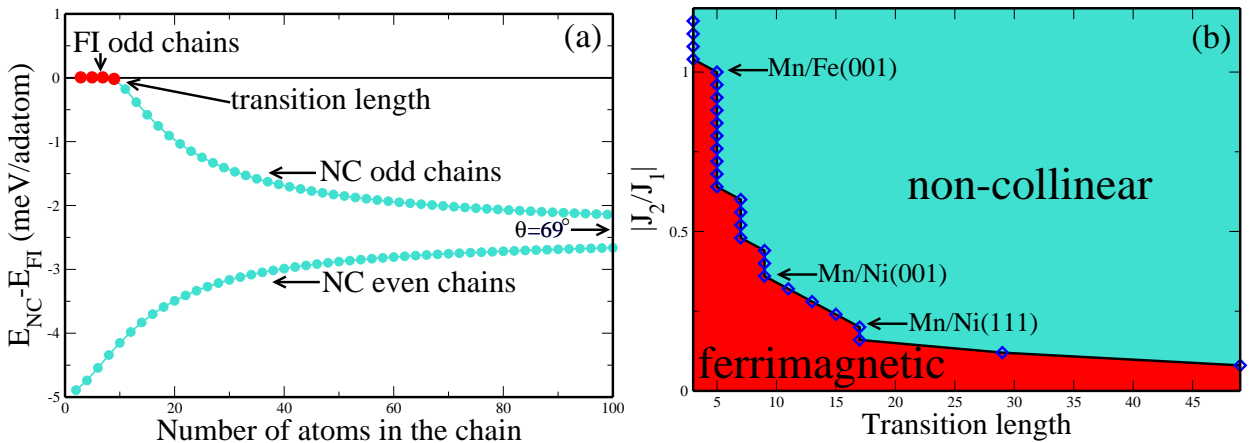

\hspace{-1cm}
\includegraphics*[width=0.5\linewidth]{Fig_chains/energydiff.eps}
\hspace{-0.2cm}
\includegraphics*[width=0.5\linewidth]{Fig_chains/transitionslength.eps}
\caption{(a) Energy differences obtained from the Heisenberg model between
  the NC and the FI configuration for different lengths of Mn chains
  on Ni(001). The parameters used were $J_1=-138$~meV and
  $J_2=52$~meV. (b) Magnetic phase diagram of the odd chains 
showing the effect of the ratio ${J_2}/{J_1}$ on the transition 
lengths. If the length of the 
chains is smaller than the transition length, the FI configuration is 
the ground state. Adapted with permission from 
Ref.\cite{lounis_prl}.}
\label{energydiff-chains}
\end{figure}

We discuss now the Heisenberg Model results. Two different approaches
are used to solve Eq.\ref{Eq:2}.\footnote{Recently, it has been
  pointed out that Eq.\ref{Eq:2} can be solved with great numerical
  and analytical profit in terms of a two-dimensional mapping
  method~\cite{politi}} In the first approach we allow the rotation
angle $\theta_i$ to vary from site to site in the chain and in the
second we consider a constant absolute value of $\theta$ at each site.
The first, the inhomogeneous approach, requires an iterative numerical
scheme while the second, the homogeneous one, leads to a simple
analytical form.  In Fig.~\ref{energydiff-chains}(a), the energy
difference between the NC and FI states determined by the first
approach is plotted versus the length of Mn-nanochains. Negative
values refer to a NC ground state.  The model reproduces the DFT
results showing that the even chains have always a NC ground state.
Within this model, the ground state for odd chains changes from FI to
NC when the number of atoms exceeds a transition length of 9 atoms
which is smaller than what predicted from DFT.  Moreover, even beyond
this length, we notice an oscillatory behavior of the energy
differences and the magnetic structure when going from the odd- to
even-numbered chains. This behavior is easier to understand when
considering the homogeneous ansatz, which leads to a somewhat larger
transition length (19 atoms) for the odd chains. In this case, the
energy difference per chain atom is given by $\Delta
E_{\mathrm{NC-FI}} = \frac{NJ_2^2}{8(N-1)J_1}+P(N)\frac{|J_2|}{N}$,
with $P(N)$, a parity function, equals 0 if $N$ is even, and 1 if $N$
is odd. Since the first term of $\Delta E_{\mathrm{NC-FI}}$ is
negative for all lengths ($J_1<0$), NC is the ground state for all
even-chains. The second term on the other hand is positive and
provides for finite lengths an energy counterbalance allowing FI as
the lowest energy solution.  As this term decreases as $N^{-1}$, a
cross-over to non-collinearity is expected for $N\approx
8|J_1|/|J_2|$, as found in Fig.~\ref{energydiff-chains}(a).  Moreover, we
notice that for big values of $N$, $\Delta E_{\mathrm{NC-FI}}$
converges to a constant, ${J_2^2}/{(8J_1})$, which is confirmed by the
convergence of the two curves in Fig.~\ref{energydiff-chains}(a) towards
the same NC state with $\theta_{\mathrm{NC}}= 69^\circ$: if the chains
are infinite the parity induced differences vanish.

The next point is the discussion of the general behavior of the 
transition length for odd chains. Using the inhomogeneous ansatz, we 
determine for each set of parameters ($J_1$, $J_2$) 
the corresponding transition length which leads to the phase diagram 
shown in Fig.~\ref{energydiff-chains}(b). The 
obtained curve seems to decay roughly as $N^{-1}$. On 
one hand, small exchange-interaction ratios lead to very long
transition lengths. This means that odd-even effects are expected to 
last for very long chains which can be easily observed experimentally. On 
the other hand, large values of $J_2$ compared to $J_1$ lead to 
very small transition lengths. The obtained phase diagram is 
interesting and can be used to 
predict the behavior as well as the transition lengths for 
other kinds of AF chains deposited on FM substrates. This model predicts for 
example a transition length of 5 atoms for Mn/Fe(001) 
while Mn/Ni(111) is characterized by a value of 17 atoms. 
Certainly, this transition length is subject to modifications
 depending on the accuracy of the exchange interactions, spin-orbit 
coupling and geometrical relaxations. Furthermore, we 
point out that the angles $\theta$ obtained by the model are in good 
agreement with our DFT calculations, meaning that the model 
reliably describes the very-long chains. In addition, it is interesting to note 
that a recent experimental 
as well as  theoretical work revealed a similar NC behavior for a 
full-monolayer of Mn deposited on a bcc Fe(001) surface~\cite{baroni}. We note that 
recent simulations~\cite{petrilli} on Mn nanowires on bcc Fe(001) did not show the even-odd 
behavior since the magnetic exchange interactions among the nearest neighboring Mn adatoms were antiferromagnetic but 
smaller then the ferromagnetic magnetic exchange interaction with the substrate. Thus a non-collinear state is not 
expected in this situation (Eq. (12) is not fulfilled). However, interesting sinusoidal modulation of the magnetization is 
obtained 
with a period corresponding to the length of the Mn nanowires.

Recently, spin-polarized STM experiments on Mn nanowires deposited on Ni(110) 
combined with DFT calculations verified the existence of an even-odd effect 
in the magnetic ground states~\cite{wulfhekel}. Similar to Ni(001) substrate, the even-numbered wires are
 non-collinear while the odd-numbered ones are collinear. Interestingly, the ferrimagnetic contrast was observed 
experimentally for the trimer, but for the dimer and tetramer no signal was observed. This is reasonable since 
if there is no spin-orbit coupling the tilted moments of Mn atoms can rotate freely around the magnetization of the substrate. 
Thus, there is no possibility to observe a contrast from the components of the adatoms magnetic moments perpendicular to the magnetization 
of the surface while the parallel components are difficult to distinguish. In Ref.~\cite{wulfhekel}, an additional proposal is made 
for the fluctuations of the magnetization using zero-point energy motion for even-numbered chains. Indeed, although a barrier, induced by 
the spin-orbit interaction,  exists between    
equivalent orientations of the magnetic moments, the strength of the fluctuations, evaluated within the Landau-Lifschitz-Gilbert 
formalism, is larger than the barrier. More details can be found in Ref.~\cite{wulfhekel}.
\section{Compact clusters}

\subsection{Fcc(001) surfaces}

\begin{figure}[h!]
\begin{center}
(a)
\includegraphics*[width=0.35\linewidth]{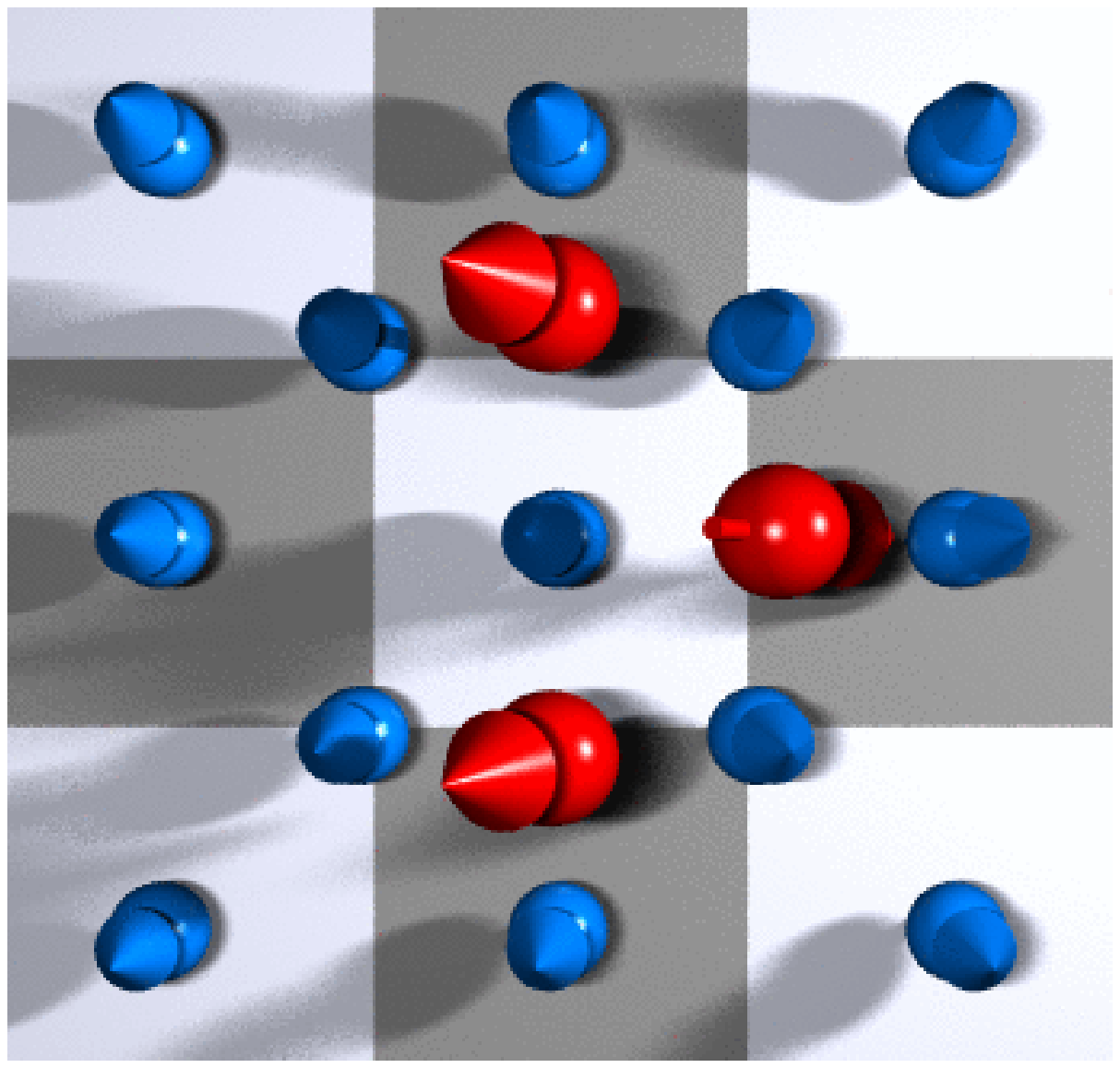}
(b)
\includegraphics*[width=0.45\linewidth]{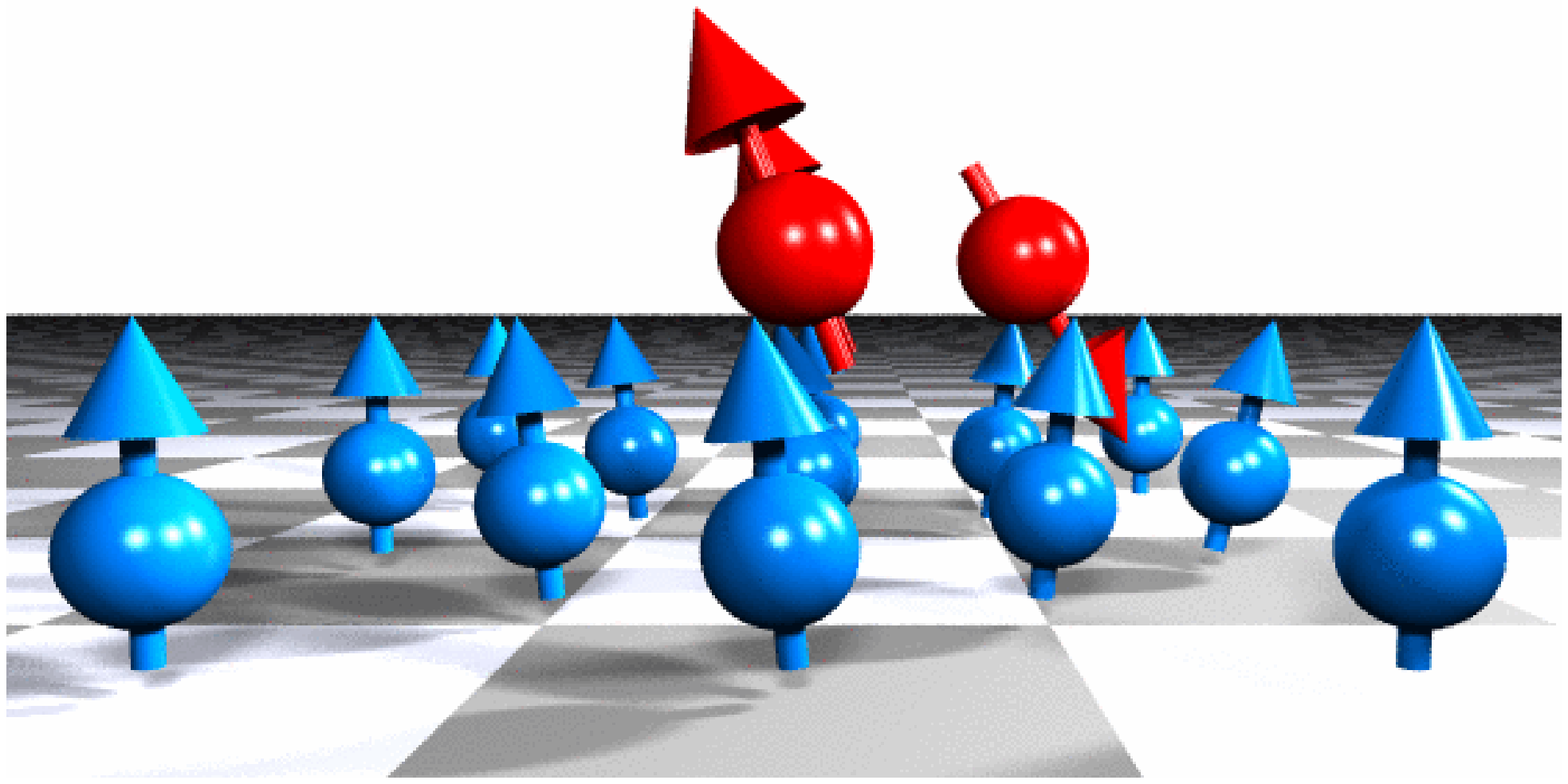}
\caption{Non-collinear state of the Mn trimer on Ni(001) surface. Side
view (a) and front view (b) are shown. This represents a local minimum
in energy, with the collinear state being the ground state (see
text). Atoms in blue correspond to the Ni substrate. Reprinted with permission from Ref.~\cite{lounis05}. 
Copyright 2005 by the American Physical Society.}
\label{mn-trimer-front}
\end{center}
\end{figure}

So far we have examined a simple one-dimensional geometry of the
magnetic nanostructure, which can be stabilized at low enough
temperatures. One expects, however, more compact structures to occur,
where the increased coordination will favor a reduction of the
cohesive energy. For example, instead of the aforementioned linear
trimer, an isosceles rectangular triangle would occur on an fcc(001)
surface (see Fig.~\ref{mn-trimer-front}).

In such a triangle on Ni(001), it is expected, and verified by
total-energy calculations, to find the $\downarrow\uparrow\downarrow$
configuration as the collinear magnetic ground state for Cr and the
$\uparrow\downarrow\uparrow$ for the Mn trimer ($\uparrow$ means an
atomic moment parallel to the substrate magnetization, $\downarrow$
means antiparallel; the middle arrow represents the direction of the
atomic moment at the right-angle corner of the triangle).

Allowing free rotation of the magnetic moments leads to no change for
the Cr trimer $\downarrow\uparrow\downarrow$---the state remains
collinear (within numerical accuracy). On the other hand, for the Mn
trimer a non-collinear solution is found (Fig.~\ref{mn-trimer-front})
with the nearest neighbors almost antiferromagnetic to each other,
but with a collective tilting angle with respect to the
substrate. This tilting angle is induced by the ferromagnetic MEI
between the central Mn atom with the substrate, competing with the
antiferromagnetic MEI with its two companions. The top view of the
surface shows that the in-plane components of the magnetic moments are
collinear.

The tilting is somewhat smaller ($21.7^{\circ}$) for the two Mn atoms
with moments up than for the Mn atom with moment down
($28.5^{\circ}$). Also the neighboring Ni--surface atoms experience
small tilting, with varying angles around $4^{\circ}-10^{\circ}$.
From the energy point of view, the ground state is collinear,
$\uparrow \downarrow \uparrow$, with an energy difference of $\Delta
E_{\mathrm{NC}-\uparrow \downarrow \uparrow} = 23$~meV with
respect to the local-minimum, non-collinear solution. 

One should note that the moments of the two first neighboring 
impurities are almost compensated in the FI solution. The third 
moment determines the total interaction between the substrate 
and the trimer which has then a net moment 
coming mainly from the additional 
impurity. This interaction is identical to the single adatom (or inatom) 
type of coupling.

\begin{figure}
\begin{center}
\hspace{-0.5cm}(a)
\includegraphics*[width=0.35\linewidth]{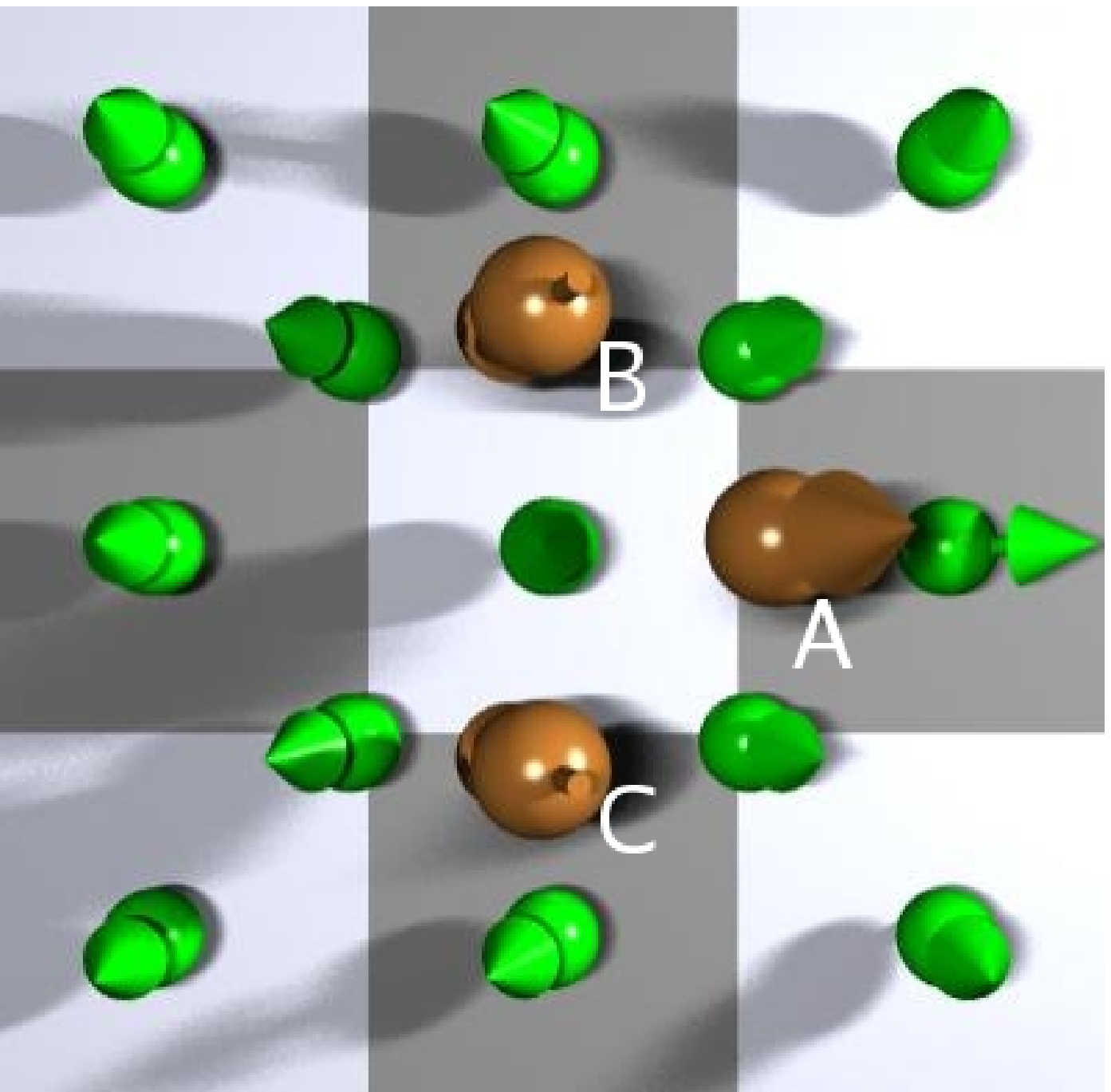}\hspace{-0.1cm}
(b)
\includegraphics*[width=0.35\linewidth]{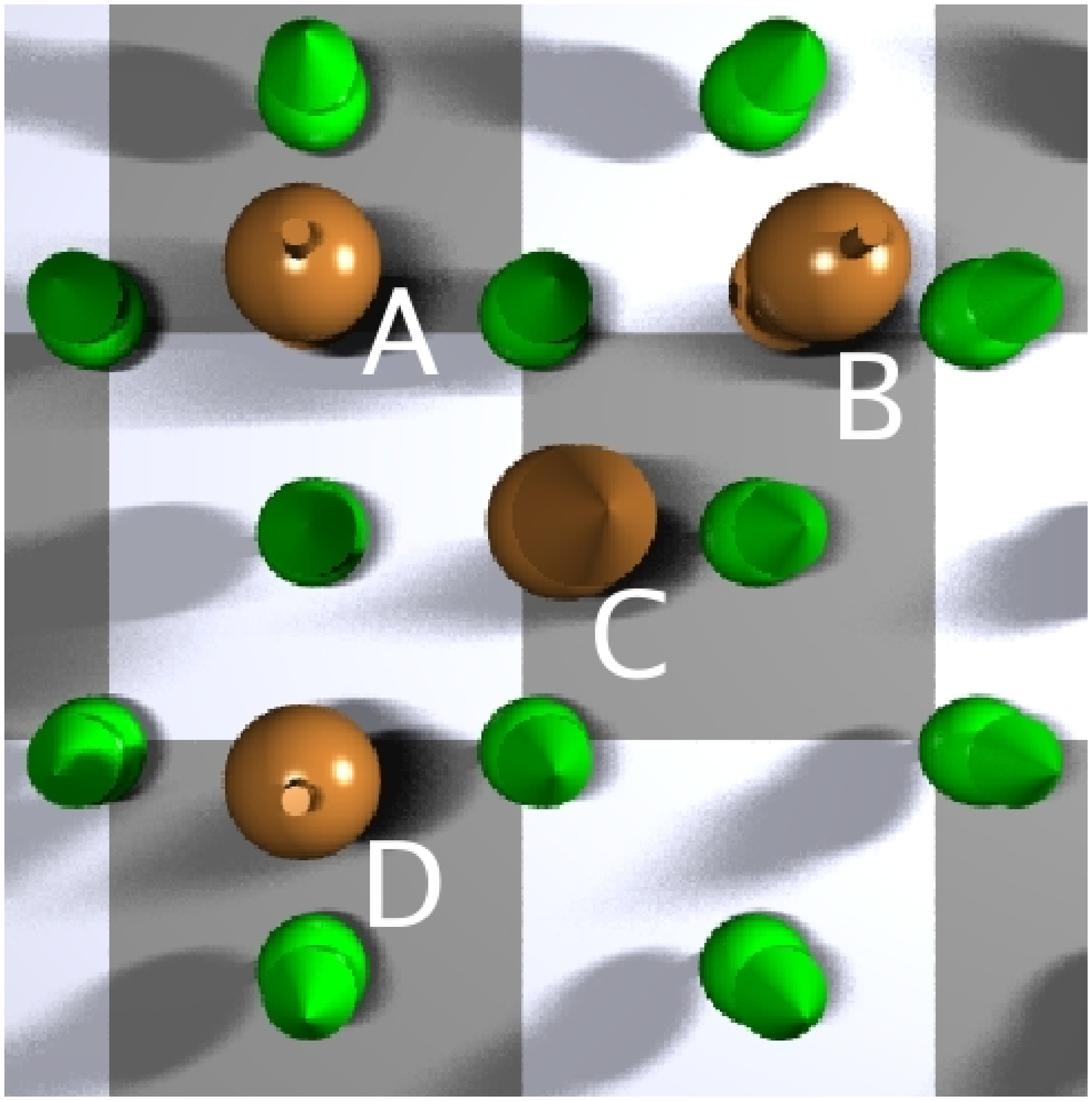}\\
\hspace{-0.5cm}(c)
\includegraphics*[width=0.35\linewidth]{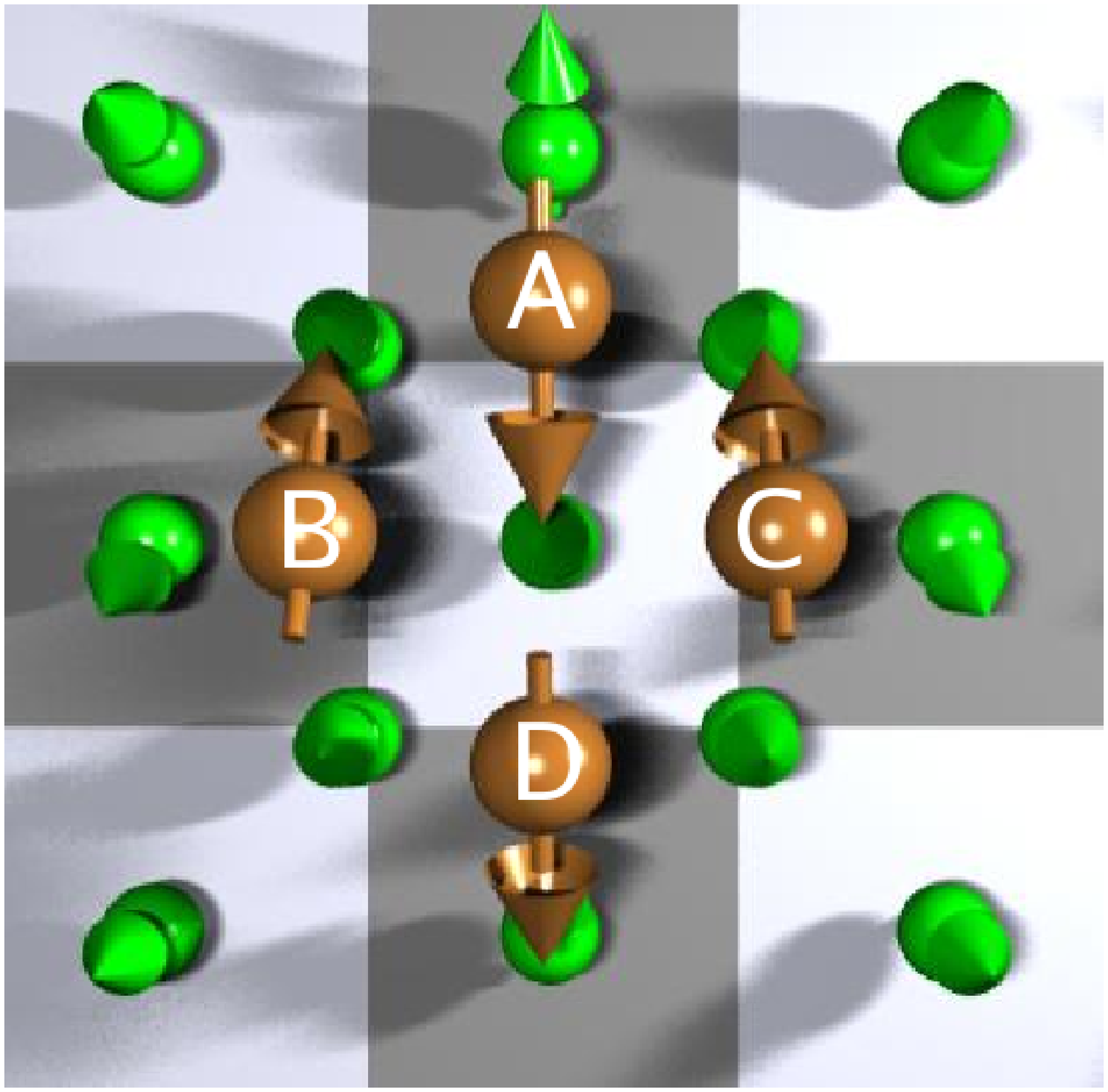}
\hspace{-0.1cm}(d)
\includegraphics*[width=0.35\linewidth]{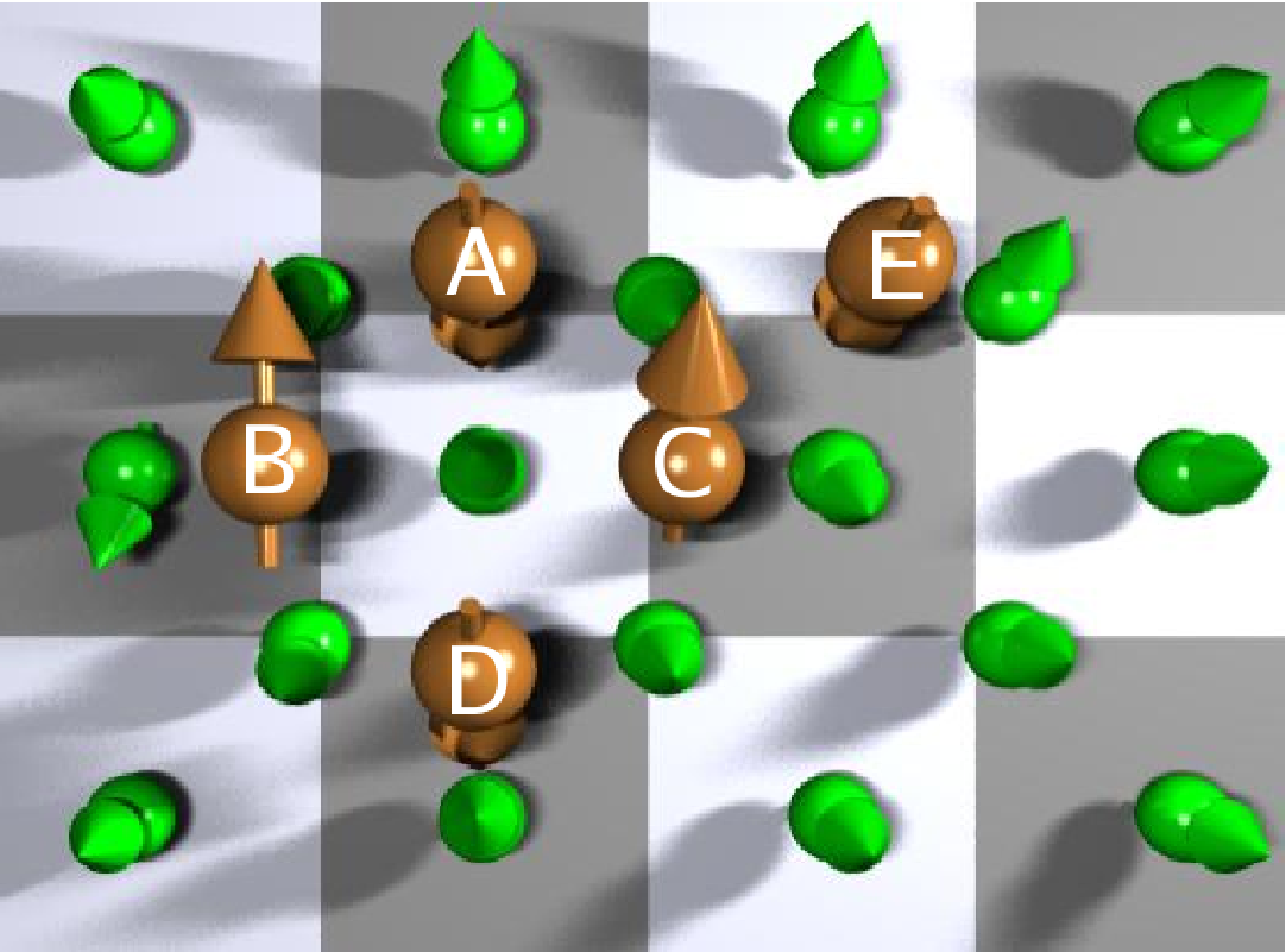}
\caption{Front views of complex magnetic states of
Cr ad-clusters on Fe$_{3\mathrm{ML}}$/Cu(001)
surface: (i) the trimer is shown in (a), (ii) the
T-shape tetramer  (tetramer 2) is presented in (b),
(iii) the tetramer 1 is depicted in (c) and finally pentamer 1 is
shown in (d).  Unlabeled atoms (in green) correspond
to the Fe substrate.  Reprinted with permission 
from Ref.\cite{lounis08}. Copyright 2008 by the IOPscience.}
\label{Cr-trimer-front}
\end{center}
\end{figure}

On Fe$_{3\mathrm{ML}}$/Cu(001) surface, the nature and type of
non-collinear structure of the trimer do not change much compared to
what is obtained on Ni(001) surface. The only difference is that,
here, the non-collinear solution is the ground state for the compact
Cr- and Mn-trimer.  The addition of a third adatom to the system
forces a rotation of the moments. The two second-neighboring Cr/Mn
impurities B and C (see Fig.~\ref{Cr-trimer-front}(a)) have a moment
tilted towards the surface by an angle of 156$^\circ$/170$^\circ$
(24$^\circ$/10$^\circ$ from the AF coupling) and adatom A moment is
tilted up/down with an angle of 77$^\circ$/20$^\circ$. As can be
noticed, the moment of the central adatom rotates by approximately
twice the rotation angle of the moments of the outer adatoms. This is
explained by the fact that the central magnetic moment experiences
twice the AF exchange coupling from its two first neighboring atoms.

We extended our study to bigger clusters, namely tetramers and
pentamer. Two types of tetramers were considered: tetramer 1 is the
most compact and forms a square (Fig.~\ref{Cr-trimer-front}(c)), while
tetramer 2 has a T-like shape (Fig.~\ref{Cr-trimer-front}(b)).
The ground state of Cr--tetramer 1 is 
non-collinear (Fig.~\ref{Cr-trimer-front}(c)) with a
 magnetic moment of 2.5~$\mu_B$ carried by each impurity. One notices
that the neighboring adatoms are almost AF coupled to each
other(the azimuthal angle $\phi$ is either equal to 0$^0$ or to
180$^0$) with all moments rotated by the angle $\theta$ = 111$^\circ$. 
Contrary to Cr, Mn--tetramer 1 has a collinear FI magnetic ground state with 
a total energy slightly lower (2.3 meV/adatom) than the energy of 
the non-collinear metastable solution. The latter is similar to the solution 
depicted in Fig.~\ref{Cr-trimer-front}(c) but with moments slightly tilted upwards.

For Cr--tetramer 2 (see Fig.~\ref{Cr-trimer-front}(b)) we obtained
several collinear magnetic configurations. The most favorable one is
characterized by an AF coupling of the three corner atoms with the
substrate. The moment of adatom C, surrounded by the remaining Cr
impurities, is then forced to orient FM to the substrate.  When we
allow for the direction of the magnetic moment to relax, we get a
non-collinear solution having a similar picture, energetically close
to the collinear one ($\Delta E_{\mathrm{col}-\mathrm{NC}} =
2.3$~meV/adatom). Adatom C has now a moment somewhat tilted by
13$^\circ$ ($\mu=$~2.31~$\mu_B$) whereas adatom A has a moment tilted
in the opposite direction by 172$^\circ$
($\mu=$~2.85~$\mu_B$). Adatoms B and D have a moment of 2.87~$\mu_B$
with an angle of 176$^\circ$.  We note that tetramer 1 with a higher
number of nearest-neighboring bonds (four instead of three for
tetramer 2) is the most stable one ($\Delta
E_{\mathrm{tet2}-\mathrm{tet1}} = 14.5$~meV/adatom) with the
non-collinear solution shown in Fig.~\ref{Cr-trimer-front}(c).

To study the pentamers, we have chosen two structural configurations: 
pentamer 1 (Fig.~\ref{Cr-trimer-front}(d)) with the highest number (five) of
first neighboring adatom bonds (NAB) and pentamer 2 which is more 
compact has only four NAB. The latter one is obtained by extending the tetramer of 
Fig.~\ref{Cr-trimer-front}(b) symmetrically with an additional adatom forming 
an X-shaped cluster. 
The pentamer 1 consists on a
tetramer of type 1 plus an adatom (E) and is characterized by a
non-collinear ground state.  Let us understand the solution obtained
in this case by starting from tetramer 1 (Fig.~\ref{Cr-trimer-front}(d)), which 
is characterized by a non-collinear almost in-plane magnetic 
configuration. As we have seen, a single adatom is strongly AF coupled to the
substrate. When attached to the tetramer it
affects primarily the first neighboring impurity (adatom C) by tilting
the magnetic moment from 111$^\circ$ to 46$^\circ$. Adatom E is also
affected by this perturbation and experiences a tilting of its moment
from 180$^\circ$ to 164$^\circ$. As a second effect, the second
neighboring adatom, A, is also affected and suffers a moment rotation
from 111$^\circ$ to 138$^\circ$.  The AF coupling between first
neighboring adatoms is always stable, thus adatom D has also a moment
rotated opposite to the magnetization direction of the substrate with
an angle of 155$^\circ$ ($\mu =$~ 2.48~$\mu_B$). As adatom B tends
to couple AF to its neighboring Cr adatoms, its magnetic moment tilts
into the positive direction with an angle of 85$^\circ$.

On the other hand, pentamer 2 is characterized by a non-collinear 
solution which is very close to the collinear one: the 
outer adatoms are AF coupled to the surface magnetization while the moment of 
the central adatom is forced to be oriented FM to the substrate. 
Surprisingly, this ad-cluster which has less 
first NAB (four instead of five) has a lower energy compared 
to pentamer 1. Here, energy difference between the two structural configurations is 
about 37 meV/adatom. The strength of the second NAB seem to be as important or 
stronger than the first NAB.

\subsubsection{Experiment} 

\begin{figure}
\begin{center}
\includegraphics*[width=.6\linewidth]{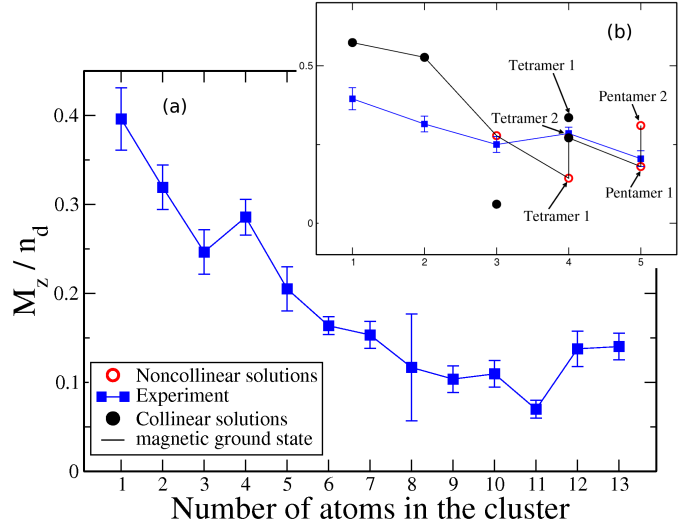}
\caption{Spin moment per number of d-holes per atom versus cluster
  size for Cr clusters on a Fe$_{3\rm ML}$/Cu(001) ferromagnetic
  substrate. Panel (a) shows experimental results for cluster-sizes up
  to 13 adatoms while panel (b) gives the comparison with {\it ab
    initio} results for cluster-sizes up to 5 adatoms. Blue squares
  describe the experimental values, black diamonds show represent
  collinear configurations, red circles correspond to non-collinear
  configurations and a black line connects the ratios obtained in the
  magnetic ground states. Reprinted with permission 
from Ref.\cite{lounis08}. Copyright 2008 by the IOPscience.} \label{seff-curve}
\end{center}
\end{figure}

It is interesting to compare the aforementioned theoretical results to
measurements determined using X-ray magnetic circular dichroism
(XMCD). This type of experiments allows to determine the of spin or
moment, $M_z$, per number of d-holes, $n_d$. The experiments were
performed by the group of Wurth~\cite{lounis08} on
size-selected Cr ad-clusters deposited by soft-landing techniques on
Fe$_{3\rm ML}$/Cu(001). The nature of the experiment is such that a
large surface area is sampled, and also only the size of the clusters
is known, but not the shape. Therefore the experimental result is a
statistical average over a number of (unknown) cluster shapes. After
applying XMCD sum rules~\cite{Laan99}, the Hamburg group of
Wurth~\cite{lounis08} arrived at the result shown in
Fig.~\ref{seff-curve}.

One notices the strong decrease of $M_{z}/n_d$ with increasing cluster
size which is due to the appearance of antiferromagnetic or
non-collinear structures as calculated by theory. The qualitative and
quantitative trends observed in the experimental results agree well
with the theoretical results (Fig.~\ref{seff-curve}(b)) for Cr-atoms
to Cr-pentamers. Although the theoretical values for Cr-atoms and
dimers lie somewhat higher than the experimental values (including
error bars) the agreement can still be judged to be very good in view
of the remaining experimental uncertainties. Even details as the
increase of spin magnetic moment from trimer to tetramer can be
addressed. In order to understand the peak formed for the tetramer we
compare in Fig.~\ref{seff-curve}.(b) the ratio $M_z/n_d$ between the
moment along the $z$-direction (defined by the magnetization of the
substrate) and number of holes per atom obtained by theory for the
different geometrical and spin structures. Black circles show the
ratio calculated by taking into account the collinear solutions and
the red ones show the ratio calculated from the non-collinear
solutions. The black line connects the ratio obtained in the magnetic
ground states.  The non-collinear tetramer 1 has a much lower value
than what was seen experimentally whereas the collinear tetramer 1 and
tetramer 2 give a better description of the kink seen
experimentally. With regards to the small energy difference ($\Delta
E=14.5V$~meV) between the two tetramers considered, one could
interpret the experimental value as resulting from an average of
non-collinear tetramers 1, collinear and non-collinear tetramer 2. We
believe that this explains why the tetramer ratio value is higher than
the one obtained for a trimer.  The trimer and the pentamers are
clearly well described by the theory and fit to the experimental
measurements.

\subsection{Fcc(111) surfaces}

Before discussing the complex spin structures induced by a magnetic surface with 
triangular symmetry, it is instructive to look at the simpler cases of non-magnetic 
substrates. 

\subsubsection{Non-magnetic surfaces}
\begin{figure}[h]
\begin{center}
\includegraphics*[width=0.15\textwidth]{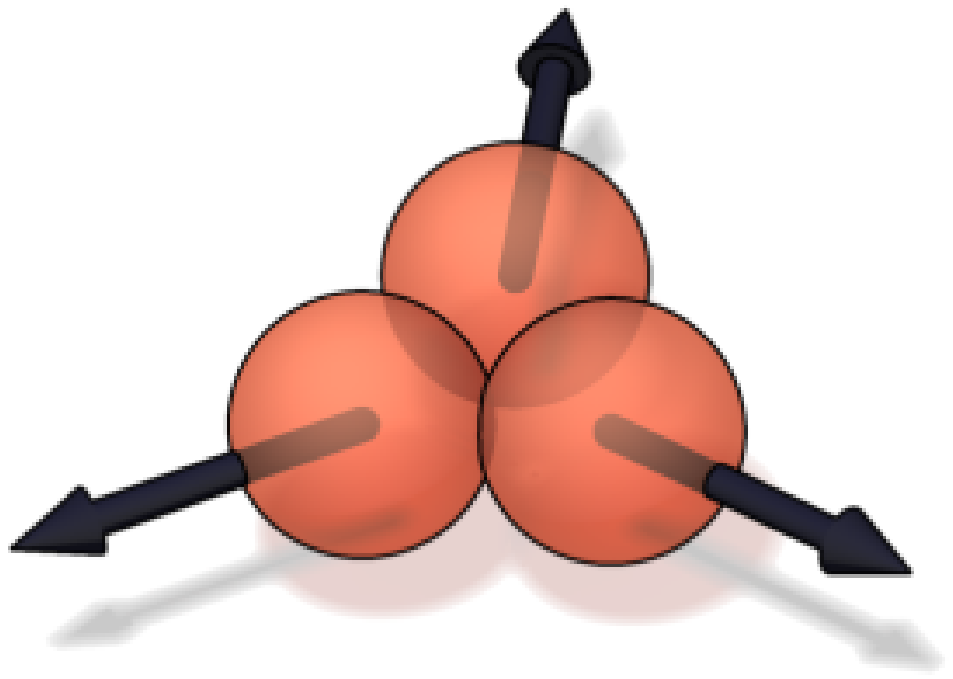}
\includegraphics*[width=0.15\textwidth]{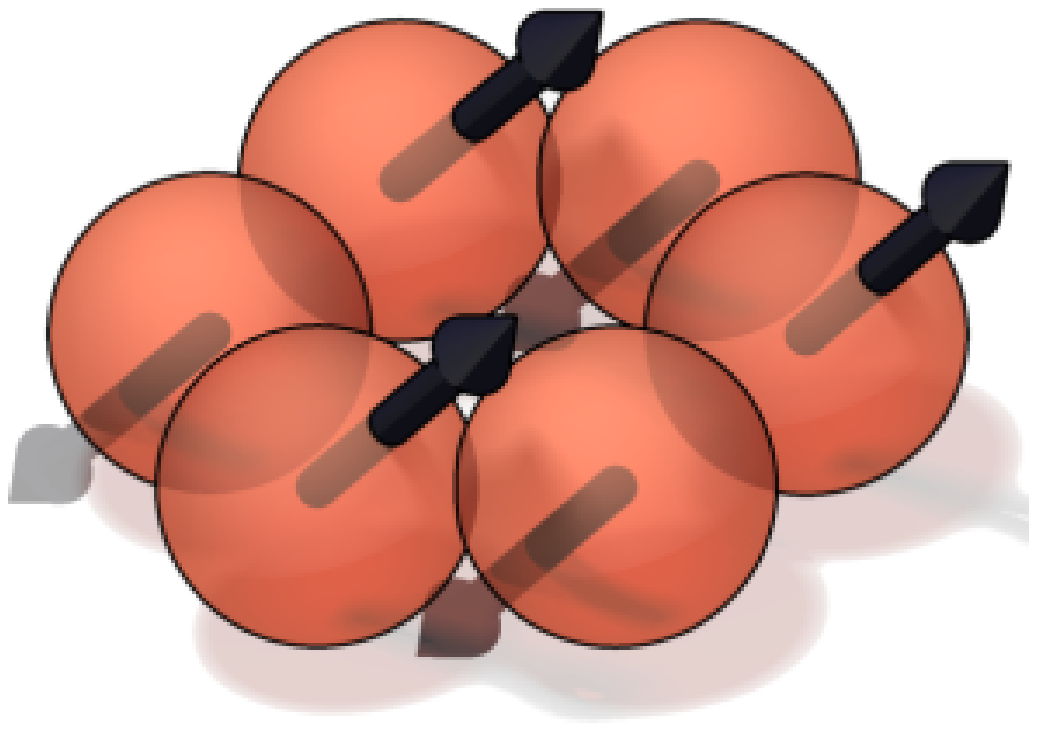}
\includegraphics*[width=0.15\textwidth]{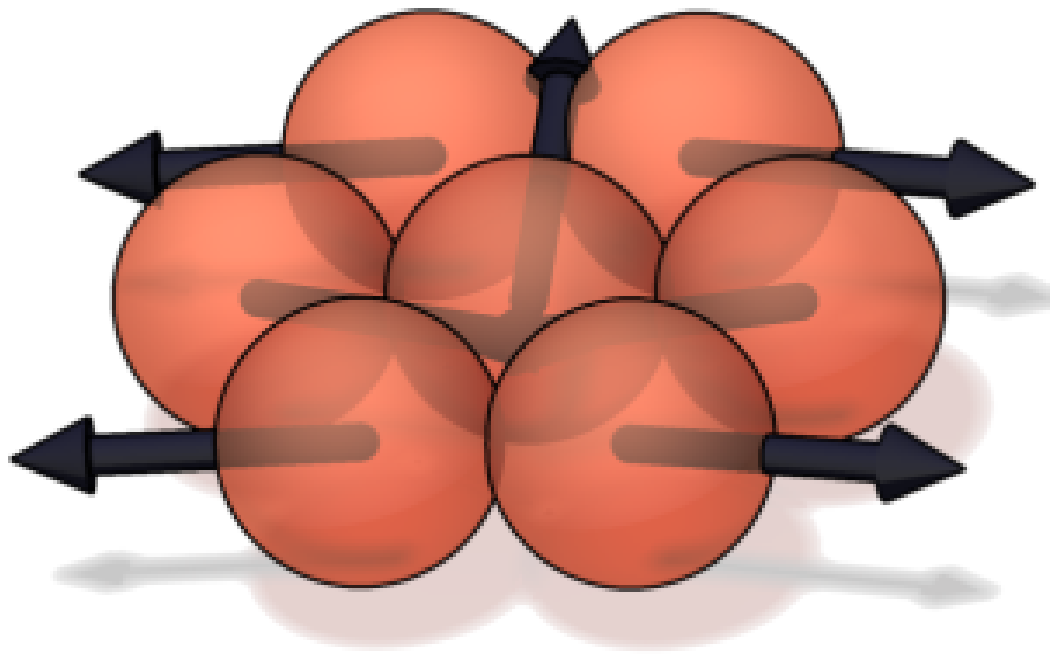}
\caption{\label{bergman} The magnetic ordering for  Mn clusters on a
  Cu(111) surface according to calculations by Bergman et al.~\cite{bergman} (Adapted with permission - Copyright American Physical Society (2006)).}
\end{center}
\end{figure}

Using the RS-LMTO-ASA method, Bergman and co-workers \cite{bergman} found that the ground state for 
the most compact Cr and Mn trimers is the Neel state with a rotation angle of 
120$^\circ$ between the magnetic moments (Fig.~\ref{bergman}(a)). Such a non-collinear structures was also reported from calculations on Cr clusters, with the same geometry, 
supported on Au(111) surface~\cite{gotsis,bergman_gold}.

In Figs.~\ref{bergman}(b) and \ref{bergman}(c) more interesting
structures are explored \cite{bergman}. First six Mn atoms forming a
hexagonal ring structure were studied [Fig~\ref{bergman}(b)]. The
antiferromagnetic nearest neighbor interaction causes a magnetic
order where every second atom has its magnetic moment pointing up and
every other has a moment pointing down, and the magnetic order is
collinear. However, for the cluster with one extra atom in the center
of the hexagonal ring [Fig.~\ref{bergman}(c)] frustration occurs
leading to a different magnetic order with a non-collinear
component. The atoms at the edge of the cluster have a canted
anti-ferromagnetic profile, with a net moment pointing antiparallel to
the magnetization direction of the atom in the center of the
cluster. The magnetic moment of the central atom is almost
perpendicular ($\sim$ 100 $^{\circ}$) to the atoms at the edge of the
cluster and with a magnetic moment of 2.7~$\mu_B$. The edge atoms have 
a magnetic moment of 4~$\mu_B$ each that has an angle of $\sim$
165$^{\circ}$ to neighboring edge atoms and is parallel to its
second nearest neighbors. Interestingly, the non-collinear behavior
does not occur for a Cr sextamer whereas the heptamer is non-collinear.

On the Au(111) surface, Antal and coworkers~\cite{antal} investigated
the magnetic behavior of Cr ad-clusters using a the
fully-relativistic KKR method. They also found the equilateral trimer
to exhibit a frustrated 120$^\circ$ N\'eel type of ground state with
interestingly a small out-of-plane component of the magnetization that
is induced by relativistic effects. In the cases of a linear chain and
an isosceles trimer, collinear antiferromagnetic ground states are
obtained with the magnetization lying parallel to the surface.

An interesting investigation was carried out by Ribeiro {\it et al.}~\cite{klautau_pt} 
for Mn corrals deposited on Pt(111) surface. In the considered structures, the coupling between 
nearest neighbors Mn adatoms is of antiferromagnetic nature and the magnetic ground state is found to be 
collinear and ferrimagnetic. But as soon as an additional Mn adatom is attached 
to the corral, non-collinear orientations of the magnetic 
moments are induced along the whole corrals similarly to the simulations presented in Ref.~\cite{nikos}.

\subsubsection{Magnetic surface: Ni(111)}

\begin{figure}
\begin{center}
{(a)}
\includegraphics*[width=0.3\linewidth]{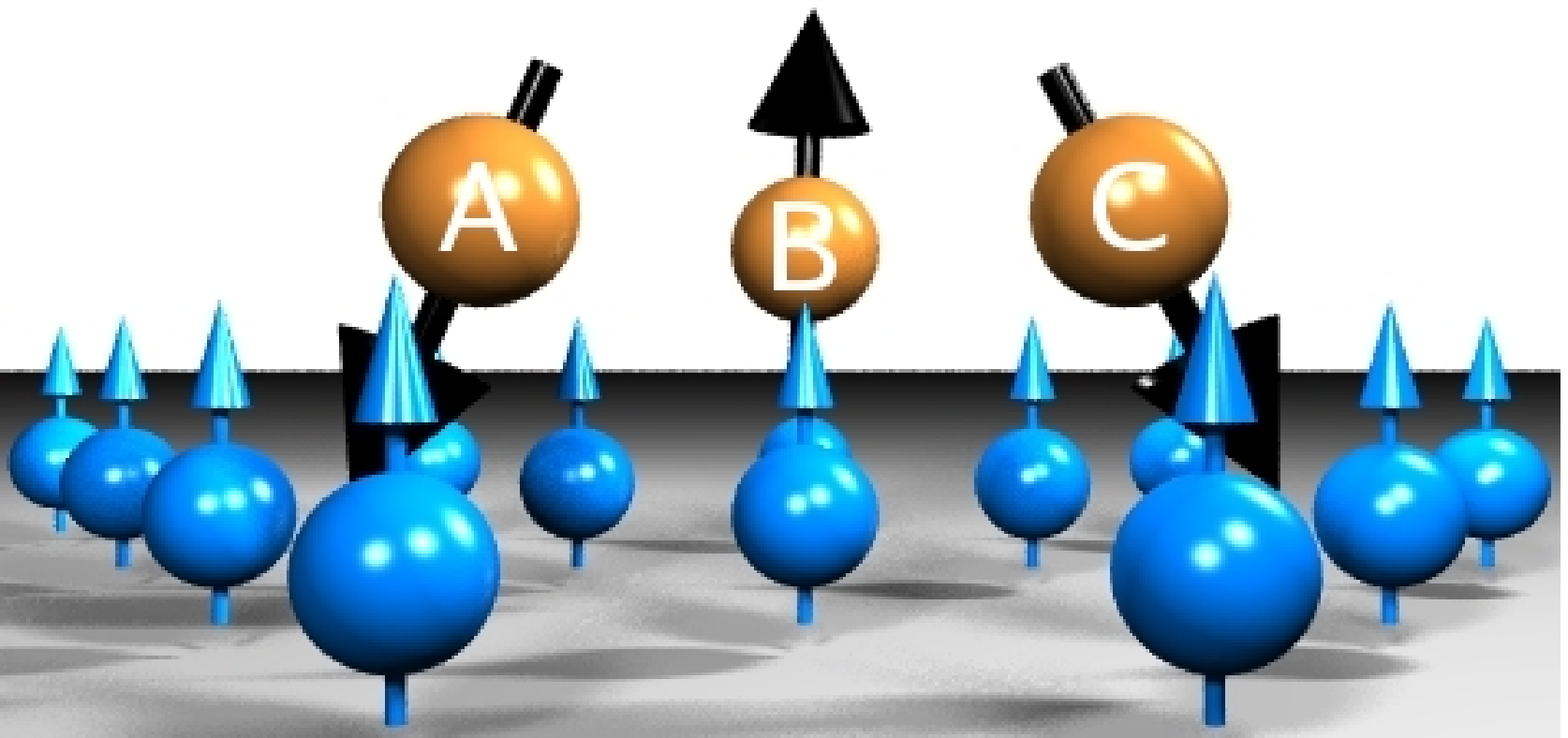}
\hspace{-0.2cm}
{(b)}
\includegraphics*[width=0.3\linewidth]{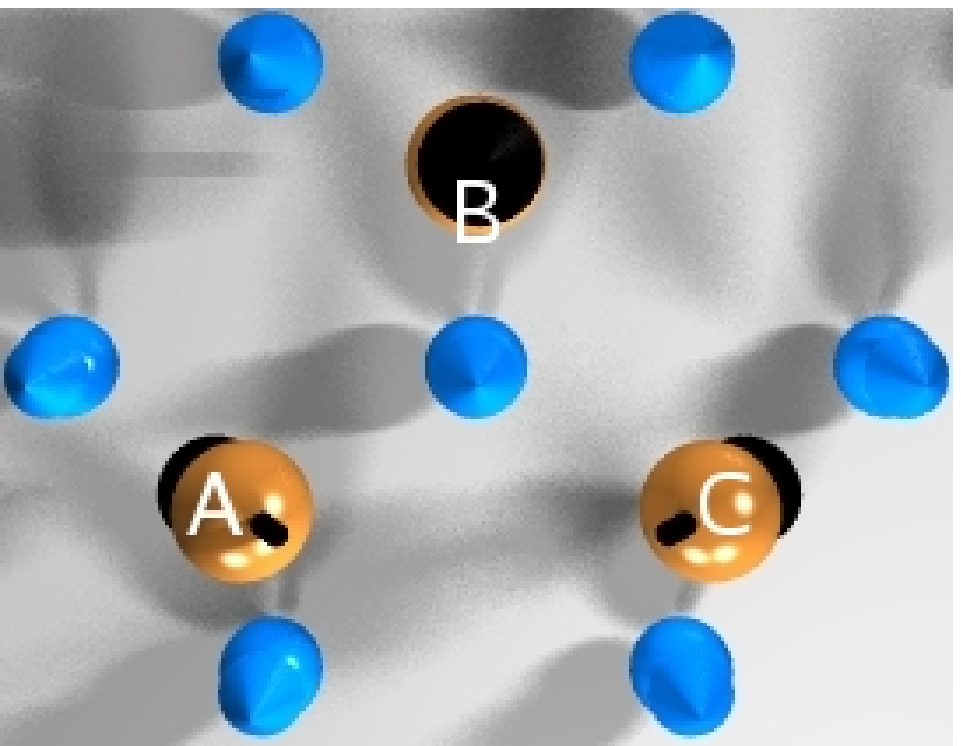}
\\
{(c)}
\includegraphics*[width=0.3\linewidth]{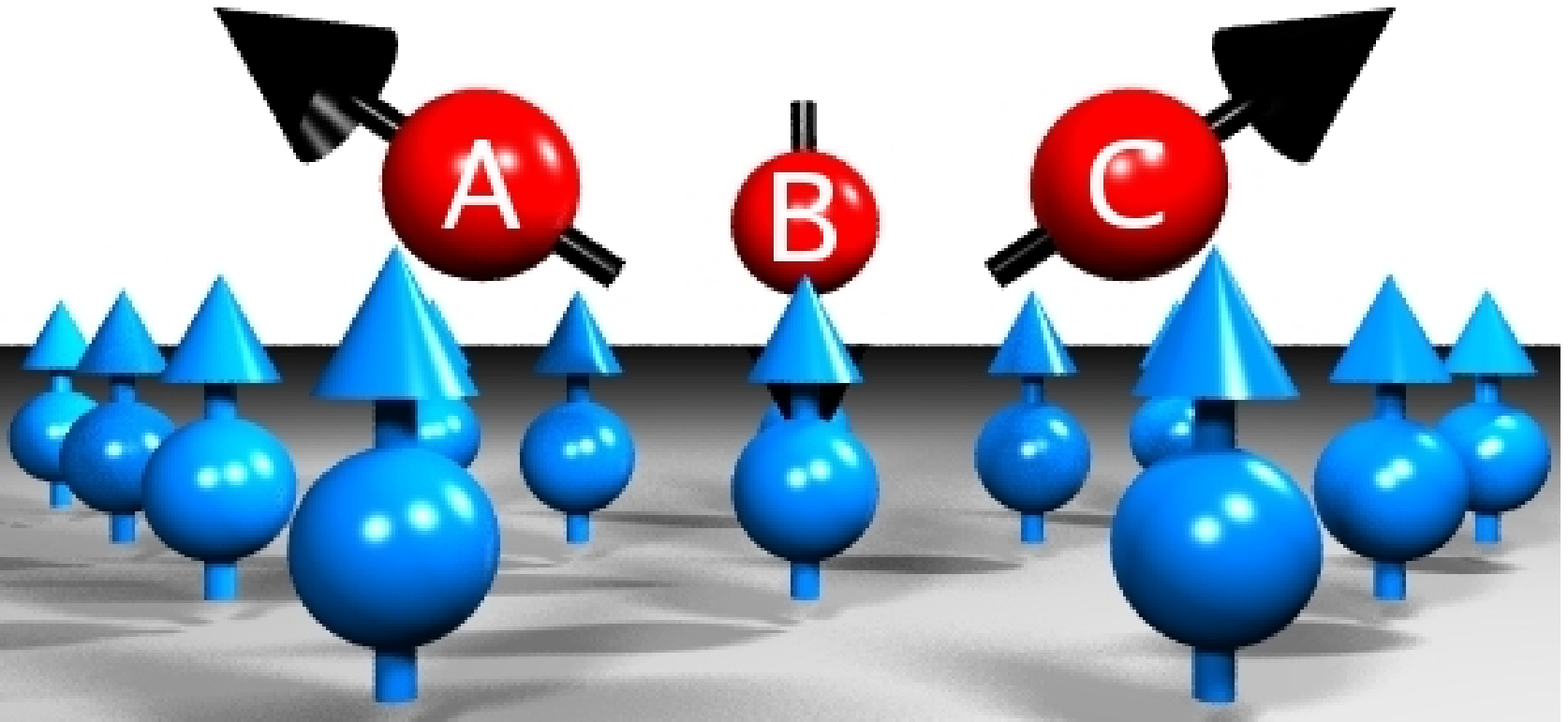}
\hspace{-0.2cm}
{(d)}
\includegraphics*[width=0.3\linewidth]{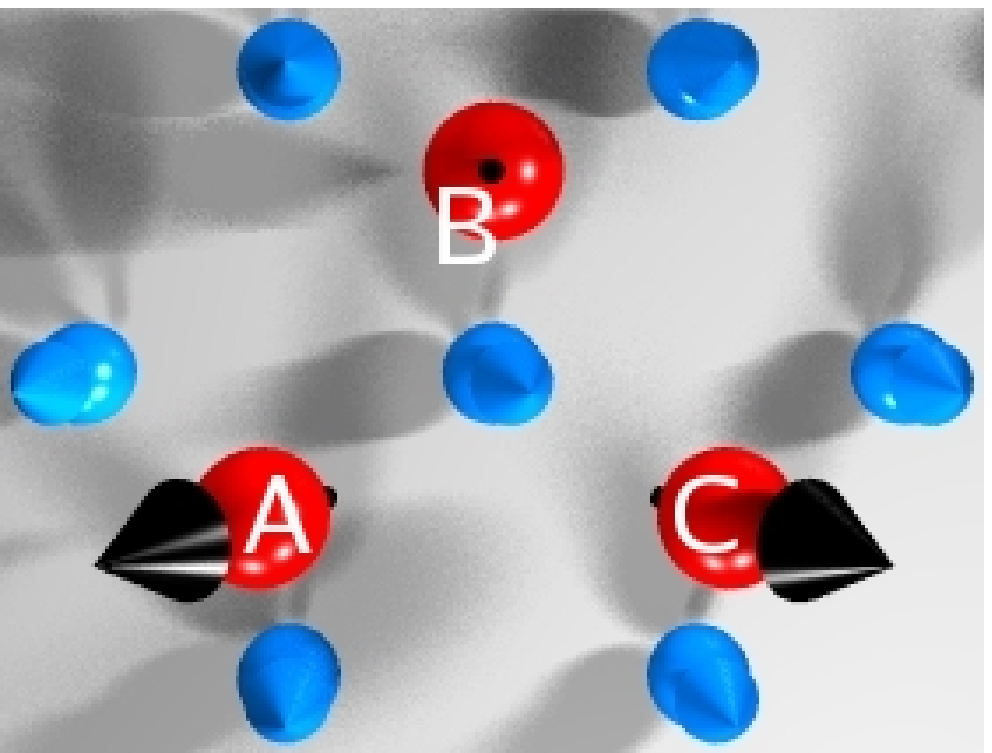}
\\
{(e)}
\includegraphics*[width=0.3\linewidth]{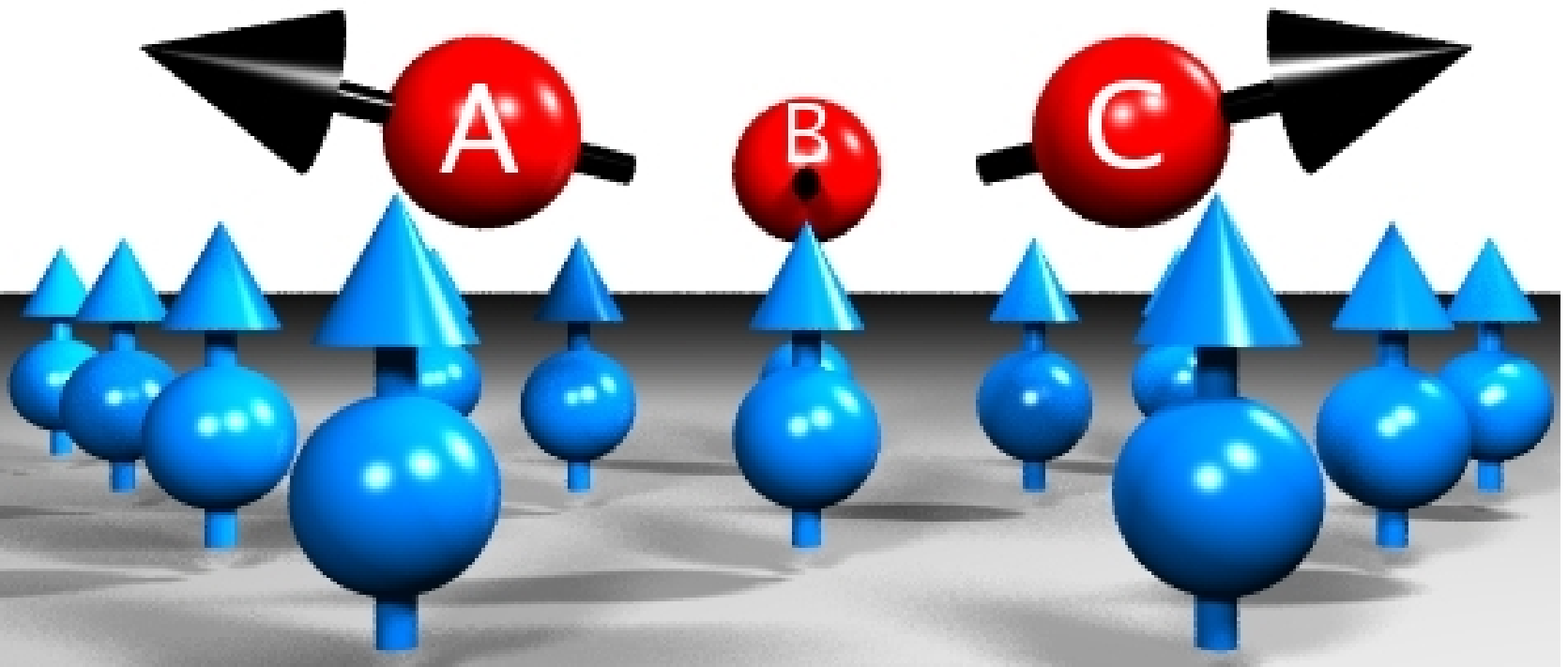}
\hspace{-0.2cm}
{(f)}
\includegraphics*[width=0.3\linewidth]{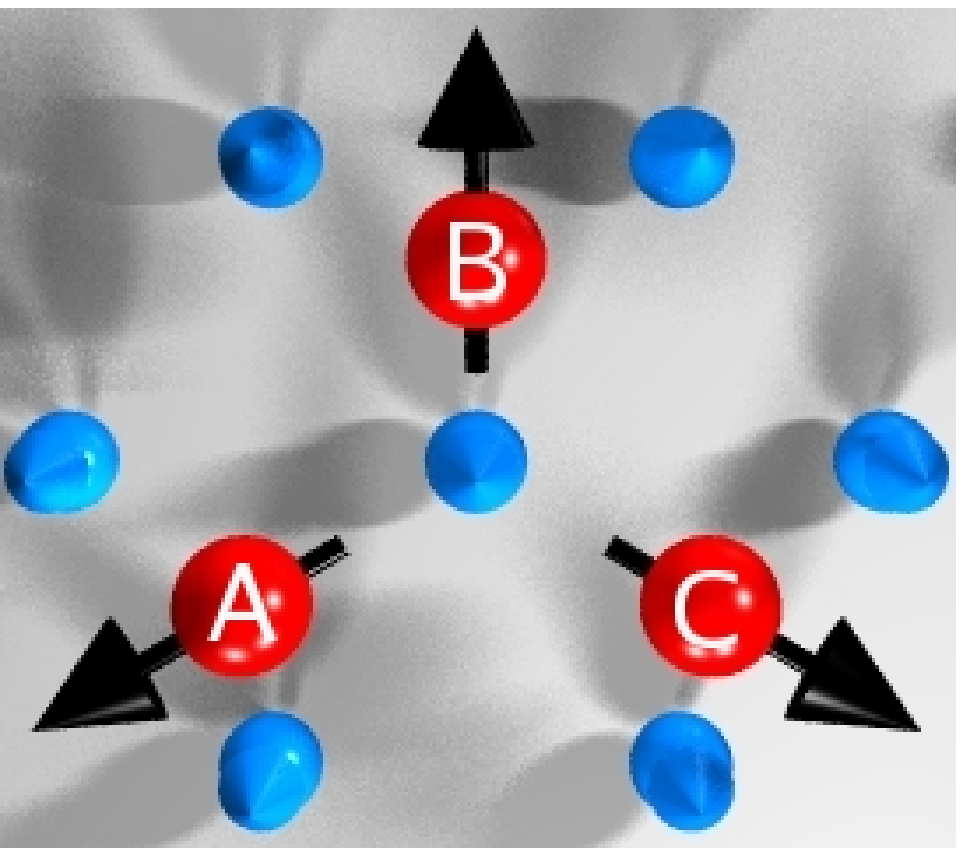}
\\
{(g)}
\includegraphics*[width=0.3\linewidth]{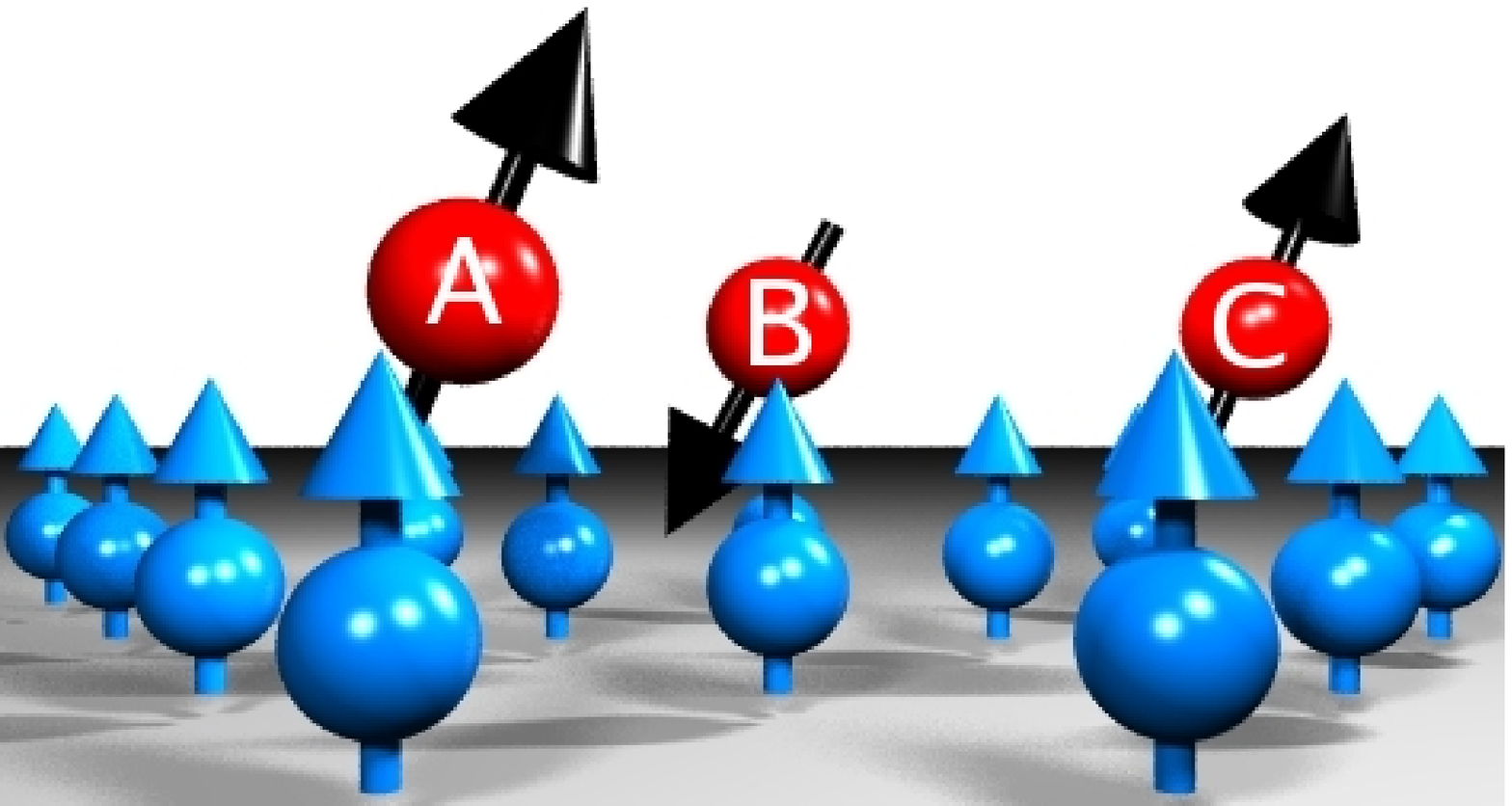}
\hspace{-0.2cm}
{(h)}
\includegraphics*[width=0.4\linewidth]{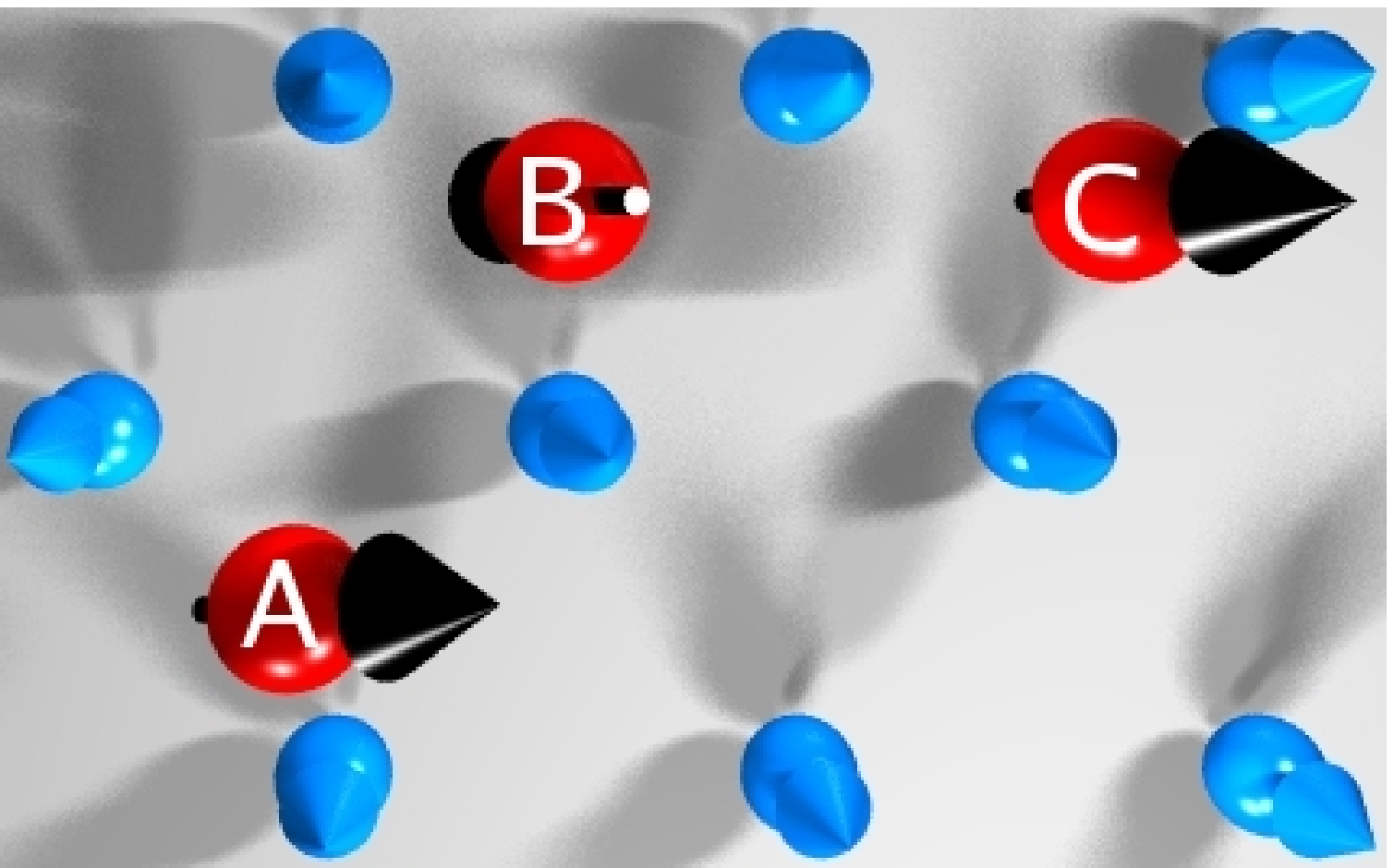}
\caption{Side view (a) and top view (b) for the most stable
  configuration of Cr compact trimer on Ni(111); this corresponds also
  to the NC1 configuration of the Mn compact trimer. (c) and (d)
  represent the side view and top view of the ground state (NC2) of Mn
  compact trimer on Ni(111) while (e) and (f) depict an almost
  degenerate state (NC3) of the same Mn trimer. Finally, the side view
  (g) and top view (h) are shown for the most stable configuration of
  Mn corner trimer on Ni(111). The adatoms are labeled by A, B, and
  C. Unlabeled atoms (in blue) correspond to the Ni
  substrate.  Reprinted with permission 
from Ref.\cite{lounis07}. Copyright 2007 by the American Physical Society.}
\label{trimer-cr-ni111}
\end{center}
\end{figure}

As mentioned previously, trimers in equilateral triangle geometry are,
in the presence of antiferromagnetic interactions, prototypes for
non-collinear magnetism, with the magnetic moments of the three atoms
having an angle of 120$^{\circ}$ to each other. In our case, the
120$^{\circ}$ state is perturbed by the exchange interaction with the
substrate, and therefore the magnetic configuration is expected to be
more complicated.

Let us start with a Cr dimer (Mn dimer) that we approach by a single
Cr adatom (Mn adatom). Three different types of trimers can be formed:
i) the compact trimer with an equilateral shape, ii) the corner trimer
with an isosceles shape and iii) the linear trimer. The adatoms are
named A, B and C, as shown in Fig.~\ref{trimer-cr-ni111} (see Ref.~\cite{lounis07}).

In the most compact trimer, the distance between the three adatoms is
the same leading to a strong intra-cluster frustration.  This is
attested for the Cr case for which we had difficulties finding a
collinear solution. Our striking result, as depicted in
fig.~\ref{trimer-cr-ni111}a-b, is that the non-collinear 120$^{\circ}$
configuration is conserved with a slight modification.  Indeed, our
self-consistent ($\theta$, $\phi)$-angles are ($2^{\circ}, 0^{\circ}$)
for adatom B and ($126^{\circ}, 0^{\circ}$) for adatom A and
($122^{\circ}, 180^{\circ}$) for adatom C. The angle between B and A
is equal to the angle between B and C (124$^{\circ}$) while the angle
between A and C is 112$^{\circ}$. The small variation from the
prototypical 120$^\circ$ configuration is due to the additional
exchange interaction with Ni atoms of the surface.  The coupling with
the substrate leads thus to a deviation from the prototype 120$^\circ$
state, with an additional rotation of $2^{\circ}$ for the FM adatom
and of 4$^{\circ}$ for the two other adatoms. In the picture of the
120$^\circ$-state in a non-magnetic substrate [e.g.,
Fig.~\ref{bergman}(a)], the moments are usually shown as if they were
parallel to the surface. However, any rotation is allowed (if
spin-orbit coupling is neglected), as long as it is the same for all
moments so that their directions relative to each other are the same.
Here the situation is different: the substrate magnetization forces a
particular absolute choice of directions, while the relative angle
between the trimer moments is not changed much.

For the compact Mn trimer, three non-collinear configurations were
obtained: As in the case of the compact Cr trimer, the free Mn trimer
must be in a 120$^{\circ}$ configuration. Nevertheless, the magnetism
of the substrate changes this coupling taking into account the single
adatom behavior: Mn adatoms prefer a FM coupling to the substrate and
an AF coupling with their neighboring Mn adatom. 

The first non-collinear magnetic configuration (NC1) is similar to the
Cr one (Fig.~\ref{trimer-cr-ni111}(a)-(b)), {\it i.e.} the moment of
adatom B is oriented FM (3.61 $\mu_B$) to the substrate moments while
the moments A (3.67 $\mu_B$) and C (3.67 $\mu_B$) are rotated into the
opposite direction with an angle of 114$^{\circ}$ between B and A and
between B and C.

The second non-collinear configuration (NC2) has the opposite magnetic
picture (Fig.~\ref{trimer-cr-ni111}(c)-(d)) as compared to compact Cr
trimer.  Moments A and C are oriented FM to the substrate, with a
tilting of $\theta$ = 49$^{\circ}$, $\phi$ = 0$^{\circ}$ for atom A
and $\theta$ = 51$^{\circ}$, $\phi$ = 180$^{\circ}$ for atom C; each
of them carries a moment of 3.62 $\mu_B$. The AF interaction of atom B
with A and C forces it to an AF orientation with respect to the
substrate, characterized by $\theta$ = 179$^{\circ}$, $\phi = 0$, and
a moment of 3.70~$\mu_B$. Thus the trimer deviates from the
$120^\circ$-configuration: the angles between A and B moments and B
and C moments are about $130^\circ$, and the angle between A and C is
about $100^\circ$.

In the third magnetic configuration (NC3) the three moments (3.65
$\mu_B$) are almost in-plane and perpendicular to the substrate
magnetization (see Fig.~\ref{trimer-cr-ni111}(e)-(f)). They are also
slightly tilted in the direction of the substrate magnetization
($\theta = 86^{\circ}$) due to the weak FM interaction with the Ni
surface atoms. Within this configuration, the 120$^{\circ}$ angle
between the adatoms is almost sustained.  Total energy calculations
show that the NC2 configuration is the ground state which is almost
degenerate with NC1 and NC3 ($\Delta
\mathrm{E}_{\mathrm{NC1}-\mathrm{NC2}} = 1.3$ meV/adatom and $\Delta
\mathrm{E}_{\mathrm{NC3}-\mathrm{NC2}} = 5.6$ meV/adatom). Thus
already at low temperatures trimers might be found in all three
configurations; in fact the spin arrangement might fluctuate among
these three 120$^{\circ}$ configurations or among the three possible
degenerate configurations corresponding to each of NC1 and NC2 states,
produced by interchanging the moments of atoms A, B and C. Compared to
the collinear state energy of the compact trimer, the NC2 energy is
lower by 138 meV/adatom. This very high energy difference is due to
frustration, even higher than breaking a bond as shown in the next
paragraphs. Contrary to this, the corner trimer shows a collinear
ground state because it is not frustrated.

The next step is to move adatom C and increase its distance with
respect to A in order to reshape the trimer into an isosceles triangle
(what we call ``corner trimer'' in Fig.~\ref{trimer-cr-ni111}(g)-(h)).
By doing this, the trimer loses the frustration and is characterized,
thus, by a collinear FI ground state: the moments of adatoms A and C
are AF oriented to the substrate (following the AF Ni-Cr exchange),
while the moment of the central adatom B is FM oriented to the
substrate, following the AF Cr-Cr coupling to its two neighbors. The
magnetic moments do not change much compared to the compact trimer.

While the non-collinear state is lost for the corner Cr trimer, it is
present for the corner Mn trimer as a local minimum with a tiny energy
difference of 4.8 meV/adatom higher than the FI ground state.
This value is equivalent to a temperature of $\sim$ 56~K, meaning that
at room temperature both configurations co-exist. Here FI means
that the central adatom B is AF oriented to the substrate with a
magnetic moment of 3.71 $\mu_B$, forced by its two FM companions A and
C (moment of 3.83 $\mu_B$) which have only one first neighboring
adatom and are less constrained. The total moment of the ad-cluster is
also high ($3.95 \mu_B$) compared to the compact trimer value,
reaching the value of the non-interacting system (with the third atom
of the trimer far away from the other two).

The FI solution is just an extrapolation of the non-collinear
solution shown in Fig.~\ref{trimer-cr-ni111}(g)-(h) (with magnetic moments
similar to the collinear ones) in which the central adatom B (3.70
$\mu_B$) tends to orient its moment also AF to the substrate ($\theta$
= 152$^{\circ}$, $\phi$ = 0$^{\circ}$) and the two other adatoms with
moments of 3.83 $\mu_B$ tend to couple FM to the surface magnetization
with the same angles ($\theta$ = 23$^{\circ}$, $\phi$ =
180$^{\circ}$).  It is important to point out that the AF coupling
between these two latter adatoms is lost by increasing the distance
between them. Indeed, one sees in Fig.~\ref{trimer-cr-ni111}(g)-(h) that
the two moments are parallel.

It is interesting to compare the total energies of the three trimers
we investigated. The compact trimer has more first neighboring bonds
and is expected to be the most stable trimer. The energy differences
confirm this statement. Indeed the total energy of the Cr compact
trimer is 119 meV/adatom lower than the total energy of the corner
trimer. Similarly, the Mn compact trimer has a lower energy of 53
meV/adatom compared to the corner trimer.

\begin{figure}[ht!]
\begin{center}
{(a)}
\includegraphics*[width=0.35\linewidth]{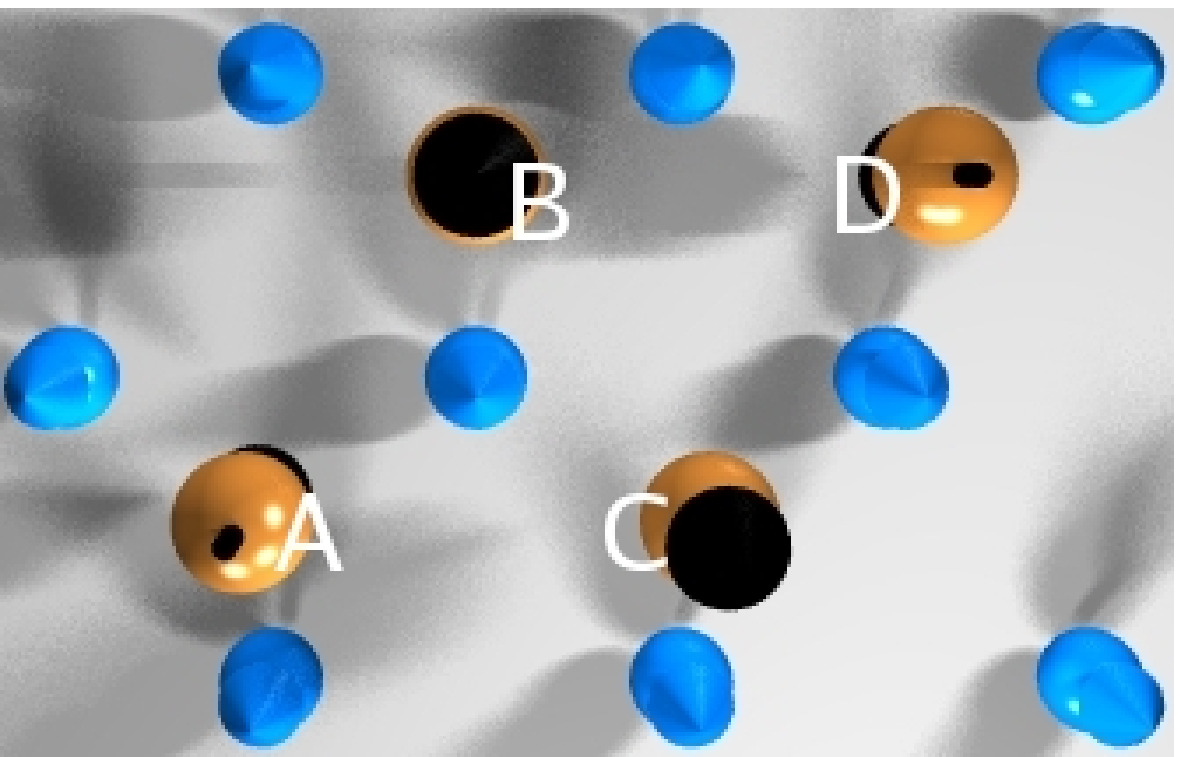}\\
{(b)}
\includegraphics*[width=0.35\linewidth]{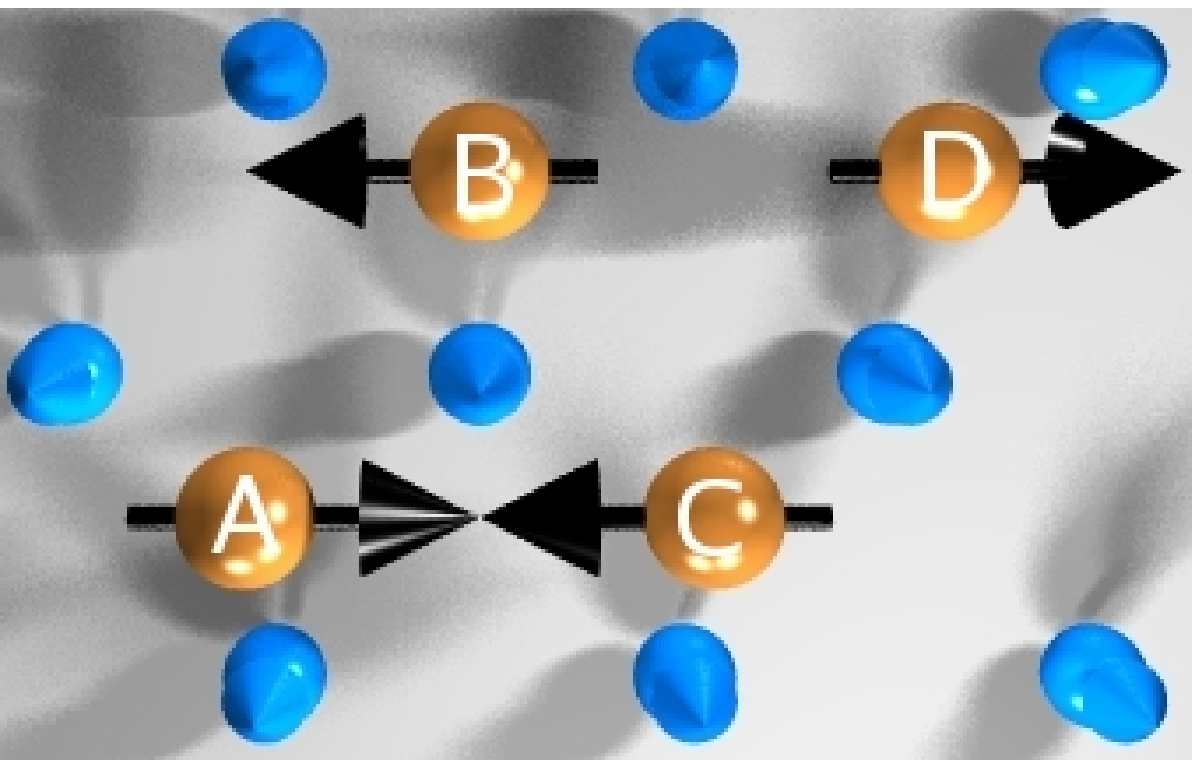}
\\
{(c)}
\includegraphics*[width=0.35\linewidth]{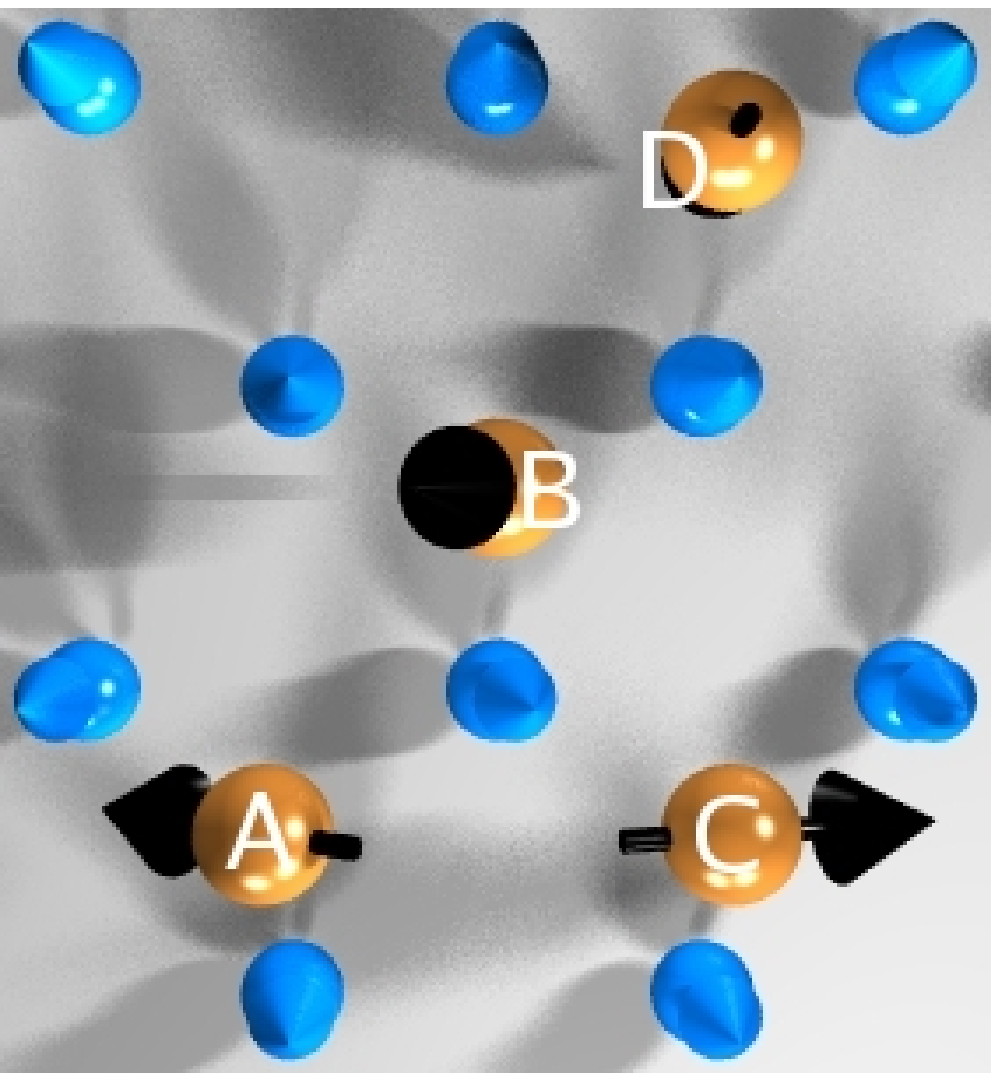}
\end{center}
\caption{Top view of the collinear most stable solution
(a) and the non-collinear metastable configuration (b) of compact
Cr~tetramer on Ni(111). In (c) is depicted the Cr~tetramer-b magnetic
ground state on Ni(111), which basically consists of the non-collinear
trimer state of Fig.\ref{trimer-cr-ni111}(a) coupled
antiferromagnetically to the fourth adatom. The adatoms are labeled
by A, B, C, and D. Unlabeled atoms (in blue) correspond
to the Ni substrate. Reprinted with permission 
from Ref.\cite{lounis07}. Copyright 2007 by the American Physical Society.}
\label{tetramer-cr-ni111}
\end{figure}

Finally we discuss the case of tetramers. We consider two types of
tetramers, formed by adding a Cr or Mn adatom (atom D in
Fig.~\ref{tetramer-cr-ni111}) to the compact trimer. We begin with the
compact tetramer (Fig.~\ref{tetramer-cr-ni111}(a)-(b)). For both
elements Cr and Mn, the FI solution is the ground state
(Fig.~\ref{tetramer-cr-ni111}(a)). The Cr tetramer, in particular,
shows also a non-collinear configuration
(Fig.~\ref{tetramer-cr-ni111}(b)) as a local minimum which has,
however, a slightly higher energy of $\Delta
\mathrm{E}_{\mathrm{NC}-\mathrm{FI}} = 1$ meV/adatom. Within this
configuration the AF coupling between the adatoms is observed.
However, the four moments are almost in-plane perpendicular to the
substrate magnetization.

An additional manipulation consists in moving adatom D and forming a
tetramer-b [Fig.~\ref{tetramer-cr-ni111}(c)]. For such a structure,
the collinear solution for the Cr tetramer is only a local minimum. In
this structure, atom D has less neighboring adatoms compared to A, B,
and C. In the non-collinear solution which is the magnetic ground
state, the moment of adatom D (3.34 $\mu_B$) is almost AF oriented to
the substrate ($\theta$ = 178$^{\circ}$, $\phi$ = 0$^{\circ}$). The
remaining adatoms form a compact trimer in which the closest adatom to
D, {\it i.e.} B, tends to orient its moment FM (2.45 $\mu_B$) to the
substrate ($\theta$ = 19$^{\circ}$, $\phi$ = 0$^{\circ}$) while the
moments of A (2.90 $\mu_B$) and C (2.80 $\mu_B$) tend to be oriented
AF ($\theta_{\mathrm{A}}$ = 124$^{\circ}$, $\phi_{\mathrm{A}}$ =
0$^{\circ}$) and ($\theta_{\mathrm{C}}$ = 107$^{\circ}$,
$\phi_{\mathrm{C}}$ = 180$^{\circ}$). In the (metastable) collinear
solution for this tetramer, the moment of adatom B is oriented FM to
the substrate while the moments of all remaining adatoms are oriented
AF to the surface atoms. The total energy difference between the two
configurations is equal to 49 meV/adatom. Compared to the total
energy of the compact tetramer, our calculations indicate that the
tetramer-b has a higher energy (109 meV/adatom).

Let us now turn to the case of the Mn tetramer-b. Also here, the
non-collinear solution is the ground state while the collinear one is
a local minimum. The energy difference between the two solutions is
very small (2.8 meV/adatom). The moments are now rotated to the
opposite direction compared to the Cr case, in order to fulfill the
magnetic tendency of the single Mn adatom which is FM to the
substrate. The Mn atom with less neighboring adatoms, {\it i.e.} D,
has a moment of 3.84 $\mu_B$ rotated by ($\theta$ = 27$^{\circ}$,
$\phi$ = 0$^{\circ}$), while its closest neighbor, atom B with a
moment of 3.44 $\mu_B$, is forced by the neighboring companions to
orient its moment AF ($\theta$ = 140$^{\circ}$, $\phi$ = 180$^{\circ}$). The
adatoms A and C with similar magnetic moments (3.63 $\mu_B$) obtain
a FM orientation with the following angles: ($\theta$ = 81$^{\circ}$, $\phi$
= 0$^{\circ}$) and ($\theta$ = 34$^{\circ}$, $\phi$ = 0$^{\circ}$). As
in the case of Cr tetramer-b, the converged collinear solution is just
the extreme extension of the non-collinear one: The ``central'' moment
of the tetramer is forced by its FM Mn neighboring atoms to be AF to
the substrate. The magnetic regime is similar to the one of Cr
tetramer-b, {\it i.e.} high, with a total magnetic moment of 4.37
$\mu_B$. As expected, the most stable tetramer is the compact one,
with an energy of 52 meV/adatom lower than tetramer b.

\section{Summary and concluding remarks}

In summary, we have reviewed recent work on the ab-initio
investigation of complex spin-structures in ad-clusters deposited on
magnetic surfaces. We have discussed prototype systems where different
kinds of frustration exist: (i) frustration within the cluster and (ii)
frustration arising from antiferromagnetic coupling between the
adatoms in the cluster and competing magnetic interactions between the
clusters and the surface atoms. We see that frustration results in
non-collinear magnetic configuration on a length scale of
nearest-neighbor distances. The energy scale of the directional
relaxation of the magnetic moments with respect to the frustrated
state can be comparable to the cohesive energy of the cluster.

In most of these cases, the present local density functional
calculations give more than one energy minima, corresponding to
different non-collinear states, that can be energetically very close
(with differences of a few meV/atom). In these situations the system
can easily fluctuate between these states. Naturally the relative
energy values that were shown here can change if one corrects for the
approximations that we used (neglect of spin-orbit coupling and
structural relaxations, and use the local spin density approximation
for the exchange-correlation energy), in particular as regards energy
differences of the order of a few meV. However, the conclusion of
existence of multiple almost degenerate magnetic states is expected to
hold. 

It is demonstrated in several occasions that the position of a single
adatom within a nanostructure or the addition of an atom to a
nanostructure provides a strong magnetic perturbation to the whole
nano-entity.  One could even envision adatoms acting as local magnetic
switches, which via the local magnetic exchange field of the single
adatom allow to switch the total moment on and off, and which
therefore might be of interest for magnetic storage. This mechanism
has been recently used experimentally to build magnetic logic
gates~\cite{alex_science}. Thus, magnetic frustration could be useful
for future nanosize information storage.

\section*{Acknowledgments}

It is a pleasure to thank the contributors to most of the work
presented in this review: Ph. Mavropoulos, R. Zeller, P. H. Dederichs and S. Bl\"ugel.
We thank the contribution of our experimental colleagues: M. Reif, L. Glaser, M. Martins and W. Wurth 
on the XMCD measurements of Cr clusters on 
Fe$_{3ML}$/Cu(001) surface. 
Also, we would like to thank A. Bergman for providing his figures and
results on clusters deposited on Cu(111) surface and acknowledge the
support of the HGF-YIG Programme VH-NG-717 (Functional Nanoscale Structure and Probe 
Simulation Laboratory, Funsilab).

\end{document}